%% file: FusionMaster.tex
\documentstyle[12pt]{article}
\input epsf 
\renewcommand{\title}[1]{\large\bf \mbox{}\\ \mbox{}\\ \mbox{}\\ \mbox{}\\
     #1\bigskip\medskip\\} 
\renewcommand{\author}[1]{\large #1\\ \smallskip}
\newcommand{\address}[1]{{\narrower\normalsize\it #1\\}\bigskip}
\newcommand{\zr}[1]{\mbox{\hspace*{#1em}}}
\newcommand{\Z}{\mbox{\sf Z\zr{-0.45}Z}}
\newcommand{\hs}[1]{\hspace*{#1cm}}

\newtheorem{theorem}{Theorem}
\newtheorem{lemma}{Lemma}
\def\as{\quad}
\def\and{\quad\mbox{and}\quad}
\def\ade{$A$--$D$--$E$\space}
\def\wt#1#2#3#4#5#6{#1\!\!\mbox{
 $\left(\matrix{#5&#4\cr#2&#3\cr}\biggm|\mbox{$#6$}\right)$}}
  
  \def\dddots{\mathinner{\mkern1mu\raise1pt\vbox{\kern1pt\hbox{.}}
                       \mkern2mu\raise4pt\hbox{.}
                       \mkern2mu\raise7pt\hbox{.}\mkern1mu}}
\def\wf#1#2#3#4#5#6#7#8#9{#1\mbox{$\left(
   \matrix{#5&#8&#4\cr#9&&#7\cr a&#6&#3\cr}\Biggm|\mbox{$#2$}\right)$}}
\def\Wf#1#2#3#4#5#6#7#8#9{W_{m\times n}\mbox{$\left(
   \matrix{#5&#8&#4\cr#9&&#7\cr #1&#6&#3\cr}\biggm|\mbox{$#2$}\right)$}}
\def\Wfml#1#2#3#4#5#6#7#8#9{W_{m\times l}\mbox{$\left(
   \matrix{#5&#8&#4\cr#9&&#7\cr #1&#6&#3\cr}\biggm|\mbox{$#2$}\right)$}}
\def\Wfln#1#2#3#4#5#6#7#8#9{W_{l\times n}\mbox{$\left(
   \matrix{#5&#8&#4\cr#9&&#7\cr #1&#6&#3\cr}\biggm|\mbox{$#2$}\right)$}}
\def\row#1#2#3#4#5#6#7#8{#1\mbox{$\left(
   \matrix{#6&#8&#5\cr #3&#7&#4\cr}\biggm|\mbox{$#2$}\right)$}}

\def\mat {\pmatrix}
\def\smat#1{\mbox{\small $\mat{#1}$}}
\def\ba {\begin{array}}
\def\ea {\end{array}}
\def\be {\begin{equation}}
\def\ee {\end{equation}}
\def\bea {\begin{eqnarray}}
\def\eea {\end{eqnarray}} 
\def\no{\nonumber}
\def\disp{\displaystyle}
\def\ol#1{\overline{#1}}
\def\sc{\scriptsize}
\def\ra{\raisebox{0pt}[0pt][0pt]}
\def\-{\!-\!}
\def\+{\!+\!}

\def\for{\;\;{\rm for}\;}
\def\h{\hspace*{0.5cm}}
\def\mh{\hspace*{-0.5cm}}
\def\v{\vspace*{-0.1cm}}
\def\val#1{\mathop{\rm val}(#1)}
\def\path{p}
\def\sym#1{\mathop{\rm sym}(#1)}
\def\smaller{\small}
\def\face#1#2#3#4#5{
  \begin{picture}(40,30)(-5,13)
  \thicklines
  \put(15,5){\makebox(0,0)[b]{\line(1,0){20}}} 
  \put(5,5){\makebox(0,0)[b]{\line(0,1){20}}}   
  \put(15,25){\makebox(0,0)[b]{\line(-1,0){20}}} 
  \put(25,25){\makebox(0,0)[b]{\line(0,-1){20}}} 
  \put(5,2){\makebox(0,0)[rb]{\sc $#1$}} \put(26,2){\sc $#2$} 
  \put(25,25){\sc $#3$}\put(5,25){\makebox(0,0)[rb]{\sc $#4$}} 
  \put(13,13){\sc $#5$} 
 \end{picture}}

\def\diagface#1#2#3#4#5#6#7{
\setlength{\unitlength}{.0943in}
\rule[-4.3\unitlength]{0in}{8.6\unitlength}
\begin{picture}(9,4)(-#6,-#7)
\put(1.5,0.5){\line(1,-1){3}}
\put(4.5,3.5){\line(1,-1){3}}
\put(4.5,3.5){\line(-1,-1){3}}
\put(7.5,0.5){\line(-1,-1){3}}
\put(1.2,0.5){\makebox(0,0)[r]{\smaller \mbox{$#1$}}}
\put(4.5,-2.8){\makebox(0,0)[t]{\smaller \mbox{$#2$}}}
\put(7.8,0.5){\makebox(0,0)[l]{\smaller \mbox{$#3$}}}
\put(4.5,3.8){\makebox(0,0)[b]{\smaller \mbox{$#4$}}}
\put(4.5,0.5){\makebox(0,0){\smaller \mbox{$#5$}}}
\end{picture}}

\def\YBR#1#2#3#4#5#6#7#8#9{\raisebox{-.5in}{
\begin{picture}(300,85)(-10,-15)
\thicklines
\multiput(60,60)(0,-60){2}{\line(1,0){40}} 
\multiput(60,60)(60,-30){2}{\line(-2,-3){20}}
\multiput(40,30)(60,30){2}{\line(2,-3){20}} 
\put(40,30){\line(1,0){40}} \put(100,60){\line(-2,-3){20}} 
\put(100,0){\line(-2,3){20}}\put(80,30){\circle*{5}}
\put(140,30){$=$}
\multiput(190,60)(0,-60){2}{\line(1,0){40}} 
\multiput(190,60)(60,-30){2}{\line(-2,-3){20}}
\multiput(170,30)(60,30){2}{\line(2,-3){20}} 
\put(190,60){\line(2,-3){20}} \put(190,0){\line(2,3){20}} 
\put(250,30){\line(-1,0){40}}
\put(210,30){\circle*{5}}  
\multiput(57,-7)(130,0){2}{\makebox(0,0)[lb]{\ra{\twlrm \sc $#1$}}}
\multiput(100,-7)(130,0){2}{\makebox(0,0)[lb]{\ra{\twlrm \sc $#2$}}}
\multiput(120,28)(130,0){2}{\makebox(0,0)[lb]{\ra{\twlrm \sc $#3$}}}
\multiput(101,63)(130,0){2}{\makebox(0,0)[lb]{\ra{\twlrm \sc $#4$}}}
\multiput(57,63)(130,0){2}{\makebox(0,0)[lb]{\ra{\twlrm \sc $#5$}}}
\multiput(35,28)(130,0){2}{\makebox(0,0)[lb]{\ra{\twlrm \sc $#6$}}}
\put(68,43){\makebox(0,0)[lb]{\ra{\twlrm \sc $#7$}}} 
\put(95,28){\makebox(0,0)[lb]{\ra{\twlrm \sc $#8$}}} 
\put(68,13){\makebox(0,0)[lb]{\ra{\twlrm \sc $#9$}}} 
\put(218,43){\makebox(0,0)[lb]{\ra{\twlrm \sc $#9$}}} 
\put(185,28){\makebox(0,0)[lb]{\ra{\twlrm \sc $#8$}}} 
\put(218,13){\makebox(0,0)[lb]{\ra{\twlrm \sc $#7$}}} 
\end{picture}}}

\def\onebytwoface#1#2#3#4#5#6#7#8#9{\vspace{-.2in}
\begin{picture}(50,20)(-5,13)
\thicklines
\put(9,13){\makebox(10,0)[b]{\sc $#7$}}
\put(33,13){\makebox(10,0)[b]{\sc $#8$}}
\multiput(5,5)(0,20){2}{\line(1,0){40}} 
\multiput(5,5)(20,0){3}{\line(0,1){20}}
\multiput(25,5)(0.40000,-0.40000){9}{\makebox(0.4444,0.6667)#9}
\multiput(25,5)(-0.40000,0.40000){9}{\makebox(0.4444,0.6667)#9}
\multiput(25,5)(0.40000,0.40000){9}{\makebox(0.4444,0.6667)#9}
\multiput(25,5)(-0.40000,-0.40000){9}{\makebox(0.4444,0.6667)#9}
\put(3,2){\makebox(0,0)[rb]{\sc $#1$}} \put(45,2){\sc $#2$} 
\put(45,26){\sc $#3$} \put(3,26){\makebox(0,0)[rb]{\sc $#4$}}
\put(21,-2){\makebox(10,0)[b]{\sc $#5$}}
\put(21,28){\makebox(10,0)[b]{\sc $#6$}}  \put(48,14){}
\end{picture}}
\def\onerowprop#1#2#3{\vspace*{.5cm}
\setlength{\unitlength}{0.0125in}%
\begin{picture}(270,60)(18,775)
\thicklines
\put(180,795){\line( 0,-1){ 30}}
\put(210,795){\line( 0,-1){ 30}}
\put(150,795){\line( 0,-1){ 30}}
\put( 90,795){\line( 0,-1){ 30}}
\put( 90,765){\line( 1, 0){195}}
\put(285,765){\line( 0, 1){ 30}}
\put(285,795){\line(-1, 0){195}}
\put( 84,756){\makebox(0,0)[lb]{\raisebox{0pt}[0pt]
[0pt]{\twlrm \sc $a$}}}
\put(288,753){\makebox(0,0)[lb]{\raisebox{0pt}[0pt]
[0pt]{\twlrm \sc $b$}}}
\put(288,795){\makebox(0,0)[lb]{\raisebox{0pt}[0pt]
[0pt]{\twlrm \sc $c$}}}
\put( 83,795){\makebox(0,0)[lb]{\raisebox{0pt}[0pt]
[0pt]{\twlrm \sc $d$}}}
\put( 18,774){\makebox(0,0)[lb]{\raisebox{0pt}[0pt]
[0pt]{\twlrm  $\disp{\sum_j^{P_{(a,\!b)}^{(n)}}} 
\disp{\phi^{(j,\alpha)}_{(a,b,n)}}$}}}
\put(192,777){\makebox(0,0)[lb]{\raisebox{0pt}[0pt]
[0pt]{\twlrm \sc $u$}}}
\put(155,777){\makebox(0,0)[lb]{\raisebox{0pt}[0pt]
[0pt]{\twlrm \sc $u\-\lambda$}}}
\put(170,798){\makebox(0,0)[lb]{\raisebox{0pt}[0pt]
[0pt]{\twlrm \sc $#2$}}}
\put(158,798){\makebox(0,0)[rb]{\raisebox{0pt}[0pt]
[0pt]{\twlrm \sc $#1$}}}
\put(201,798){\makebox(0,0)[lb]{\raisebox{0pt}[0pt]
[0pt]{\twlrm \sc $#3$}}}
\put(161,755){\makebox(0,0)[lb]{\raisebox{0pt}[0pt]
[0pt]{\twlrm \tiny $p(a,\!b,\!n)_{j,i}$}}}
\put(115,755){\makebox(0,0)[lb]{\raisebox{0pt}[0pt]
[0pt]{\twlrm \tiny $p(a,\!b,\!n)_{j,i\+1}$}}}
\put(200,755){\makebox(0,0)[lb]{\raisebox{0pt}[0pt]
[0pt]{\twlrm \tiny $p(a,\!b,\!n)_{j,i\-1}$}}}
\end{picture}}
\def\onerowfusion#1#2{
\setlength{\unitlength}{0.0125in}%
\begin{picture}(270,63)(18,774)
\thicklines
\put( 90,795){\line( 0,-1){ 30}}
\put( 90,765){\line( 1, 0){195}}
\put(285,765){\line( 0, 1){ 30}}
\put(285,795){\line(-1, 0){195}}
\put(255,795){\line( 0,-1){ 30}}
\put(225,795){\line( 0,-1){ 30}}
\put(120,795){\line( 0,-1){ 30}}
\put( 84,756){\makebox(0,0)[lb]{\raisebox{0pt}
[0pt][0pt]{\twlrm \sc $a$}}}
\put(288,753){\makebox(0,0)[lb]{\raisebox{0pt}
[0pt][0pt]{\twlrm \sc $b$}}}
\put(288,795){\makebox(0,0)[lb]{\raisebox{0pt}
[0pt][0pt]{\twlrm \sc $c$}}}
\put( 84,795){\makebox(0,0)[lb]{\raisebox{0pt}
[0pt][0pt]{\twlrm \sc $d$}}}
\put(237,750){\makebox(0,0)[lb]{\raisebox{0pt}
[0pt][0pt]{\twlrm \sc $p(a,b,n)_{i\!,\!n}$}}}
\put(105,750){\makebox(0,0)[lb]{\raisebox{0pt}
[0pt][0pt]{\twlrm \sc $p(a,b,n)_{i\!,\!2}$}}}
\put(256,774){\makebox(0,0)[lb]{\raisebox{0pt}
       [0pt][0pt]{\twlrm \tiny $u\!\+\!(\!n$-$\!1\!)\lambda$}}}
\put(226,774){\makebox(0,0)[lb]{\raisebox{0pt}
       [0pt][0pt]{\twlrm \tiny $u\!\+\!(\!n$-$\!2\!)\lambda$}}}
\put(102,775){\makebox(0,0)[lb]{\raisebox{0pt}
       [0pt][0pt]{\twlrm \sc $u$}}}
\put(156,774){\makebox(0,0)[lb]{\raisebox{0pt}
[0pt][0pt]{\twlrm \sc $\cdots$}}}
\put(105,801){\makebox(0,0)[lb]{\raisebox{0pt}
[0pt][0pt]{\twlrm \sc $#1$}}}
\put(237,801){\makebox(0,0)[lb]{\raisebox{0pt}
[0pt][0pt]{\twlrm \sc $#2$}}}
\put( 18,774){\makebox(0,0)[lb]{\raisebox{0pt}
[0pt][0pt]{\twlrm  $\disp{\sum_i^{P_{(a,b)}^{(n)}}} 
\disp{\phi^{(i,\alpha)}_{(a,b,n)}}$}}}
\end{picture}}

\def\cella#1#2#3#4#5#6#7{\begin{picture}(40,30)(-5,13)
\setlength{\unitlength}{0.0180in}%
\thicklines
\multiput(5,5)(0,20){2}{\vector(1,0){13}} 
\multiput(25,5)(0,20){2}{\line(-1,0){10}} 
\multiput(5,5)(20,0){2}{\line(0,1){10}}   
\multiput(5,25)(20,0){2}{\vector(0,-1){13}} 
\put(1,2){\scriptsize $#1$} \put(25,2){\scriptsize $#2$} 
\put(25,25){\scriptsize $#3$} \put(1,25){\scriptsize $#4$} 
\put(26,13){\tiny $ #5 $} 
\put(13,0){\tiny $#6$}  \put(13,27){\tiny $#7$}
\setlength{\unitlength}{0.0120in}
\end{picture}}

\def\cellp#1#2#3#4{\begin{picture}(40,30)(-5,13)
\setlength{\unitlength}{0.0180in}%
\thicklines
\multiput(5,5)(0,20){2}{\line(1,0){13}} 
\multiput(25,5)(0,20){2}{\vector(-1,0){13}} 
\multiput(5,5)(20,0){2}{\line(0,1){10}}   
\multiput(5,25)(20,0){2}{\vector(0,-1){13}} 
\put(1,2){\scriptsize $#1$} \put(25,2){\scriptsize $#2$} 
\put(25,25){\scriptsize $#3$} \put(1,25){\scriptsize $#4$} 
\setlength{\unitlength}{0.0120in}
\end{picture}}

\begin{document}

\begin{center}
\title{FUSION OF \ade LATTICE MODELS}
\author{Yu-kui Zhou\footnote{Email: ykzhou@mundoe.maths.mu.oz.au} and 
   Paul A. Pearce\footnote{Email: pap@mundoe.maths.mu.oz.au}}
\address{Mathematics Department, University of Melbourne,\\
            Parkville, Victoria 3052, Australia }

\begin{abstract}
\begin{quotation}
Fusion hierarchies of \ade face models are constructed. The fused 
critical $D$, $E$ and elliptic $D$ models yield new solutions of the 
Yang-Baxter equations with bond variables on the edges of faces in 
addition to the spin variables on the corners. It is shown directly 
that the row transfer matrices of the fused models satisfy special 
functional equations. Intertwiners between the fused \ade models are
constructed by fusing the cells that intertwine the elementary face
weights. As an example, we calculate explicitly the fused $2\times 2$ 
face weights of the 3-state Potts model associated with the $D_4$ 
diagram as well as the fused intertwiner cells for the $A_5$--$D_4$ 
intertwiner. Remarkably, this $2\times 2$ fusion yields the face weights 
of both the Ising model and 3-state CSOS models. 
\end{quotation}
\end{abstract}  
\end{center} 

\section{ Introduction}

The fusion procedure is very useful in studying two-dimensional solvable
vertex
and face models \cite{KuReSk:81,DJKMO:86,DJKMO:88}. Essentially, fusion
enables
the construction of new solutions to the Yang-Baxter equations from a given
fundamental solution. Among \ade lattice models
\cite{ABF:84,Pasquier:87,KuniY,PearS}, much effort has been focused on the
fusion
of the $A$ models
\cite{DJKMO:88,WadatiTakagi}. By contrast, fusion  of the $D$ and $E$
models has
received no attention. The fusion procedure is important because it plays a
key
role in the solution of these lattice models. Specifically, it leads to
solvable
functional equations for the fusion hierarchy of commuting transfer
matrices
\cite{BaRe:89,KlPe:92}. Indeed, it has been argued \cite{Paul:92} that the
fusion
and inversion hierarchies of functional equations for the  $D$ and $E$
models are
exactly the same as those for the associated  $A$ model related to it by an
intertwining relation \cite{FeGi:89,Roch:90,FrZu:90,PeZh:93}. 

Here we extend the fusion procedure to all the critical \ade and the elliptic $D$
lattice models. In particular, we establish the fusion and inversion hierarchies 
directly for the classical $D$, $E$ and the elliptic $D$ models. We also extend 
the construction of intertwiners to
the fusion \ade models. In this paper, for simplicity, we focus on the
classical
\ade models although similar arguments apply for the affine and dilute \ade
models. The paper is organized as follows. In the next section we define
the critical classical \ade lattice models and the elliptic $D_L$ models and 
modify the face weights by an appropriate gauge transformation. The modified 
face weights satisfy a group of  special properties which ensure that they 
can be taken as the elementary blocks for fusion. 
In  section~3 we give the procedure for constructing the fused \ade
face weights. This is accomplished by introducing parities for the fusion
projectors.
In section~4 we derive directly the fusion hierarchies satisfied by the
fused \ade
row transfer matrices. The intertwiners between the fused $A$ and the fused
$D$ or
$E$ models are presented in  section~5. Also, in this section, we find the
gauge
transformation to obtain the symmetric fused face weights. In section~6, as
an
example, we give explicitly the fused $D_4$ face weights and the fused
cells that
intertwine them with the fused $A_5$ face weights. Finally, after a brief
conclusion, we present in the appendices a comprehensive table of the
adjacency
diagrams for the classical \ade fusion models as well as the parities of
the first
four fusion levels of the $E_6$ model.


\input{Fusion2.tex}

\input{Fusion3.tex}

\input{Fusion4.tex}

\input{Fusion5.tex}

\input{Fusion6.tex}

\section*{Acknowledgements} This research is supported by the
Australian Research Council. We thank Ole Warnaar for a critical reading of
the
manuscript.

\clearpage
\appendix

\section{Appendix: Fused \ade Adjacency Graphs}
\label{Appendix}

\subsection{Adjacency graphs of the fused $D_7$ models}

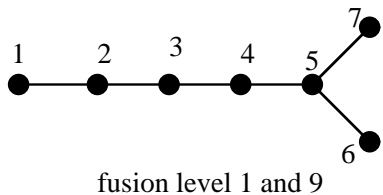
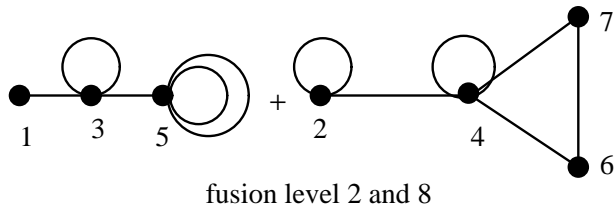
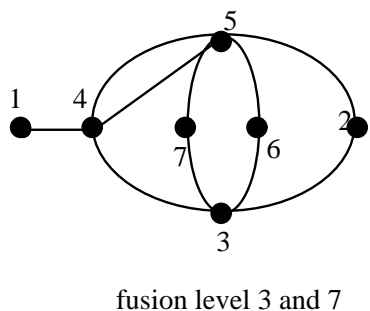
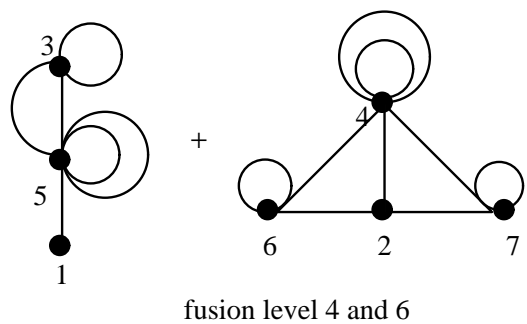
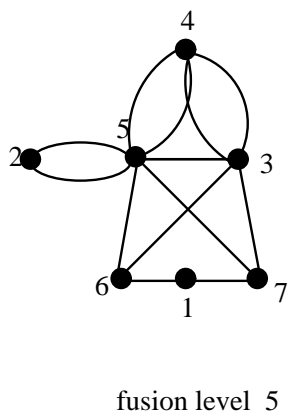
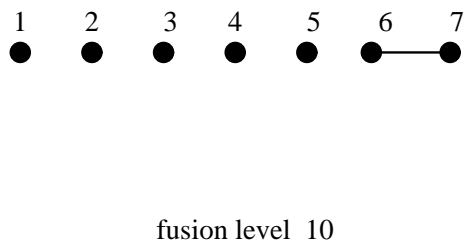
\[\epsfbox{D7Dynkin.eps}\]
\clearpage

\subsection{Adjacency graphs of the fused $E_6$ models}

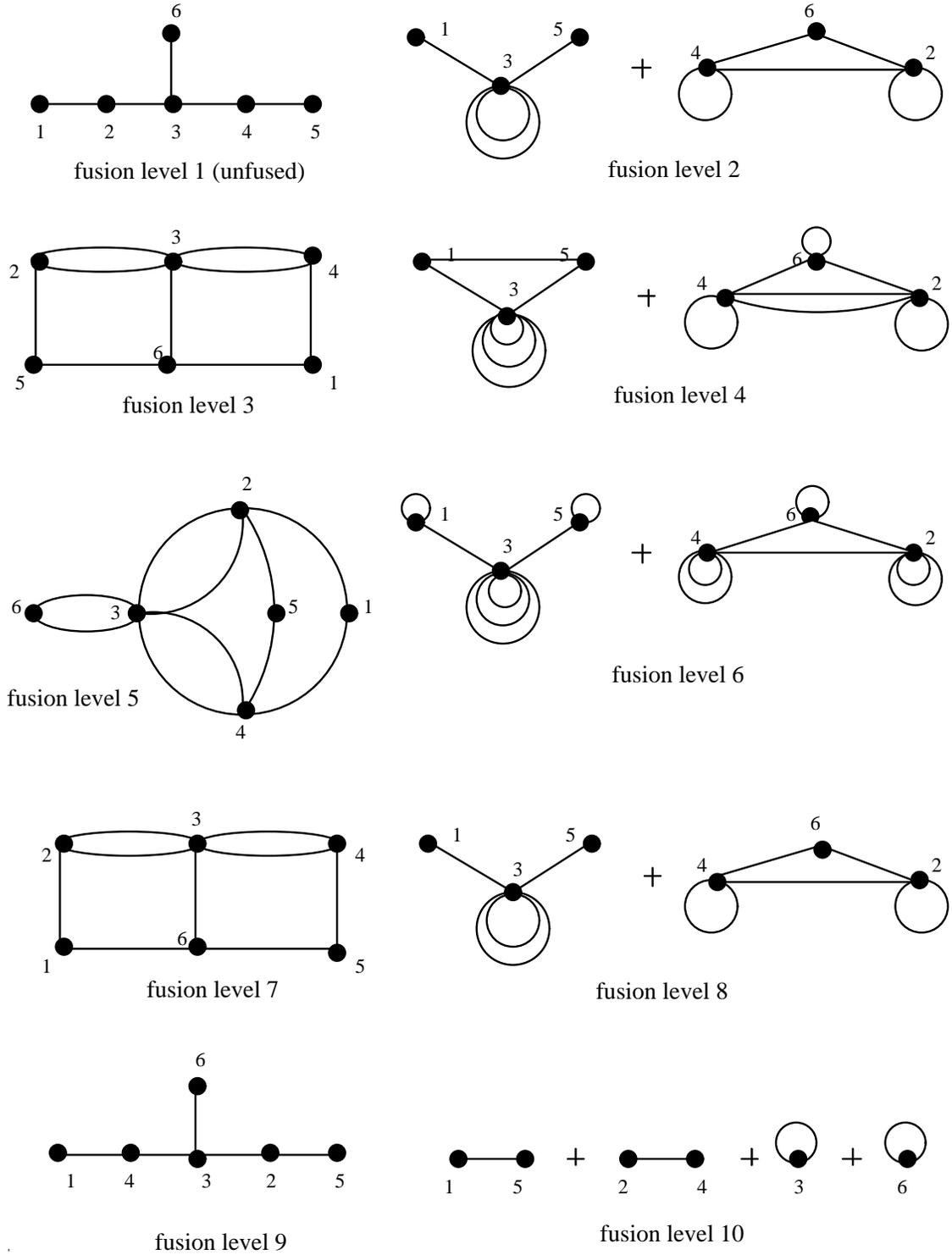
\[ \epsfbox{E6Dynkin.eps}\]
\clearpage

\subsection{Adjacency graphs of the fused $E_7$ models}

\[\epsfbox{E7Dynkin.eps} \]
\clearpage

\subsection{Adjacency graphs of the fused $E_8$ models}

\[ \hspace*{-0.9cm}\epsfbox{E8aDynkin.eps}\]
\clearpage

\[\hspace*{-0.7cm}\epsfbox{E8bDynkin.eps}\]
\clearpage

\section{Appendix: Parities $\phi$ of the $E_6$}

{\bf Fusion level 2:}\vspace{1.0cm}

\begin{tabular}{lcccl}
\begin{tabular}{|l|c|}\hline  
                 &  {$\alpha(2,2,2)$}          \\\cline{2-2}
 $p(2,2,2)_i$    &  $\alpha=1$                 \\ 
                 &   (2,1,2)                   \\ \hline\cline{2-2}
  (2,1,2)        &      1                      \\ \hline
  (2,3,2)        &      1                      \\ \hline
\end{tabular}
& & & & \begin{tabular}{|l|c|}\hline 
                 &  {$\alpha(4,4,2)$}          \\\cline{2-2}
  $p(4,4,2)_i$   &  $\alpha=1$             \\ 
                 &   (4,3,4)               \\ \hline\cline{2-2}
  (4,3,4)        &      1                  \\ \hline
  (4,5,4)        &      1                  \\ \hline
\end{tabular} \\ & & & & \\
\begin{tabular}{|l|c|c|}\hline 
      & \multicolumn{2}{c|} {$\alpha(3,3,2)$}     \\ \cline{2-3}
$p(3,3,2)_i$     &  $\alpha=1$  & $\alpha=2$      \\ 
                 &   (3,2,3)    &    (3,4,3)      \\ \hline\cline{2-3}
  (3,2,3)        &      1        &      0         \\ \hline
  (3,4,3)        &      0        &      1         \\ \hline
  (3,6,3)        &      1        &      1         \\ \hline
\end{tabular}
& & & & $\begin{array}{l}
\mbox{$\phi_{(a,b,2)}^{(i,\alpha)}=\phi_{(a,b,2)}^{(1,1)}=1$ } \\
\mbox{for other paths because }\\ 
\mbox{$L_{(a,b)}^{(2)}=1$. }  \end{array}$
\end{tabular}
\vspace{1.0cm}

{\bf Fusion level 3:}\vspace{1.0cm}

\begin{tabular}{lcccl}
\begin{tabular}{|l|c|c|}\hline 
      & \multicolumn{2}{c|} {$\alpha(2,3,3)$}     \\ \cline{2-3}
$p(2,3,3)_i$     &  $\alpha=1$  & $\alpha=2$    \\ 
                 &   (2,3,2,3)  &    (2,3,4,3)  \\ \hline\cline{2-3}
  (2,1,2,3)      &      1       &      0        \\ \hline
  (2,3,2,3)      &      1       &      0        \\ \hline
  (2,3,4,3)      &      0       &      1        \\ \hline
  (2,3,6,3)      &      1       &      1        \\ \hline
\end{tabular}
&&&&\begin{tabular}{|l|c|}\hline  
                 &  {$\alpha(3,6,3)$}          \\\cline{2-2}
 $p(3,6,3)_i$    &  $\alpha=1$                 \\ 
                 &   (3,2,3,6)                   \\ \hline\cline{2-2}
  (3,2,3,6)        &      1                      \\ \hline
  (3,4,3,6)        &     $-1$                      \\ \hline
\end{tabular}\\& & & & \\
\begin{tabular}{|l|c|c|}\hline 
      & \multicolumn{2}{c|} {$\alpha(4,3,3)$}     \\ \cline{2-3}
$p(4,3,3)_i$     &  $\alpha=1$  & $\alpha=2$    \\ 
                 &   (4,3,2,3)  &    (4,3,4,3)  \\ \hline\cline{2-3}
  (4,3,2,3)      &      1       &      0        \\ \hline
  (4,3,4,3)      &      0       &      1        \\ \hline
  (4,3,6,3)      &      1       &      1        \\ \hline
\end{tabular} 
&&&&
$\begin{array}{l}
\mbox{$\phi_{(3,2,3)}^{(i,\alpha)}=\phi_{(2,3,3)}^{(i,\alpha)}$,
     $\phi_{(3,4,3)}^{(i,\alpha)}=\phi_{(4,3,3)}^{(i,\alpha)}$,}\\ 
\mbox{$\phi_{(a,b,3)}^{(i,\alpha)}=\phi_{(a,b,3)}^{(1,1)}=1$ 
                                          $\;$  for $\;$ other  }\\ 
\mbox{paths because $L_{(a,b)}^{(3)}=1$.}
\end{array}$
\end{tabular}

\clearpage
\vspace{1.0cm}

{\bf Fusion level 4:}\vspace{.5cm}
\begin{center}
\begin{tabular}{|l|c|c|c|}\hline 
            & \multicolumn{3}{c|}{$\alpha(3,3,4)$}       \\ \cline{2-4}
$p(3,3,4)_i$&  $\alpha=1$ & $\alpha=2$   & $\alpha=3$      \\ 
            &  (3,2,3,2,3)&(3,2,3,4,3)   & (3,4,3,2,3)    \\
\hline\cline{2-4}
(3,2,1,2,3) &        1    &        0     &     0          \\ \hline
(3,2,3,2,3) &        1    &        0     &     0     \\ \hline
(3,2,3,6,3) &        1    &        0     &     0     \\ \hline
(3,6,3,2,3) &        1    &        0     &     0     \\ \hline\cline{1-1}
(3,2,3,4,3) &        0    &        1     &     0     \\ \hline
(3,2,3,6,3) &        0    &        1     &     0     \\ \hline
(3,6,3,4,3) &        0    &        1     &     0     \\ \hline\cline{1-1}
(3,4,3,2,3) &        0    &        0     &     1     \\ \hline
(3,4,3,6,3) &        0    &        0     &     1     \\ \hline
(3,6,3,2,3) &        0    &        0     &     1     \\ \hline\cline{1-1}
(3,4,5,4,3) &       $-1$    &       $-1$     &    $-1$       \\ \hline
(3,4,3,4,3) &       $-1$    &       $-1$     &    $-1$     \\ \hline
(3,4,3,6,3) &       $-1$    &       $-1$     &    $-1$     \\ \hline
(3,6,3,4,3) &       $-1$    &       $-1$     &    $-1$     \\ \hline
\end{tabular}

\begin{tabular}{|l|c||}\hline  
       &  {$\alpha(1,3,4)$}\\\cline{2-2}
 $p(1,3,4)_i$    &  $\alpha=1$       \\ 
                 &   (1,2,3,4,3)       \\ \hline\cline{2-2}
  (1,2,3,4,3)    &      1            \\ \hline
  (1,2,3,6,3)    &      1            \\ \hline
\end{tabular}
\begin{tabular}{|l|c|}\hline  
       &  {$\alpha(5,3,4)$}\\\cline{2-2}
 $p(5,3,4)_i$    &  $\alpha=1$       \\ 
                 &   (5,4,3,2,3)     \\ \hline\cline{2-2}
  (5,4,3,2,3)    &         1        \\ \hline
  (5,4,3,6,3)    &         1       \\ \hline
\end{tabular}

\begin{tabular}{|l|c||}\hline  
       &  {$\alpha(2,2,4)$}\\\cline{2-2}
 $p(2,2,4)_i$    &  $\alpha=1$       \\ 
                 &   (2,3,4,3,2)     \\ \hline\cline{2-2}
  (2,3,4,3,2)    &         1        \\ \hline
  (2,3,6,3,2)    &         1       \\ \hline
\end{tabular}
\begin{tabular}{|l|c|}\hline  
       &  {$\alpha(4,4,4)$}\\\cline{2-2}
 $p(4,4,4)_i$    &  $\alpha=1$       \\ 
                 &   (4,3,2,3,4)     \\ \hline\cline{2-2}
  (4,3,2,3,4)    &         1        \\ \hline
  (4,3,6,3,4)    &         1       \\ \hline
\end{tabular}

\begin{tabular}{|l|c||}\hline  
       &  {$\alpha(2,4,4)$}\\\cline{2-2}
 $p(2,4,4)_i$    &  $\alpha=1$       \\ 
                 &   (2,3,2,3,4)     \\ \hline\cline{2-2}
  (2,1,2,3,4)    &         1       \\ \hline
  (2,3,2,3,4)    &         1        \\ \hline
  (2,3,6,3,4)    &         1       \\ \hline
\end{tabular}
\begin{tabular}{|l|c|}\hline  
       &  {$\alpha(2,6,4)$}\\\cline{2-2}
 $p(2,6,4)_i$    &  $\alpha=1$       \\ 
                 &   (2,3,2,3,6)     \\ \hline\cline{2-2}
  (2,1,2,3,6)    &         1        \\ \hline
  (2,3,2,3,6)    &         1        \\ \hline\cline{1-1}
  (2,3,4,3,6)    &        $-1$        \\ \hline
\end{tabular}

\begin{tabular}{|l|c|}\hline  
       &  {$\alpha(4,6,4)$}\\\cline{2-2}
 $p(4,6,4)_i$    &  $\alpha=1$       \\ 
                 &   (4,3,2,3,6)     \\ \hline\cline{2-2}
  (4,3,2,3,6)    &         1        \\ \hline\cline{1-1}
  (4,3,4,3,6)    &        $-1$        \\ \hline
  (4,5,4,3,6)    &        $-1$       \\ \hline
\end{tabular} 
\begin{tabular}{|l|c|}\hline  
       &  {$\alpha(6,6,4)$}\\\cline{2-2}
 $p(6,6,4)_i$    &  $\alpha=1$       \\ 
                 &   (6,3,2,3,6)     \\ \hline\cline{2-2}
  (6,3,2,3,6)    &         1        \\ \hline\cline{1-1}
  (6,3,4,3,6)    &        $-1$       \\ \hline
\end{tabular}\vspace{0.5cm}
\end{center}
\mbox{The others are given by
$\phi_{(a,b,4)}^{(i,\alpha)}=\phi_{(b,a,4)}^{(i,\alpha)}$ }
\goodbreak


\end{document}

%% file: Fusion2.tex
\section{Properties of the Face Weights}
\setcounter{equation}{0}

The \ade lattice models \cite{Pasquier:87,OwBa:87,Paul:90} are
interaction-round-a-face or IRF models
\cite{Baxter:82} that generalize the restricted solid-on-solid (RSOS) 
models of Andrews, Baxter and Forrester \cite{ABF:84}. 
At criticality, these models are given by solutions of the Yang-Baxter
equation \cite{Baxter:82} based on the Temperley-Lieb algebra and
are associated with the classical and affine \ade Dynkin diagrams shown in
Figure~1.  States at adjacent sites of the square lattice must be adjacent
on the 
Dynkin diagram. The face weights of faces not satisfying this adjacency 
condition for each pair of adjacent sites around a face vanish. 

\begin{figure}[htb]
\begin{center}
\begin{picture}(400,280)(-20,-180)
\put(50,103){\bf Classical } \put(260,103){\bf Affine }

\put(-10,80){$A_L$} 
\multiput(20,80)(20,0){6}{\line(1,0){20}}
\multiput(20,80)(20,0){7}{\circle*{5}}
\put(20,85){\tiny 1} \put(40,85){\tiny 2} \put(60,85){\tiny 3}
\put(140,85){\tiny $L$}

\put(175,80){$A_{L-1}^{(1)}$} 
\multiput(220,80)(20,0){6}{\line(1,0){20}}
\put(280,50){\line(2,1){60}}\put(280,50){\line(-2,1){60}}
\multiput(220,80)(20,0){7}{\circle*{5}} \put(280,50){\circle*{5}}
\put(220,85){\tiny 1} \put(240,85){\tiny 2} \put(260,85){\tiny 3}
\put(340,85){\tiny $L-1$}
\put(275,38){\tiny $L$}

\put(-10,10){$D_L$} 
\multiput(20,10)(20,0){5}{\line(1,0){20}} 
\put(120,10){\line(4,3){20}} \put(120,10){\line(4,-3){20}}
\multiput(20,10)(20,0){6}{\circle*{5}}   
\multiput(140,25)(0,-30){2}{\circle*{5}}
\put(20,15){\tiny 1} \put(40,15){\tiny 2} \put(60,15){\tiny 3} 
\put(143,28){\tiny $L$}\put(143,-12){\tiny $L-1$}

\put(175,10){$D_{L-1}^{(1)}$} 
\multiput(240,10)(20,0){4}{\line(1,0){20}} 
\put(240,10){\line(-4,3){20}} \put(240,10){\line(-4,-3){20}}
\put(320,10){\line(4,3){20}}  \put(320,10){\line(4,-3){20}}
\multiput(240,10)(20,0){5}{\circle*{5}}
\multiput(220,25)(0,-30){2}{\circle*{5}}
\multiput(340,25)(0,-30){2}{\circle*{5}}
\put(220,30){\tiny 2} \put(240,15){\tiny 3} \put(260,15){\tiny 4}
\put(344,25){\tiny $L-1$}
\put(344,-12){\tiny $L$} \put(222,-12){\tiny 1}

\put(-10,-60){$E_6$} 
\multiput(20,-60)(20,0){4}{\line(1,0){20}}
\multiput(20,-60)(20,0){5}{\circle*{5}}
\put(60,-40){\line(0,-1){20}} \put(60,-40){\circle*{5}}
\put(20,-55){\tiny 1} \put(40,-55){\tiny 2} \put(80,-55){\tiny 4}
\put(100,-55){\tiny 5}
\put(58,-71){\tiny 3} \put(58,-35){\tiny 6}

\put(175,-60){$E_6^{(1)}$} 
\multiput(220,-60)(20,0){4}{\line(1,0){20}}
\multiput(220,-60)(20,0){5}{\circle*{5}}
\multiput(260,-40)(0,20){2}{\line(0,-1){20}}
\multiput(260,-40)(0,20){2}{\circle*{5}}
\put(220,-55){\tiny 1} \put(240,-55){\tiny 2} \put(280,-55){\tiny 4}
\put(300,-55){\tiny 5}
\put(258,-71){\tiny 3} \put(262,-39){\tiny 6}  \put(258,-15){\tiny 7}

\put(-10,-120){$E_7$} 
\multiput(20,-120)(20,0){5}{\line(1,0){20}}
\multiput(20,-120)(20,0){6}{\circle*{5}}
\put(80,-100){\line(0,-1){20}} \put(80,-100){\circle*{5}}
\put(20,-115){\tiny 1} \put(40,-115){\tiny 2} \put(60,-115){\tiny 3}
\put(100,-115){\tiny 5}
\put(120,-115){\tiny 6} \put(78,-131){\tiny 4} \put(78,-95){\tiny 7}

\put(175,-120){$E_7^{(1)}$} 
\multiput(220,-120)(20,0){6}{\line(1,0){20}}
\multiput(220,-120)(20,0){7}{\circle*{5}}
\put(280,-100){\line(0,-1){20}} \put(280,-100){\circle*{5}}
\put(220,-115){\tiny 1} \put(240,-115){\tiny 2} \put(260,-115){\tiny 3}
\put(300,-115){\tiny 5}
\put(320,-115){\tiny 6} \put(340,-115){\tiny 7} \put(278,-131){\tiny 4}
\put(278,-95){\tiny 8}

\put(-10,-180){$E_8$} 
\multiput(20,-180)(20,0){6}{\line(1,0){20}}
\multiput(20,-180)(20,0){7}{\circle*{5}}
\put(100,-160){\line(0,-1){20}} \put(100,-160){\circle*{5}}
\put(20,-175){\tiny 1} \put(40,-175){\tiny 2} \put(60,-175){\tiny 3}
\put(80,-175){\tiny 4}
\put(120,-175){\tiny 6} \put(140,-175){\tiny 7} \put(98,-191){\tiny 5}
\put(98,-154){\tiny 8}

\put(175,-180){$E_8^{(1)}$} 
\multiput(220,-180)(20,0){7}{\line(1,0){20}}
\multiput(220,-180)(20,0){8}{\circle*{5}}
\put(320,-160){\line(0,-1){20}} \put(320,-160){\circle*{5}}
\put(220,-175){\tiny 1} \put(240,-175){\tiny 2} \put(260,-175){\tiny 3}
\put(280,-175){\tiny 4}
\put(300,-175){\tiny 5} \put(340,-175){\tiny 7} \put(360,-175){\tiny 8} 
\put(318,-191){\tiny 6} \put(318,-154){\tiny 9}
\end{picture}
\end{center}
\caption{Dynkin diagrams of the classical and affine \ade Lie algebras}
\end{figure}
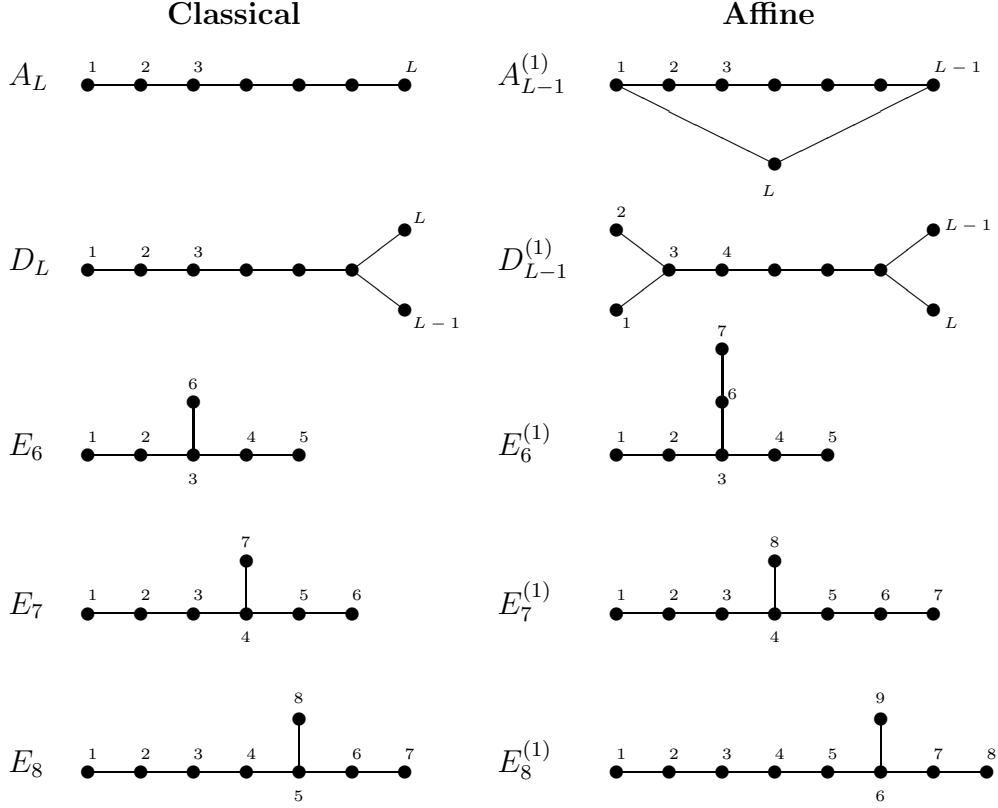

In this paper we will restrict our attention to the classical \ade models.
The face
weights of the  classical \ade models at criticality are given  by
\cite{Pasquier:87} 
\begin{equation}
\wt Wabcdu=\face abcdu =
 {\sin (\lambda -u) \over \sin \lambda }\;\delta_{a,c}A_{a,b}A_{a,d} +
       {\sin u \over \sin \lambda } \sqrt{S_a S_c \over S_b S_d}\;
       \delta_{b,d}A_{a,b}A_{b,c} 
\label{eq:cADEface} 
\end{equation} 
where $u$ is the spectral parameter and $\lambda=\pi/h $ is the crossing
parameter. Here  
\be
h=\cases{L+1, &for $A_L$\cr
         2L-2,&for $D_L$\cr
         12,18,30,&for $E_L=E_{6,7,8}$}
\ee
is the Coxeter number and $S_a$ are the elements of the Perron-Frobenius
eigenvector $S$ of the
adjacency  matrix $A$ with elements 
\begin{equation}
A_{a,b}=\left\{ \ba{ll}  
             1,\qquad & \mbox{$(a,b)$ adjacent} \\
             0,\qquad & \mbox{otherwise.}
                 \ea  \right.
\end{equation}

In analogy to the classical $A$ models, we modify the \ade face weights 
(\ref{eq:cADEface}) by a gauge transformation as follows
\be 
\face abcdu  \mapsto\   {g(d,c) g(c,b)\over g(d,a) g(a,b)}\ \;\face abcdu=\
\ 
\sqrt{S_c\over S_a}\;{f_c\over f_a}\ \;\face abcdu
\label{gauge}
\ee
where we set $g(a,b)=g_ag_b$ with $g_a=S_a^{1/4}f_a^{1/2}$ and
\goodbreak
\begin{eqnarray}
 &f_a=\left.\begin{array}{ll}
\ (-1)^{a\over 2}, &\ \ \;\,\mbox{for $a=1,2,\cdots ,L$}  \end{array}
\right.
             \;\;\;\;\;\;\;\;\;\;\;\;\;\;\;\;&\;A_L\\
&f_a=\left\{ \begin{array}{ll}
      (-1)^{a\over 2}, & \mbox{for $a=1,2,\cdots ,L-1$} \\
      (-1)^{L-1\over 2 }, & \mbox{for $a=L$}\end{array} \right.
                \;\;\;\;\;\;\;\;\;\;&\;D_L  \label{f-D}\\
 &f_a=\left\{ \begin{array}{ll}
              (-1)^{a\over 2},   & \mbox{for $a=1,2,\cdots ,L-3,L-1$}  \\ 
              (-1)^{L-4\over 2}, & \mbox{for $a=L-2$} \\ 
              (-1)^{L-2\over 2}, & \mbox{for $a=L$}
              \end{array} \right.
                        &\;E_L=E_{6,7,8}
\end{eqnarray}
In this gauge, the modified face weights are given by  
\be
\face abcdu =
{\sin (\lambda -u)\over \sin\lambda }\;\delta_{a,c}
      A_{a,b}A_{a,d} +{\sin u \over \sin \lambda }{S_c\over S_b }
                 \epsilon_{a,c}\;\delta_{b,d}A_{a,b} A_{b,c} 
\label{eq:element} 
\ee
where we have introduced the symmetric sign symbol
\be
\epsilon_{a,c}=\epsilon_{c,a}={f_c\over f_a}=
\cases{1,&$a=c$\cr
       1,&$(a,c)=(L-1,L)$ or $(L,L-1)$ for $D_L$\cr
       1,&$(a,c)=(L-4,L-2)$ or $(L-2,L-4)$ for $E_L$\cr
       -1, &otherwise.}
\ee
The face weights (\ref{eq:cADEface}) or (\ref{eq:element}) satisfy the
Yang-Baxter
equations
\be 
 \YBR abcdefu{u\-v}v \label{eq:YBR}
\ee
where the solid circles indicate sums over the central spins. 

Each node $a$ of the \ade Dynkin diagrams has a coordination number or
valence
$\val a=1,2,3$. Specifically, the valence $\val a=2$ except for the
endpoints with $\val a=1$ and branch points with $\val a=3$.
In the modified gauge (\ref{gauge}) the face weights acquire the following
properties:  
\bea \hspace*{-1.0cm}
 \face abcd0 &=&\delta_{a,c} \label{eq:g1p1}\\
 \face abcd\lambda &=&0,\qquad b\not=d\label{eq:g1p2}\\
 \face cbcb\lambda \!&A_{a,b} &=\;\epsilon_{c,a} \face abcb\lambda
\ =\ {S_c\over S_b}A_{a,b}A_{b,c} \h \val b >1.
\label{eq:g1p3}
\eea
Moreover, at $u=-\lambda$, the face weights also satisfy the properties:
\bea
\face abab{\-\lambda} &=&0 \hspace*{4.2cm}  \val
b=1\label{eq:g2p1}\\&&\no\\
\face {a\!\pm\!1}a{a\!\pm\!1}a{\-\lambda} &=&\face 
    {a\!\pm\!1}a{a\!\mp\!1}a{\-\lambda}\hspace*{2.95cm}\val a=2
                                                \label{eq:g2p2}\\&&\no\\
\face a{L\-2}{L\-3}{L\-2}{\-\lambda}&=&\face
            a{L\-2}{L\-1}{L\-2}{\-\lambda}\;\;+\;\;\face 
      a{L\-2}L{L\-2}{\-\lambda}\hspace*{1.cm}\for D_L
                                               \label{eq:g2p3}\\&&\no\\
\face a{L\-3}L{L\-3}{\-\lambda}&=&\face a{L\-3}{L\-4}{L\-3}{\-\lambda}
\;\;+\;\;\face a{L\-3}{L\-2}{L\-3}{\-\lambda}\hspace*{1.cm}\for E_L
                                              \label{eq:g2p4}\\&&\no
\eea
These properties are useful for constructing the fused face weights.
However, to 
study the fusion hierarchy we also need the additional properties:
\bea
\sum_a \face abcb{\-\lambda} &=& 2\cos\lambda \;A_{b,c}
                             \qquad \val b=2
                      \label{eq:g3p1} \\&&\no\\
\sum_{{\ol a}\in \sym a}\face {\ol a}bcb{\-\lambda} 
      &=&2\cos\lambda  \;A_{b,c}\left\{\begin{array}{ll}
    (\delta_{a,c}+\delta_{c,L-3}) & \mbox{$\for D_L$}\\
    (\delta_{a,c}+\delta_{c,L})   & \mbox{$\for E_L$}
    \end{array}
                                     \right.\quad \val b=3
      \label{eq:g3p2}\\&&\no\\ 
\face abab\lambda &=&2\cos\lambda\; A_{a,b}  
       \hspace*{0.9cm} \val a=3,\ \val b=1\; \label{eq:g3p3}\\&&\no\\
\face abcd\lambda &+&\face abcd{\-\lambda} =\;2\cos\lambda\; A_{b,c}A_{d,c}
     \delta_{a,c} \hspace*{.9cm} \label{eq:g3p4}\\&&\no\\
\face {b\pm 1}b{b\pm 1}b\lambda &-&\face {b\pm 1}b{b\mp 1}b\lambda =
           2\cos\lambda\;  \hspace*{0.7cm} \val b=2 
                                            \label{eq:g3p5}\\&&\no\\
\face a{L\-2}{L\-3}{L\-2}\lambda &-&
              \face a{L\-2}{L\-1}{L\-2}\lambda
             \; \;-\;\; \face a{L\-2}L{L\-2}\lambda \no\\&&\no\\
&=&2\cos\lambda (\delta_{a,L-3}-\delta_{a,L-1}
  -\delta_{a,L})\qquad \for D_L\label{eq:g3p8}\\
\face a{L\-3}L{L\-3}\lambda &-&
               \face a{L\-3}{L\-2}{L\-3}\lambda
                  \; \;-\;\;\face a{L\-3}{L\-4}{L\-3}\lambda \no\\&&\no\\
&=&2\cos\lambda (\delta_{a,L} -\delta_{a,L-2}-
      \delta_{a,L-4})\qquad \for E_L \label{eq:g3p7}
\eea
where the symmetric sum is over
\bea
\sym a&=&\cases{\{L-3,L-1\},&$\;a=L-1$\cr
                \{L-3,L\},&$\;a=L$}
       \qquad \for D_L\\
\sym a&=&\cases{\{L-4,L\},&\qquad$a=L-4$\cr
                \{L-2,L\},&\qquad$a=L-2$}\qquad\for E_L
\eea
We will introduce the corresponding antisymmetric sums in Section~3.

 However the fusion procedure constructed in Section~3 is described by 
studying the classical $ADE$ models. In fact it works also for the 
elliptic $D_L$ models with the nonzero face weights $W_D$ 
\cite{Pasquier:87} which are related to the face weights $W_A$
of the elliptic $A_{2L-3}$ models by orbifold duality \cite{FeGi:89,Roch:92}
\bea
&& \wt {W_D}{L-3}{L-2}{L-1}{L-2}u     \no \\
&&\h=\wt {W_D}{L-3}{L-2}{L}{L-2}u = {1\over \sqrt{2}} 
     \wt {W_A}{L-3}{L-2}{L-1}{L-2}u                     \\
&&\wt {W_D}{L-1}{L-2}{L-1}{L-2}u \no\\
&&\h=\wt {W_D}{L}{L-2}{L}{L-2}u \no\\
&&\h={1\over 2}\wt {W_A}{L-1}{L-2}{L-1}{L-2}u+{1\over 2}
    \wt {W_A}{L-1}{L}{L-1}{L-2}u     \\
&&\wt {W_D}{L-1}{L-2}{L}{L-2}u \no\\
&&\h=\wt {W_D}{L}{L-2}{L-1}{L-2}u \no\\
&&\h={1\over 2}\wt {W_A}{L-1}{L-2}{L-1}{L-2}u-{1\over 2}\wt {W_A}{L-1}{L}{L-1}{L-2}u \\
&&\wt {W_D}{L-2}{L-1}{L-2}{L-1}u \no\\
&&\h=\wt {W_D}{L-2}{L}{L-2}{L}u    \no\\
&&\h=\wt {W_A}{L-2}{L-1}{L-2}{L-1}u-\wt {W_A}{L}{L-1}{L-2}{L-1}u  \\
&&\wt {W_D}{L-2}{L}{L-2}{L-1}u \no\\
&&\h=\wt {W_D}{L-2}{L-1}{L-2}{L}u  \no\\
&&\h=\wt {W_A}{L-2}{L-1}{L-2}{L-1}u+\wt {W_A}{L}{L-1}{L-2}{L-1}u    \\
&&\wt {W_D}{L-3}{L-2}{L-3}{L-2}u =\wt {W_A}{L-3}{L-2}{L-3}{L-2}u   \\
&&\wt {W_D}{L-3}{L-4}{L-3}{L-2}u =\wt {W_A}{L-3}{L-4}{L-3}{L-2}u    \\    
&&\wt {W_D}{a}{b}{c}{d}u = \wt {W_A}{a}{b}{c}{d}u\hs{0.7} \mbox{if $d\not=L-2,L-1,L$.}
\eea  
Here the nonzero face weights $W_A$ are given by \cite{ABF:84}
\bea
&&\wt {W_A}{a+1}{a}{a-1}{a}u =\no   \\
&\h&\wt {W_A}{a-1}{a}{a+1}{a}u =h(u)\sqrt{h(w_{a\-1})h(w_{a\+1})} /h(w_a)      \no\\
&&\wt {W_A}{a}{a+1}{a}{a-1}u =\no\\
&\h &\wt {W_A}{a}{a-1}{a}{a+1}u =h(\lambda-u) \no\\
&&\wt {W_A}{a+1}{a}{a+1}{a}u =h(\lambda)h(w_a+u)/h(w_a)    \no\\
&&\wt {W_A}{a-1}{a}{a-1}{a}u =h(\lambda)h(w_a-u)/h(w_a)
\eea
where $h(u)=\theta_1(u)\theta_4(u)$, $w_a=a\lambda$ and $\theta_1,\theta_4$ are 
the usual theta  functions of nome $p$. 

We have the same properties as (\ref{eq:g1p1})--(\ref{eq:g2p3}) if in
the gauge transformation (\ref{gauge}) we set  
$g_a=h^{1/4}_af_a^{1/2}$ where
\be
h_a=\cases{2^{-1/4}h(w_{L-1}) ,&$a=L-1,L$\cr
           2^{1/4}h(w_a), &otherwise.}
\ee
and $f_a$ is given by (\ref{f-D}). With these changes,
the fusion of the elliptic $D_L$ models proceeds as for
the critical \ade models.

%% file: Fusion3.tex
\section{Elementary Fusion}
\setcounter{equation}{0}

The Temperley-Lieb \ade models are related to the six-vertex model and
hence to the spin algebra $su(2)$. The higher-spin representations of this 
algebra are obtained by taking tensor products of the fundamental
representation. The analog of this process for the \ade face models is 
fusion. Starting with a fundamental $A$,
$D$ or $E$ solution of the Yang-Baxter equations  it is possible to obtain
a hierarchy of ``higher-spin" solutions by fusing blocks of faces together.
The fused $A$ models  have been discussed by a number of authors 
\cite{DJKMO:86,DJKMO:88,BaRe:89,WadatiTakagi,KlPe:92}. In this section, we
extend the fusion procedure to the classical $D$, $E$ models and the elliptic 
$D$ models. We focus on the critical $ADE$ models and the arguments apply for 
the elliptic  $D$ models by replacing all $\sin u$ functions with the elliptic
functions $h(u)$.
  
\subsection{Admissibility} 

The adjacency matrices $A^{(n)}$ of the level $n$ fused models  are
determined by the $su(2)$ fusion rules \cite{PeZh:93} truncated at level 
$h-2$
\begin{eqnarray} & & A^{(n)} A^{(1)}=A^{(n+1)}+A^{(n-1)},\quad
n=1,2,3,\cdots ,h-2
\nonumber \\ & & A^{(0)}=I,\quad A^{(1)}=A ,\quad A^{(n)}=0,\quad n>h-2 ,
\label{eq:adjfusion} \\ & & A^{(h-2)}=\left\{ \begin{array}{ll}
                I , & \mbox{for$\;D_{2L}, \; E_7\;$and$\;E_8 $}  \\
                Y , & \mbox{for$\;A_L\;$, $D_{2L-1}$ and$\;E_6$}
                \end{array} \right. \nonumber 
\end{eqnarray} 
where $I$ is the identity matrix, $h$ is the Coxeter number and 
$Y$ is the corresponding height reflection operator defined by
\be   
Y_{a,b}=\delta_{a,r(b)}  
\ee  where 
\begin{eqnarray}
r(b)&=& h-b  \qquad \mbox{for $A_L$} \\ 
r(b)&=&\cases{6-b      &\mbox{if $b<6$}\cr
6      &\mbox{if $b=6$}}\qquad \mbox{for $E_6$}\\ 
r(b)&=&\cases{b &\mbox{if $b<2L-2$}\cr
{2L-1}&\mbox{if $b=2L-2$}\cr
2L-2    &\mbox{if $b={2L-1}$}}\qquad \mbox{for $D_{2L-1}$}
\end{eqnarray}
Here $A^{(1)}=A$ is the adjacency matrix for the elementary classical \ade
model. As  examples, we draw the  adjacency diagrams describing the allowed
or
admissible states of adjacent sites of the fused $D_7$ and $E_L$ models in
Appendix~A. In contrast to fusing the $A_L$ models, the elements of
$A^{(n)}$ can in
general be nonnegative integers greater than one. In this case we
distinguish the
edges of the adjacency diagram joining two given sites by bond variables
$\alpha,\beta=1,2,\ldots$ If there is just one edge then the corresponding
bond
variable is $\alpha=1$.
\begin{figure}[b]
\begin{center}
\setlength{\unitlength}{0.0195in}%
\onebytwoface abcd\alpha{\beta(c')}{u}{u\+\lambda\;}{{\sevrm .}} 
\end{center}
\caption{Elementary fusion of two faces. The cross denotes a symmetric sum
labelled by $\alpha=1,2$ as defined in lemma~1. The other spins are fixed. 
If $\val c=\val d=3$ we assume  that $c'\not=L-3$ for $D_L$ and $c'\not=L$ 
for $E_L$. For clarity both the spin $c'$ and the bond variable $\beta$ 
are indicated.}    
\end{figure}

\subsection{One by two fusion}

We implement the elementary fusion of a one by two block of face weights.
The
properties of this elementary fusion then suffice to establish the fusion
of general
$m\times n$ blocks of face weights. Notice that in the level 2 fused $D$
and $E$
models, the occurrence of bond variables on the edges of the
fused face weights only arises when both adjacent sites are branch points
with
valence $\val a=3$. 

\begin{lemma}[Elementary Fusion] \label{lem1} If $(a,b)$ and $(d,c)$ are
admissible
edges at fusion level two  we define the $1\times 2$ fused weights by
\begin{equation}
\wf {W_{12}}ubcd{\alpha}{}{\beta}{\vspace*{-0.5cm}} = 
 {\disp \sum_{a'}}\;\;\;\wt Wa{a'}{c'}d{u} \wt W{a'}bc{c'}{u\+\lambda}
\label{eq:lem1} 
\end{equation} where the sum over $a'$ is over all possible spins  (i.e. a
normal
sum with $\alpha=1$) if $a$ and $b$ are not both of valence 3. If $a$ and
$b$ are
both of valence 3, the  sum is accomplished in two different ways by
summing over
$\sym{a'}$. Explicitly, for $D_L$ (resp. $E_L$) we sum over $L-3$ and $L-1$
(resp.
$L-4$ and $L$) if the bond variable $\alpha =1$ and over
$L-3$ and $L$ (resp. $L-2$ and $L$) if the bond variable $\alpha=2$. Then
it follows
that:

(i) The RHS is independent of $c'$ except for its dependence on the bond
variable
\be
\beta(c')=\cases{2,&$c=d=L-2$ and $c'=L;\phantom{\mbox{}-2}$\qquad $D_L$\cr
                 2,&$c=d=L-3$ and $c'=L-2$;\,\qquad $E_L$\cr
                 1,&otherwise.}
\ee

(ii) For all a,b,c,d we have
$\wf{W_{12}}{0}bcd{\alpha}{\vspace*{-0.5cm}}{\beta}{}=0$.  
\end{lemma} {\sl Proof:} To establish (i) it is enough to consider the case
$c=d$,
otherwise 
$c'$ is uniquely determined by the adjacency conditions. Setting $v=\lambda
$ and
$c=e$ in the Yang-Baxter equation (\ref{eq:YBR}) we have  
\be
\setlength{\unitlength}{0.0110in}%
\YBR{\hspace*{-.15cm}a'}bc{c'}cau{\hspace*{-.2cm}u\-\lambda}\lambda 
\label{eq:proofYBR} 
\ee

If $a=b$, then take the special sum over  $a'$ in (\ref{eq:proofYBR}).
Owing to 
(\ref{eq:g1p3}), the special summation over $a'$ with each  fixed $c'$
vanishes in
the LHS. Therefore for any $(a,b)$ we always have
\bea
\onebytwoface abcc{\alpha}{\beta}{u\-\lambda}u{{\sevrm .}}\h&=&\h 0  
              \hspace*{6.2cm}\val c=1 \label{eq:1by2fu1}\\&&\no\\
\onebytwoface abcc{\alpha}{1(c\-1)}{u\-\lambda}u{{\sevrm .}}\h&=&\h
\onebytwoface abcc{\alpha}{1(c\+1)}{u\-\lambda}u{{\sevrm .}} 
      \hspace*{4.5cm} \val c=2 \label{eq:1by2fu2}\\&&\no\\
\onebytwoface ab{L\-2}{L\-2}{}{L\-3}{u\-\lambda}u{{\sevrm .}} \h&=&\h 
\onebytwoface ab{L\-2}{L\-2}{\alpha}{1(L\-1)}{u\-\lambda}u{{\sevrm .}} \h
+\h 
\onebytwoface ab{L\-2}{L\-2}{\alpha}{2(L)}{u\-\lambda}u{{\sevrm .}} 
                     \h\h\h \for D_L \label{eq:1by2fu3}\\&&\no\\
\onebytwoface ab{L\-3}{L\-3}{}{L}{u\-\lambda}u{{\sevrm .}} \h &=&\h 
\onebytwoface ab{L\-3}{L\-3}{\alpha}{1(L\-4)}{u\-\lambda}u{{\sevrm .}} \h
+\h 
\onebytwoface ab{L\-3}{L\-3}{\alpha}{2(L\-2)}{u\-\lambda}u{{\sevrm .}}
                       \h \h\h\for E_L \label{eq:1by2fu4}\\&&\no
\eea These equations imply part $(i)$ of the lemma.  Part $(ii)$ follows by
(\ref{eq:g1p1}) if $c'\not=a$ and by (\ref{eq:g1p3}) if $c'=a$. 

Lemma~1 gives the $1\times 2$ fused face weights incorporating the level
two fusion 
adjacency conditions. A bond variable $\alpha$ has been added between each
pair
$(a,b)$ of adjacent spins to form edges with states $(a,\alpha,b)$. The
adjacency
condition for bond variables is that $\alpha=1,2$ if $a=b=L-2$ (resp.
$L-3$) for
$D_L$ (resp. $E_L$) and otherwise the bond variable takes the fixed value
$\alpha=1$. Similarly, the spin variables are constrained by 
$|a-b|=0,2$ and $2<|a+b|<2L-4$ (resp. $2L-2$) or $(a,b)=(L-1,L)$ 
(resp. $(a,b)=(L-4,L)$) for $D_L$ (resp. $E_L$). Observing
properties (\ref{eq:g2p1})--(\ref{eq:g2p4}) we find that this adjacency is
completely
determined by the operator $P(1,-\lambda)$  with elements
\be P(1,-\lambda)^{d,c,b}_{d,a,b}= 
\diagface abcd{-\lambda}00
\label{eq:p1}
\ee
So it can be considered as the projector of level 2 fusion. 

\subsection{Operator P(n,u)}
 
Let us define graphically 
\be 
P(n,u)^{a,a_1,a_2,\cdots ,b}_{a,b_1,b_2,\cdots ,b}=
\setlength{\unitlength}{0.0105in}%
\begin{minipage}{3.5in}
\begin{picture}(316,210)(47,580)
\thicklines
\put(210,780){\line(-1,-1){150}}
\put(210,780){\line( 1,-1){150}}
\put(210,780){\line(-1,-1){150}}
\put( 60,630){\line( 1,-1){ 30}}
\put( 90,600){\line( 1, 1){150}}
\put(120,690){\line( 1,-1){ 90}}
\put(210,600){\line( 1, 1){ 90}}
\put(150,720){\line( 1,-1){120}}
\put(270,600){\line( 1, 1){ 60}}
\put(180,750){\line( 1,-1){150}}
\put(330,600){\line( 1, 1){ 30}}
\put( 90,660){\line( 1,-1){ 60}}
\put(150,600){\line( 1, 1){120}}
\put( 47,627){\makebox(0,0)[lb] {\raisebox{0pt}[0pt][0pt]{\twlrm \sc
$b_n$}}}
\put( 90,581){\makebox(0,0)[lb] {\raisebox{0pt}[0pt][0pt]{\twlrm \sc $b$}}}
\put(330,584){\makebox(0,0)[lb] {\raisebox{0pt}[0pt][0pt]{\twlrm \sc $b$}}}
\put(208,786){\makebox(0,0)[lb] {\raisebox{0pt}[0pt][0pt]{\twlrm \sc $a$}}}
\put(170,755){\makebox(0,0)[lb] {\raisebox{0pt}[0pt][0pt]{\twlrm \sc
$b_1$}}}
\put(141,724){\makebox(0,0)[lb] {\raisebox{0pt}[0pt][0pt]{\twlrm \sc
$b_2$}}}
\put(239,755){\makebox(0,0)[lb] {\raisebox{0pt}[0pt][0pt]{\twlrm \sc
$a_1$}}}
\put(266,724){\makebox(0,0)[lb] {\raisebox{0pt}[0pt][0pt]{\twlrm \sc
$a_2$}}}
\put(306,626){\makebox(0,0)[lb] {\raisebox{0pt}[0pt][0pt]{\twlrm \tiny
$u\!\+\!(n\!-\!1)\lambda$}}}
\put(363,630){\makebox(0,0)[lb] {\raisebox{0pt}[0pt][0pt]{\twlrm \sc
$a_n$}}}
\put(208,749){\makebox(0,0)[lb] {\raisebox{0pt}[0pt][0pt]{\twlrm \sc $u$}}}
\put(159,717){\makebox(0,0)[lb] {\raisebox{0pt}[0pt][0pt]{\twlrm \sc
$\!\-\!(n\!-\!1)\lambda$}}}
\put(226,717){\makebox(0,0)[lb] {\raisebox{0pt}[0pt][0pt]{\twlrm \sc
$u\+\lambda$}}}
\put(227,653){\makebox(0,0)[lb] {\raisebox{0pt}[0pt][0pt] {\twlrm \sc
$\-2\lambda$}}}
\put(107,654){\makebox(0,0) [lb]{\raisebox{0pt}[0pt][0pt]{\twlrm \sc
$\-2\lambda$}}}
\put( 85,622){\makebox(0,0) [lb]{\raisebox{0pt}[0pt][0pt]{\twlrm \sc
$\-\lambda$}}}
\put(260,625){\makebox(0,0) [lb]{\raisebox{0pt}[0pt][0pt]{\twlrm \sc
$\-\lambda$}}}
\end{picture} 
\end{minipage}
\label{eq:projdef}
\ee

\noindent
Then the operator $P(n,-n \lambda)$ is the projector of level $n+1$ fusion.

For $n=1$ it is the face
weight of an elementary  block. For $n=2$ it produces the 1 by 2 fusion
presented in the last  section. This follows from the properties
(\ref{eq:g2p1})--(\ref{eq:g2p4}) and  (\ref{eq:p1}) we have
\begin{eqnarray} P(2,u)^{d,c',c,b}_{d,a,a_1,b}  =\left\{ \begin{array}{ll}
  P(1,-\lambda)^{a,L\-4,b}_{a,a_1,b}
       \wf {W_{12}}ubcd{1}{}{\beta}{\vspace*{-0.5cm}}+ & \\
  \;\;\;\;\;P(1,-\lambda)^{a,L\-2,b}_{a,a_1,b}
      \wf {W_{12}}ubcd{2}{}{\beta}{\vspace*{-0.5cm}}  &
             \mbox{if$\;a$=$b$=$L\-3\;$for$\;E_L$}       \\
  P(1,-\lambda)^{a,L\-1,b}_{a,a_1,b}
       \wf {W_{12}}ubcd{1}{}{\beta}{\vspace*{-0.5cm}}+ & \\
  \;\;\;\;\;P(1,-\lambda)^{a,{L\-2},b}_{a,a_1,b}
        \wf {W_{12}}ubcd{2}{}{\beta}{\vspace*{-0.5cm}} &
                 \mbox{if$\;a$=$b$=$L\-2\;$for$\;D_L$}   \\
  P(1,-\lambda)^{a,a',b}_{a,a_1,b} 
       \wf {W_{12}}ubcd{1}{}{\beta}{\vspace*{-0.5cm}} &
                                          \mbox{otherwise}         
\end{array} \right.  \label{eq:p2} 
\end{eqnarray} 
where $a'$ is determined by the adjacency condition
$A_{a,a'}=A_{a',b}=1$. 

We now study the operator $P(n,-n\lambda)$
for level $n+1$ fusion. With the help of Yang-Baxter equation
(\ref{eq:YBR}) we can show 
that this operator satisfies
\vspace*{-2.cm}
\be
\setlength{\unitlength}{0.0115in}%
\begin{picture}(222,156)(36,726)
\thicklines
\put( 74,675){\line( 1, 1){103.500}}
\put(178,778){\line(-1, 1){ 26}}
\put(152,804){\line(-1,-1){103.500}}
\put( 48,701){\line( 1,-1){25.75}}
\put(229,675){\line(-1, 1){103.500}}
\put(152,804){\line( 1,-1){103}}
\put(255,701){\line(-1,-1){ 25.75}}
\put(100,752){\line( 1,-1){ 77.500}}
\put(178,675){\line( 1, 1){ 51.500}}
\put(203,752){\line(-1,-1){ 77.500}}
\put(125,675){\line(-1, 1){ 51.500}}
\put( 64,696){\makebox(0,0)[lb]{\raisebox{0pt}[0pt][0pt]{\twlrm \sc
$-\lambda$}}}
\put(166,696){\makebox(0,0)[lb]{\raisebox{0pt}[0pt][0pt]{\twlrm \sc
$-\lambda$}}}
\put(148,776){\makebox(0,0)[lb]{\raisebox{0pt}[0pt][0pt]{\twlrm \sc $u$}}}
\put(107,750){\makebox(0,0)[lb]{\raisebox{0pt}[0pt][0pt]{\twlrm \sc
$-(n\-1)\lambda$}}}
\put( 69,659){\makebox(0,0)[lb]{\raisebox{0pt}[0pt][0pt]{\twlrm \sc $b$}}}
\put(231,659){\makebox(0,0)[lb]{\raisebox{0pt}[0pt][0pt]{\twlrm \sc $b$}}}
\put(148,807){\makebox(0,0)[lb]{\raisebox{0pt}[0pt][0pt]{\twlrm \sc $a$}}}
\put(120,780){\makebox(0,0)[lb]{\raisebox{0pt}[0pt][0pt]{\twlrm \sc
$b_1$}}}
\put( 36,699){\makebox(0,0)[lb]{\raisebox{0pt}[0pt][0pt]{\twlrm \sc
$b_n$}}}
\put(258,696){\makebox(0,0)[lb]{\raisebox{0pt}[0pt][0pt]{\twlrm \sc
$a_n$}}}
\put(183,777){\makebox(0,0)[lb]{\raisebox{0pt}[0pt][0pt]{\twlrm \sc
$a_1$}}}
\put(207,699){\makebox(0,0)[lb]{\raisebox{0pt}[0pt][0pt]{\twlrm \tiny
$u\+(n\-1)\lambda$}}}
\put(207,753){\makebox(0,0)[lb]{\raisebox{0pt}[0pt][0pt]{\twlrm \sc
$a_2$}}}
\end{picture}=
\setlength{\unitlength}{0.0115in}%
\begin{picture}(222,151)(48,724)
\thicklines
\put( 86,786){\line( 1,-1){103.500}}
\put(164,657){\line(-1, 1){103.500}}
\put( 60,760){\line( 1, 1){25.75}}
\put(241,786){\line(-1,-1){103.500}}
\put(164,657){\line( 1, 1){103}}
\put(267,760){\line(-1, 1){25.75}}
\put(112,709){\line( 1, 1){ 77.500}}
\put(190,786){\line( 1,-1){ 51.500}}
\put(215,709){\line(-1, 1){ 77.500}}
\put(137,786){\line(-1,-1){ 51.500}}
\put(160,685){\makebox(0,0)[lb]{\raisebox{0pt}[0pt][0pt]{\twlrm \sc $u$}}}
\put(161,650){\makebox(0,0)[lb]{\raisebox{0pt}[0pt][0pt]{\twlrm \sc $b$}}}
\put( 84,793){\makebox(0,0)[lb]{\raisebox{0pt}[0pt][0pt]{\twlrm \sc $a$}}}
\put(243,793){\makebox(0,0)[lb]{\raisebox{0pt}[0pt][0pt]{\twlrm \sc $a$}}}
\put(226,759){\makebox(0,0)[lb]{\raisebox{0pt}[0pt][0pt]{\twlrm \sc
$-\lambda$}}}
\put( 63,759){\makebox(0,0)[lb]{\raisebox{0pt}[0pt][0pt]{\twlrm \tiny
$u\+(n\-1)\lambda$}}}
\put(130,762){\makebox(0,0)[lb]{\raisebox{0pt}[0pt][0pt]{\twlrm \sc
$-\lambda$}}}
\put(167,708){\makebox(0,0)[lb]{\raisebox{0pt}[0pt][0pt]{\twlrm \sc
$-(n\-1)\lambda$}}}
\put(195,675){\makebox(0,0)[lb]{\raisebox{0pt}[0pt][0pt]{\twlrm \sc
$a_n$}}}
\put(126,672){\makebox(0,0)[lb]{\raisebox{0pt}[0pt][0pt]{\twlrm \sc
$b_n$}}}
\put( 48,756){\makebox(0,0)[lb]{\raisebox{0pt}[0pt][0pt]{\twlrm \sc
$b_1$}}}
\put(270,759){\makebox(0,0)[lb]{\raisebox{0pt}[0pt][0pt]{\twlrm \sc
$a_1$}}}
\end{picture} \label{eq:updown}
\ee

\be
\setlength{\unitlength}{0.0115in}%
\begin{picture}(222,156)(36,726)
\thicklines
\put( 74,675){\line( 1, 1){103.500}}
\put(178,778){\line(-1, 1){ 26}}
\put(152,804){\line(-1,-1){103.500}}
\put( 48,701){\line( 1,-1){25.75}}
\put(229,675){\line(-1, 1){103.500}}
\put(152,804){\line( 1,-1){103}}
\put(255,701){\line(-1,-1){25.75}}
\put(100,752){\line( 1,-1){ 77.500}}
\put(178,675){\line( 1, 1){ 51.500}}
\put(203,752){\line(-1,-1){ 77.500}}
\put(125,675){\line(-1, 1){ 51.500}}
\put( 64,696){\makebox(0,0)[lb]{\raisebox{0pt}[0pt][0pt]{\twlrm \sc
$-\lambda$}}}
\put(166,696){\makebox(0,0)[lb]{\raisebox{0pt}[0pt][0pt]{\twlrm \sc
$-\lambda$}}}
\put(148,776){\makebox(0,0)[lb]{\raisebox{0pt}[0pt][0pt]{\twlrm \sc $u$}}}
\put(107,750){\makebox(0,0)[lb]{\raisebox{0pt}[0pt][0pt]{\twlrm \sc
$-(n\-1)\lambda$}}}
\put( 69,659){\makebox(0,0)[lb]{\raisebox{0pt}[0pt][0pt]{\twlrm \sc $b$}}}
\put(231,659){\makebox(0,0)[lb]{\raisebox{0pt}[0pt][0pt]{\twlrm \sc $b$}}}
\put(148,807){\makebox(0,0)[lb]{\raisebox{0pt}[0pt][0pt]{\twlrm \sc $a$}}}
\put(120,780){\makebox(0,0)[lb]{\raisebox{0pt}[0pt][0pt]{\twlrm \sc
$b_1$}}}
\put( 36,699){\makebox(0,0)[lb]{\raisebox{0pt}[0pt][0pt]{\twlrm \sc
$b_n$}}}
\put(258,696){\makebox(0,0)[lb]{\raisebox{0pt}[0pt][0pt]{\twlrm \sc
$a_n$}}}
\put(183,777){\makebox(0,0)[lb]{\raisebox{0pt}[0pt][0pt]{\twlrm \sc
$a_1$}}}
\put(207,699){\makebox(0,0)[lb]{\raisebox{0pt}[0pt][0pt]{\twlrm \tiny
$u\+(n\-1)\lambda$}}}
\put(207,753){\makebox(0,0)[lb]{\raisebox{0pt}[0pt][0pt]{\twlrm \sc
$a_2$}}}
\end{picture}=
\setlength{\unitlength}{0.0115in}%
\begin{picture}(219,150)(39,723)
\thicklines
\put( 74,675){\line( 1, 1){103.500}}
\put(152,804){\line(-1,-1){103.500}}
\put( 48,701){\line( 1,-1){25.75}}
\put(229,675){\line(-1, 1){103.500}}
\put(152,804){\line( 1,-1){103}}
\put(255,701){\line(-1,-1){25.75}}
\put(203,752){\line(-1,-1){ 77.500}}
\put(125,675){\line(-1, 1){ 51.500}}
\put(100,752){\line( 1,-1){ 77.500}}
\put(178,675){\line( 1, 1){ 51.500}}
\put(109,699){\makebox(0,0)[lb]{\raisebox{0pt}[0pt][0pt]{\twlrm \sc
$-\lambda$}}}
\put( 70,698){\makebox(0,0)[lb]{\raisebox{0pt}[0pt][0pt]{\twlrm \sc $u$}}}
\put(181,780){\makebox(0,0)[lb]{\raisebox{0pt}[0pt][0pt]{\twlrm \sc $a$}}}
\put(153,804){\makebox(0,0)[lb]{\raisebox{0pt}[0pt][0pt]{\twlrm \sc
$b_1$}}}
\put( 39,698){\makebox(0,0)[lb]{\raisebox{0pt}[0pt][0pt]{\twlrm \sc $b$}}}
\put( 69,663){\makebox(0,0)[lb]{\raisebox{0pt}[0pt][0pt]{\twlrm \sc
$a_n$}}}
\put(258,699){\makebox(0,0)[lb]{\raisebox{0pt}[0pt][0pt]{\twlrm \sc
$a_{n-1}$}}}
\put(228,663){\makebox(0,0)[lb]{\raisebox{0pt}[0pt][0pt]{\twlrm \sc
$a_n$}}}
\put(214,699){\makebox(0,0)[lb]{\raisebox{0pt}[0pt][0pt]{\twlrm \sc
$-\lambda$}}}
\put(129,774){\makebox(0,0)[lb]{\raisebox{0pt}[0pt][0pt]{\twlrm \tiny
$u\+(n\-1)\lambda$}}}
\put(158,750){\makebox(0,0)[lb]{\raisebox{0pt}[0pt][0pt]{\twlrm \sc
$-(n\-1)\lambda$}}}
\end{picture} 
\ee \vspace*{1.5cm} \newline 
These properties will be useful in later sections.

Using the YBE (\ref{eq:YBR}) and the relations (\ref{eq:p2}) it is easy to
see that any 
two adjacent faces with spectral parameters $u+j\lambda$ and 
$u+(j-1)\lambda$ in (\ref{eq:projdef}) can be considered as an instance of
1 by 2 fusion.
So the properties (\ref{eq:1by2fu1})--(\ref{eq:1by2fu4}) imply
\begin{eqnarray} & & P(n,u)_{(a,b_1,\cdots
,b_{i-2},b_{i-1},b_i,b_{i+1},b_{i+2},
     \cdots ,b)}^{(a,a_1,\cdots ,\cdots ,a_{i-1},a_i,a_{i+1},
  \cdots ,\cdots ,b)}=0,\quad\mbox{if $\val {b_{i-1}}=\val {b_{i+1}}=1$}
\label{eq:projprop1}  \\ & &\nonumber  \\ & &
P(n,u)^{(a,a_1,\cdots ,a_{i-2},a_{i-1},a_i,a_{i+1},a_{i+2},
     \cdots ,b)}_{(a,b_1,\cdots ,\cdots ,b_{i-1},b_i,b_{i+1},
     \cdots ,\cdots ,b)}=0,\quad \mbox{if $\val {a_{i-1}}=\val
{a_{i+1}}=1$}
\label{eq:projprop2}\\ & &\nonumber  \\ & &
P(n,u)_{(a,b_1,\cdots ,\cdots ,b_{i-1},b_i,b_{i+1},
      \cdots ,\cdots ,b)}^{(a,a_1,\cdots ,L-2,L-3,L-2,
  \cdots ,b)}=P(n,u)_{(a,b_1,\cdots ,\cdots ,b_{i-1},b_i,
    b_{i+1},\cdots ,\cdots ,b)}^{(a,a_1,\cdots ,L-2,
              L-1,L-2,\cdots ,b)}   \nonumber \\  & &
\;\;+P(n,u)_{(a,b_1,\cdots
,\cdots ,b_{i-1},b_i,b_{i+1},\cdots ,
        \cdots ,b)}^{(a,a_1,\cdots ,L-2,{L},L-2,
   \cdots ,b)} \hspace*{3.1cm}\for D_L  \label{eq:projprop3}\\ & &\nonumber
 \\ & &
P(n,u)_{(a,b_1,\cdots ,\cdots ,b_{i-1},b_i,b_{i+1},\cdots 
        ,\cdots ,b)}^{(a,a_1,\cdots ,L-3,L,L-3,\cdots ,b)}=
   P(n,u)_{(a,b_1,\cdots ,b_{i-1},b_i,b_{i+1},\cdots 
 ,b)}^{(a,a_1,\cdots ,L-3,L-4,L-3,\cdots ,b)} \nonumber \\ & &
\;\;+P(n,u)_{(a,b_1,\cdots ,\cdots ,b_{i-1},b_i,b_{i+1},\cdots 
    ,\cdots ,b)}^{(a,a_1,\cdots ,L-3,L-2,L-3,\cdots ,b)} 
             \hspace*{3.3cm}\for E_L \label{eq:projprop4} \\ & &\nonumber 
\\ & &
P(n,u)_{(a,b_1,\cdots ,b_{i-1},b_i,b_{i+1},\cdots 
    ,b)}^{(a,a_1,\cdots ,a_{i-1},a_{i-1}-1,a_{i+1},
     \cdots ,b)}=P(n,u)_{(a,b_1,\cdots ,b_{i-1},b_i,
        b_{i+1},\cdots ,b)}^{(a,a_1,\cdots ,a_{i-1},
       a_{i-1}+1,a_{i+1},\cdots ,b)}\nonumber \\  & & \hspace*{5.9cm}\for
a_{i-1}=a_{i+1}\ \mbox{and}\ \val {a_{i-1}}=2
                              \label{eq:projprop5}
\end{eqnarray}

Let $\path(a,b,n)$ represent the set  of all allowed paths of $n$ steps
from $a$ to $b$  on the Dynkin diagrams excluding paths, such as in
(\ref{eq:projprop1}),
which only give zero elements for the projector.  Similarly, let
$P_{(a,b)}^{(n)}$ be the number of paths in the set $\path(a,b,n)$. For
convenience
let $\path(a,b,n)_i$ represent the $i$-th path in 
$\path(a,b,n)$ and $\path(a,b,n)_{i,j}$ be the $j$-th element of 
$\path(a,b,n)_i$. So we can rewrite the elements of the projector 
$P(n-1,u)$ to be 
  $$P(n-1,u)^{\path(a,b,n)_i}_{\path(a,b,n)_j}$$ 

The operator $P(n-1,u)$ is a square matrix and can be  written in  block
diagonal
form. By the properties (\ref{eq:projprop3})--(\ref{eq:projprop5}) we may
have
$|P(n-1,u)^{\path(a,b,n)_i}_{\path(a,b,n)_k}|=
|P(n-1,u)^{\path(a,b,n)_j}_{\path(a,b,n)_k}|$ or 
$P(n-1,u)^{\path(a,b,n)_i}_{\path(a,b,n)_k}=
P(n-1,u)^{\path(a,b,n)_j}_{\path(a,b,n)_k}+
P(n-1,u)^{\path(a,b,n)_{\ol j}}_{\path(a,b,n)_k}$ for any path
$\path(a,b,n)_k$ and $\;$suitable $\;j$ and $\;\ol j$. If so we treat the paths
$\path(a,b,n)_i$ and
$\path(a,b,n)_j$ as dependent paths. Otherwise the paths $\path(a,b,n)_i$
and
$\path(a,b,n)_j$ are independent. Suppose there are 
$m_{(a,b)}^{(n)}$ independent equations  deriving from the properties
(\ref{eq:projprop3})--(\ref{eq:projprop5}), then  there are
$A^{(n)}_{(a,b)}=P_{(a,b)}^{(n)}-m_{(a,b)}^{(n)}$  independent paths in
$\path(a,b,n)$
where $A^{(n)}_{(a,b)}$ is precisely the element of the fused adjacency
matrices
given in (\ref{eq:adjfusion}). 
We denote these independent paths  by $\alpha (a,b,n)$, $\alpha =1,2,\cdots
,A^{(n)}_{(a,b)}$. There  are many ways to choose the independent paths but
they all 
lead to equivalent fused models. The remaining paths should satisfy 
\begin{eqnarray} & &P(n-1,u)^{\path(a,b,n)_i}_{\beta(a,b,n)}=\sum_{\alpha
=1}^{A^{(n)}_{(a,b)}}
       \phi^{(i,\alpha)}_{(a,b,n)}
\;P(n-1,u)_{\beta(a,b,n)}^{\alpha(a,b,n)}; 
     \;\;i=1,2,\cdots ,m_{(a,b)}^{(n)}.\hspace*{0.5cm}\label{eq:pathexpand}
 \\ &
&P(n-1,u)^{a,a'_1,a'_2,\cdots ,a'_{n-1},b}_{a,b_1,b_2,\cdots ,b_{n-1},b}=0,
            \;\;\;\;\;\;\;\;\;\;n>h-2 \label{eq:pathclose}
\end{eqnarray} 
The value of $\phi^{(i,\alpha)}_{(a,b,n)}$ is zero if the path 
$\path(a,b,n)_i$ is independent of the path $\alpha(a,b,n)$ and is
$+1$ or $-1$ otherwise.  According to (\ref{eq:pathexpand}) we can divide
$\path(a,b,n)$ into 
$A^{(n)}_{(a,b)}$ independent sets defined by
\begin{equation} p(n,a,\alpha,b)=\{(\path(a,b,n)_i)|
\phi^{(i,\alpha)}_{(a,b,n)}\not=0\},\quad
           \alpha=1,2,\cdots ,A^{(n)}_{(a,b)}.
\end{equation} 
The first path in $p(n,a,\alpha,b)$ is $\alpha(a,b,n)$, the
$i$-th path is  denoted by $p(n,a,\alpha,b)_i$ and $p(n,a,\alpha,b)_{i,j}$
denotes the
$j$-th element  of the path $p(n,a,\alpha,b)_i$. We call
$\phi^{(i,\alpha)}_{(a,b,n)}$ the  parity of the path $\path(a,b,n)_i$
relative to the
independent path $\alpha(a,b,n)$. By (\ref{eq:updown}) it is obvious that
\begin{eqnarray}
&& \phi^{(\alpha,\alpha)}_{(a,b,n)}=\phi^{(i,i)}_{(a,b,n)}=1 ,\\
&&\phi^{(i,\alpha)}_{(a,b,n)}=\phi^{(i,\alpha)}_{(b,a,n)}. 
\end{eqnarray}
Equation (\ref{eq:pathclose}) holds because all paths in $\path(a,b,n)$
with
$n>h-2$ are related by (\ref{eq:projprop5}) to 
$P(n-1,u)^{(a,b_1,\cdots ,b_{i-2},b_{i-1},b_i,b_{i+1},b_{i+2},\cdots
,b)}_{(a,a_1,\cdots
,\cdots ,a_{i-1},a_i,a_{i+1},
\cdots ,\cdots ,b)}=0$ with $\val {b_{i-1}}=\val {b_{i+1}}=1$. As an example,
we give explicitly the parities of the first four fusion level of the $E_6$ 
model in Appendix B.

From (\ref{eq:projprop3})--(\ref{eq:projprop5}) it follows that the maximum
number
of terms on the right hand side of (\ref{eq:pathexpand}) is two. Let us set
$t^\alpha_k=P(n-1,u)^{\alpha(a,b,n)}_{\path(a,b,n)_k}$ and
$t^i_j=P(n-1,u)^{\path(a,b,n)_i}_{\path(a,b,n)_j}$. Then in general, by
(\ref{eq:projprop5}), we can divide the submatrix 
$P(n-1,u)^{\path(a,b,n)}_{\path(a,b,n)}$ of $P(n-1,u)$ into columns
$$\smat{ 
t_1^\beta &\cdots        &t_1^\beta              &\cdots\cr
t_2^\beta &\cdots        &t_2^\beta              &\cdots\cr 
 .        &              &    .                  &      \cr
 .        &              &    .                  &      \cr
 .        &              &    .                  &      \cr
t_{P_{(a,b)}^{(n)}}^\beta&\cdots             
                &t_{P_{(a,b)}^{(n)}}^\beta       &\cdots\cr
}\hspace*{1.cm} 
{\rm and}\hspace*{1.cm}
\smat{ 
t_1^j &\cdots        & t_1^j                      &\cdots\cr t_2^j
&\cdots        & t_2^j                      &\cdots\cr  .     &            
 &        
.                  &      \cr  .     &              &         .            
&      \cr  .     &              &         .                  &      \cr 
t_{P_{(a,b)}^{(n)}}^j
      &\cdots        &t_{P_{(a,b)}^{(n)}}^j       &\cdots\cr }$$ where 
$\alpha,\beta=1,2,\cdots ,A^{(n)}_{(a,b)}$ and $t_k^j$
$(1\le k\le P_{(a,b)}^{(n)})$ can be expressed as 
$$
t_k^j=\phi^{(j,\alpha)}_{(a,b,n)}t_k^\alpha +
  \phi^{(j,\beta)}_{(a,b,n)}t_k^\beta
$$ 
by (\ref{eq:projprop3})--(\ref{eq:projprop4}). 
For the $A_L$ models only the first group
appears and $A^{(n)}_{(a,b)}=1$. 
For the $D_L$ and $E_L$ models the second group is related to first group. 
It is easy to see that
$$
\det P(n-1,u)^{\path(a,b,n)}_{\path(a,b,n)}=0 \hspace*{1.cm} 
             {\rm and}\hspace*{1.cm} \det P(n-1,u)=0 \label{eq:zerodet}
$$ 
This means that the matrix $P(n-1,u)$ or
$P(n-1,u)^{\path(a,b,n)}_{\path(a,b,n)}$  is
reducible. The irreducible operator $\wp(n-1,u,a,b)$ is obtained from  the
reducible one
$P(n-1,u)^{\path(a,b,n)}_{\path(a,b,n)}$ by picking the independent
elements as follows
\be
\wp(n-1,a,b)=\smat{ t_1^1  & t_1^2 &\cdots & t_1^{A^{(n)}_{(a,b)}}     \cr
t_2^1  &
t_2^2 &\cdots & t_2^{A^{(n)}_{(a,b)}}     \cr
\vdots & \vdots&\ddots & \vdots                    \cr
t_{A^{(n)}_{(a,b)}}^1&t_{A^{(n)}_{(a,b)}}^2 &\cdots &
t_{A^{(n)}_{(a,b)}}^{A^{(n)}_{(a,b)}}  }\ee where
$t^\alpha_\beta=P(n-1,-n\lambda)^{\alpha(a,b,n)}_{\beta(a,b,n)}$. So
(\ref{eq:pathexpand}) can be written as
\be
 P(n-1,-n\lambda)^{\path(a,b,n)_j}_{\beta(a,b,n)}=\sum_{\alpha=1}^{A^{(n)}_{
(a,b)}}
\phi^{(j,\alpha)}_{(a,b,n)}\;\wp(n-1,a,b)^{\alpha(a,b,n)}_{\beta(a,b,n)}
 .\label{eq:RP}
\ee 
Finally, using (\ref{eq:RP}), the operator (\ref{eq:projdef}) can be
factorized
\vspace{-0.7cm}as
\be 
\sum_{\alpha=1}^{A^{(n)}_{(b_1,b)}}\!\wp(n-1,b_1,a)^{\alpha(b_1,b,n)}_{\beta
(b_1,b,n)}\!\sum_{i=1}^{P^{(n)}_{(b_1,b)}}\!\phi^{(i,\alpha)}_{(b_1,b,n)}
\hspace{-2.1cm}
\setlength{\unitlength}{0.0125in}%
\begin{picture}(270,63)(18,774)
\thicklines
\put( 90,795){\line( 0,-1){ 30}}
\put( 90,765){\line( 1, 0){195}}
\put(285,765){\line( 0, 1){ 30}}
\put(285,795){\line(-1, 0){195}}
\put(255,795){\line( 0,-1){ 30}}
\put(225,795){\line( 0,-1){ 30}}
\put(120,795){\line( 0,-1){ 30}}
\put( 84,756){\makebox(0,0)[lb]{\raisebox{0pt} [0pt][0pt]{\twlrm \sc
$b_1$}}}
\put(288,753){\makebox(0,0)[lb]{\raisebox{0pt} [0pt][0pt]{\twlrm \sc $b$}}}
\put(288,795){\makebox(0,0)[lb]{\raisebox{0pt} [0pt][0pt]{\twlrm \sc
$a_n$}}}
\put( 84,795){\makebox(0,0)[lb]{\raisebox{0pt} [0pt][0pt]{\twlrm \sc $a$}}}
\put(237,754){\makebox(0,0)[lb]{\raisebox{0pt} [0pt][0pt]{\twlrm \sc
$p(b_1,b,n)_{i\!,\!n}$}}}
\put(105,754){\makebox(0,0)[lb]{\raisebox{0pt} [0pt][0pt]{\twlrm \sc
$p(b_1,b,n)_{i\!,\!2}$}}}
\put(256,774){\makebox(0,0)[lb]{\raisebox{0pt}
       [0pt][0pt]{\twlrm \tiny $u\!\+\!(\!n$-$\!1\!)\lambda$}}}
\put(226,774){\makebox(0,0)[lb]{\raisebox{0pt}
       [0pt][0pt]{\twlrm \tiny $u\!\+\!(\!n$-$\!2\!)\lambda$}}}
\put(102,775){\makebox(0,0)[lb]{\raisebox{0pt}
       [0pt][0pt]{\twlrm \sc $u$}}}
\put(156,774){\makebox(0,0)[lb]{\raisebox{0pt} [0pt][0pt]{\twlrm \sc
$\cdots$}}}
\put(105,801){\makebox(0,0)[lb]{\raisebox{0pt} [0pt][0pt]{\twlrm \sc
$p(a,a_n,n)_{j\!,\!2}$}}}
\put(237,801){\makebox(0,0)[lb]{\raisebox{0pt} [0pt][0pt]{\twlrm \sc
$p(a,a_n,n)_{j\!,\!n}$}}}
\end{picture}\label{eq:how-fuse}\hspace{.1cm}
 \ee\vspace{.3cm}\newline 
This result implies that the fusion can be carried out if the
operator 
$\wp(n-1,b_1,a)$ is invertible. The existence of the inverse operator 
$\wp(n-1,b_1,a)^{-1}$ is shown  in Section~5.2.

\subsection{General Fusion }
\label{GeneralFusion}

 Let $m$ and $n$ be positive integers and define

\mbox{}\vspace{-.4in}
\bea & &\wf {W_{m\times n}}{u}bcd{\alpha}{\nu}{\beta}{\mu} =
\begin{picture}(50,50)(-5,19)\thicklines
\put(23,23){\scriptsize u}
\multiput(5,5)(0,40){2}{\line(1,0){40}} 
\multiput(5,5)(40,0){2}{\line(0,1){40}}
\put(1,2){\scriptsize $a$}\put(24,-1){\scriptsize $\alpha$}
\put(46,2){\scriptsize $b$} \put(46,24){\scriptsize $\nu$} 
\put(46,46){\scriptsize $c$}\put(24,48){\scriptsize $\beta$}
\put(0,46){\scriptsize $d$}\put(-2,24){\scriptsize $\mu$}
\end{picture}\h  =\sum_{j=1}^{L^{(m)}_{(d,a)}}\!\phi_{(a,d,m)}^{(j,\mu)}\!
 \sum_{\alpha_2,\cdots,\alpha_m}\h \no\\&&\no\\  &&\prod_{k=1}^{m} 
\row {W_{1\times n}}{u\-(m\-k)\lambda}{p(a,d,m)_{j,k}}{\nu(b,c,m)_k}{\nu(b,
     c,m)_{k+1}}{p(a,d,m)_{j,k+1}}{\hspace*{-0.3cm}\alpha_k
        \hspace*{-0.3cm}}{\hspace*{-0.3cm}\alpha_{k+1}\hspace*{-0.3cm}}.  
\label{eq:mnfusion}\eea 
Here $a=p(a,d,m)_{j,1}$, $b=\nu(b,c,m)_1$,
$c=\nu(b,c,m)_{m+1}$,
$d=p(a,d,m)_{j,m+1}$, $\alpha=\alpha_1$, $\beta=\alpha_{m+1}$ and the
summation
over $\alpha_k$ ranges over
$\alpha_k=1,\cdots,A_{(p(a,d,m)_{j,k},\nu(b,c,m)_k)}^{(n)}$.
The $1\times n$ fusion in turn is defined by
\bea &&\wf {W_{1\times n}}{u}bcd{\alpha}{\vspace*{-0.5cm}}{\beta}{}=\no\\
&&\sum_{i=1}^{L^{(n)}_{(a,b)}}\! 
\phi_{(a,b,n)}^{(i,\alpha)} \prod_{k=1}^{n} \wt 
W{\path(a,b,n)_{i,k}}{\hspace*{-0.3cm}\path(a,b,n)_{i,k+1}}{\hspace*{-0.3cm}
 \beta(d,c,n)_{k+1}}{\beta(d,c,n)_k}{\!u\+(k\-1)\lambda} \; .
\eea

\begin{figure}[tb]
\setlength{\unitlength}{0.0125in}%
\begin{picture}(365,207)(2,600)
\thicklines
\put(171,657){\circle*{6}}
\put(291,657){\circle*{6}}
\put(336,657){\circle*{6}}
\put(292,701){\circle*{6}}
\put(336,747){\circle*{6}}
\put(291,747){\circle*{6}}
\put(171,702){\circle*{6}}
\put(336,702){\circle*{6}}
\put(171,747){\circle*{6}}
\put(126,792){\line( 1, 0){255}}
\put(381,792){\line( 0,-1){180}}
\put(381,612){\line(-1, 0){255}}
\put(126,612){\line( 0, 1){180}}
\put(171,792){\line( 0,-1){ 45}}
\put(171,702){\line( 0,-1){ 90}}
\put(336,792){\line( 0,-1){ 45}}
\put(336,702){\line( 0,-1){ 90}}
\put(291,702){\line( 0,-1){ 90}}
\put(291,792){\line( 0,-1){ 45}}
\multiput(291,747)(0.00000,-8.18182){6}{\line( 0,-1){  4.091}}
\multiput(336,747)(0.00000,-8.18182){6}{\line( 0,-1){  4.091}}
\multiput(171,747)(0.00000,-8.18182){6}{\line( 0,-1){  4.091}}
\multiput(171,747)(7.74194,0.00000){16}{\line( 1, 0){  3.871}}
\multiput(171,702)(7.74194,0.00000){16}{\line( 1, 0){  3.871}}
\multiput(171,657)(7.74194,0.00000){16}{\line( 1, 0){  3.871}}
\put(126,747){\line( 1, 0){ 45}}
\put(126,657){\line( 1, 0){ 45}}
\put(126,702){\line( 1, 0){ 45}}
\put(291,702){\line( 1, 0){ 90}}
\put(291,657){\line( 1, 0){ 90}}
\put(291,747){\line( 1, 0){ 90}}
\put(291,675){\makebox(0,0)[lb]{\raisebox{0pt}[0pt][0pt]{\twlrm 
\tiny $u\!+\!(n\!-\! m)\lambda$}}}
\put(291,630){\makebox(0,0)[lb]{\raisebox{0pt}[0pt][0pt]{\twlrm 
\tiny $u\+\!(\!n\!\!-\!\!m\!\!-\!\!1\!)\lambda$}}}
\put(339,630){\makebox(0,0)[lb]{\raisebox{0pt}[0pt][0pt]{\twlrm 
\tiny $u\+(\!n\!\!-\!\!m\!)\lambda$}}}
\put(336,675){\makebox(0,0)[lb]{\raisebox{0pt}[0pt][0pt]{\twlrm 
\tiny $u\!\+\!(\!n\!\!-\!\!m\!\+\!1\!)\lambda$}}}
\put(142,766){\makebox(0,0)[lb]{\raisebox{0pt}[0pt][0pt]{\twlrm 
\sc $u$}}}
\put(128,628){\makebox(0,0)[lb]{\raisebox{0pt}[0pt][0pt]{\twlrm
 \tiny $u\!- \!(m\!- \!1)\lambda$}}}
\put(128,675){\makebox(0,0)[lb]{\raisebox{0pt}[0pt][0pt]{\twlrm
 \tiny $u\!-\!(m\!-\!2)\lambda$}}}
\put(339,766){\makebox(0,0)[lb]{\raisebox{0pt}[0pt][0pt]{\twlrm 
  \tiny $u\+(n\!- \!1)\lambda$}}}
\put(293,767){\makebox(0,0)[lb]{\raisebox{0pt}[0pt][0pt]{\twlrm 
  \tiny $u\+(n\!-\! 2)\lambda$}}}
\put(117,603){\makebox(0,0)[lb]{\raisebox{0pt}[0pt][0pt]{\twlrm 
\sc $a$}}}
\put(117,795){\makebox(0,0)[lb]{\raisebox{0pt}[0pt][0pt]{\twlrm 
\sc $d$}}}
\put(387,792){\makebox(0,0)[lb]{\raisebox{0pt}[0pt][0pt]{\twlrm 
\sc $c$}}}
\put(387,603){\makebox(0,0)[lb]{\raisebox{0pt}[0pt][0pt]{\twlrm 
\sc $b$}}}
\put(384,651){\makebox(0,0)[lb]{\raisebox{0pt}[0pt][0pt]{\twlrm 
\sc $\nu(b,c,m)_m$}}}
\put(384,696){\makebox(0,0)[lb]{\raisebox{0pt}[0pt][0pt]{\twlrm 
\sc $\nu(b,c,m)_{m- 1}$}}}
\put(384,741){\makebox(0,0)[lb]{\raisebox{0pt}[0pt][0pt]{\twlrm 
\sc $\nu(b,c,m)_2$}}}
\put( 77,651){\makebox(0,0)[lb]{\raisebox{0pt}[0pt][0pt]{\twlrm 
\sc $q(\!d,\!a,\!m\!)_{\!j,m}$}}}
\put( 77,744){\makebox(0,0)[lb]{\raisebox{0pt}[0pt][0pt]{\twlrm 
\sc $q(d,a,m)_{j,2}$}}}
\put( 74,693){\makebox(0,0)[lb]{\raisebox{0pt}[0pt][0pt]{\twlrm 
\sc $q(\!d,\!a,\!m\!)_{\!j,m\!-\!1}$}}}
\put(264,600){\makebox(0,0)[lb]{\raisebox{0pt}[0pt][0pt]{\twlrm 
\sc $p(\!a,\!b,\!n\!)_{\!i,\!n\!- \!1}$}}}
\put(150,600){\makebox(0,0)[lb]{\raisebox{0pt}[0pt][0pt]{\twlrm 
\sc $\path(a,b,n)_{i,2}$}}}
\put(321,600){\makebox(0,0)[lb]{\raisebox{0pt}[0pt][0pt]{\twlrm
 \sc $\path(a,b,n)_{i,n}$}}}
\put(321,795){\makebox(0,0)[lb]{\raisebox{0pt}[0pt][0pt]{\twlrm 
\sc $\beta(d,c,n)_n$}}}
\put(264,795){\makebox(0,0)[lb]{\raisebox{0pt}[0pt][0pt]{\twlrm
 \sc $\beta(d,c,n)_{n\!- \!1}$}}}
\put(150,795){\makebox(0,0)[lb]{\raisebox{0pt}[0pt][0pt]{\twlrm
 \sc $\beta(d,c,n)_2$}}}
\put(  2,712){\makebox(0,0)[lb]{\raisebox{0pt}[0pt][0pt]{\twlrm
 \sc $\disp{\sum_{i,j}}
\!\phi^{(i,\alpha)}_{(a,b,n)}\!\phi^{(j,\mu)}_{(a,d,m)}$}}}
\end{picture}
\caption{ Diagrammatic representation of the face weights of the $m\times
n$ fused ADE
models. Sites indicated with a solid circle are summed over all possible
spin states.}
\end{figure}
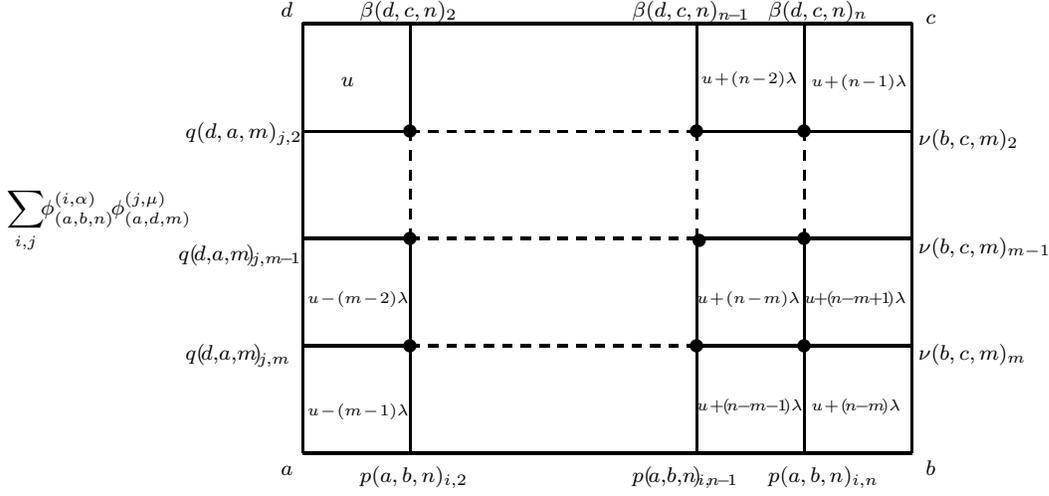

The fused face weights (\ref{eq:mnfusion}) associated with a bond state
$(a,\alpha,b)$ are
obtained by summing over the dependent paths within the set
$p(n,a,\alpha,b)$. Similar ideas have been applied to the fusion of the 
$A_n^{(1)}$ models in \cite{ZhHo:89,JKMO:88b}.
The resulting fused face weights depend on both the spin variables
$a,b,c,d$ and the
bond variables $\alpha,\beta,\mu,\nu$. For the $A_L$ models these bond
variables take only
the value $1$ whereas they take $A^{(n)}_{(a,b)}$ values for the adjacent
spins $a,b$ for the
$D_L$ and $E_L$ models. For the $A_L$ models the fused face weights do not
change at all if we change the paths $p(m,b,1,c)_1$ to
$p(m,b,1,c)_j$ and $p(n,d,1,c)_1$ to $p(n,d,1,c)_i$. But, for the $D_L$ and
$E_L$ models, we
have the following lemma:
\begin{lemma} If the path $\beta(d,c,n)$ is replaced with its dependent
path
$p(n,d,\beta,c)_j$ then the fused weight
\be
\wf {W_{m\times
n}}{u}bcd{\alpha}{\nu}{j}{\mu}=\sum_{\beta'=1}^{A^{(n)}_{(d,c)}}
\phi^{(j,\beta')}_{(d,c,n)}\;
\wf {W_{m\times n}}{u}bcd{\alpha}{\nu}{\beta'}{\mu}
\label{eq:mbynfusionprop1} \ee 
Similarly, if the path $\nu(b,c,m)$ is replaced by its 
dependent path $p(m,b,\nu,c)_j$ then
\be
\wf {W_{m\times
n}}{u}bcd{\alpha}{j}{\beta}{\mu}=\sum_{\nu'=1}^{A^{(m)}_{(b,c)}}
\phi^{(j,\nu')}_{(b,c,m)}\;
\wf {W_{m\times n}}{u}bcd{\alpha}{\nu'}{\beta}{\mu}
\label{eq:mbynfusionprop2}.\ee 
\end{lemma}

{\sl Proof:} Let us first consider $1\times n$ \vspace*{-0.7cm}fusion 
\be
\onerowprop {c_{i\+1}\;}{\;\;c_i}{\;c_{i\-1}}\label{eq:func1} 
\ee 
\vspace{0.1cm}\newline From (\ref{eq:how-fuse}) it follows that the
indices 
$(c_{i+1},c_i,c_{i-1})$ of the weight (\ref{eq:func1}) satisfy the 
properties (\ref{eq:1by2fu1})--(\ref{eq:1by2fu4}) (or 
(\ref{eq:projprop2})--(\ref{eq:projprop5})). This means that some of the 
fused weights in (\ref{eq:func1}) are dependent. In total there
are $A_{(c,d)}^{(n)}$ independent paths in the set $p(d,c,n)$.  Choosing an
independent path $c_i=\beta(d,c,n)$ we have the $1\times n$ fused face
\vspace*{-0.7cm} weight
\bea
\wf {W_{1\times n}}{u}bcd{\alpha}{\vspace*{-0.5cm}}{\beta}{}=
\onerowfusion {\beta(d,c,n)_2}{\beta(d,c,n)_n}{} \no \\ \no \\
\eea   
where $\alpha=1,2,\cdots,A_{(a,b)}^{(n)}$ and
$\beta=1,2,\cdots, A_{(c,d)}^{(n)}$. These represent the independent fused
face
weights. The others can be obtained from the independent weights via the 
\vspace*{-0.7cm} relation    
\bea
\wf {W_{1\times n}}{u}bcd{\alpha}{\vspace*{-0.5cm}}{j}{}&=&
\onerowfusion 
{p(n,\!d,\!\beta,\!c)_{j\!,\!2}}{p(n,\!d,\!\beta,\!c)_{j\!,\!n}\;}{}
\no\\&&\no\\
&=&\sum_{\beta'=1}^{A^{(n)}_{(d,c)}}\phi^{(j,\beta')}_{(d,c,n)}\; 
\wf {W_{1\times n}}{u}bcd{\alpha}{\vspace*{-0.5cm}}{\beta'}{} 
\label{eq:func2}\eea 
where $\phi^{(j,\beta')}_{(d,c,n)}$ are
the parities of the  path $p(n,d,\beta,c)_j$ relative to the dependent
paths $\beta'(d,c,n)$
with $\beta'=1,\cdots,\beta,\cdots,A^{(n)}_{(d,c)}\;$.   These properties
are exactly the
same as (\ref{eq:RP}). Furthermore, we have the following push through
\vspace*{-0.3cm}property from (\ref{eq:func2}) 
\bea  &&\sum_j^{P_{(a,\!b)}^{(n)}} 
\!\phi^{(j,\alpha)}_{(a,\!b,\!n)}
\setlength{\unitlength}{0.0125in}%
\begin{picture}(10,84)(80,731)
\thicklines
\put(201,753){\line( 0,-1){ 30}}
\put(231,753){\line( 0,-1){ 30}}
\put(171,753){\line( 0,-1){ 30}}
\put(111,753){\line( 0,-1){ 30}}
\multiput(171,753)(30,0){3}{\circle*{5}}
\put(111,723){\line( 1, 0){195}}
\put(306,723){\line( 0, 1){ 30}}
\put(306,753){\line(-1, 0){195}}
\put(309,711){\makebox(0,0)[lb]{\raisebox{0pt}[0pt][0pt] {\twlrm \sc $b$}}}
\put(210,732){\makebox(0,0)[lb]{\raisebox{0pt}[0pt][0pt] {\twlrm \sc $u$}}}
\put(175,732){\makebox(0,0)[lb]{\raisebox{0pt}[0pt][0pt] {\twlrm \sc
$u-\lambda$}}}
\put(180,711){\makebox(0,0)[lb]{\raisebox{0pt}[0pt][0pt] {\twlrm \sc
$p(a,\!b,\!n)_{j\!,\!i}$}}}
\put(132,711){\makebox(0,0)[lb]{\raisebox{0pt}[0pt][0pt] {\twlrm \sc
$p(a,\!b,\!n)_{j\!,i\-1}$}}}
\put(221,711){\makebox(0,0)[lb]{\raisebox{0pt}[0pt][0pt] {\twlrm \sc
$p(a,\!b,\!n)_{\!j\!,i\+1}$}}}
\put(111,783){\line( 0,-1){ 30}}
\put(111,753){\line( 1, 0){195}}
\put(306,753){\line( 0, 1){ 30}}
\put(306,783){\line(-1, 0){195}}
\put(171,783){\line( 0,-1){ 30}}
\put(201,783){\line( 0,-1){ 30}}
\put(231,783){\line( 0,-1){ 30}}
\put(165,786){\makebox(0,0)[lb]{\raisebox{0pt}[0pt][0pt] {\twlrm \sc
$c_{i\-1}$}}}
\put(198,786){\makebox(0,0)[lb]{\raisebox{0pt}[0pt][0pt] {\twlrm \sc
$c_i$}}}
\put(222,786){\makebox(0,0)[lb]{\raisebox{0pt}[0pt][0pt] {\twlrm \sc
$c_{i\+1}$}}}
\put(175,762){\makebox(0,0)[lb]{\raisebox{0pt}[0pt][0pt] {\twlrm \sc
$v-\lambda$}}}
\put(210,762){\makebox(0,0)[lb]{\raisebox{0pt}[0pt][0pt] {\twlrm \sc $v$}}}
\put(312,783){\makebox(0,0)[lb]{\raisebox{0pt}[0pt][0pt] {\twlrm \sc $c$}}}
\put(102,786){\makebox(0,0)[lb]{\raisebox{0pt}[0pt][0pt] {\twlrm \sc $d$}}}
\put(39,747){\makebox(0,0)[lb]{\raisebox{0pt}[0pt][0pt] {\twlrm \sc $$}}}
\put(102,747){\makebox(0,0)[lb]{\raisebox{0pt}[0pt][0pt] {\twlrm \sc $e$}}}
\put(312,747){\makebox(0,0)[lb]{\raisebox{0pt}[0pt][0pt] {\twlrm \sc $f$}}}
\put(105,714){\makebox(0,0)[lb]{\raisebox{0pt}[0pt][0pt] {\twlrm \sc $a$}}}
\end{picture} \no\\  &&
=\ \ \sum_{\beta=1}^{A^{(n)}_{(e,f)}}\ \ \ \
\setlength{\unitlength}{0.0125in}%
\begin{picture}(279,100)(68,720)
\thicklines
\put(216,717){\line( 0,-1){ 30}}
\put(246,717){\line( 0,-1){ 30}}
\put(186,717){\line( 0,-1){ 30}}
\put(126,717){\line( 0,-1){ 30}}
\put(126,687){\line( 1, 0){195}}
\put(321,687){\line( 0, 1){ 30}}
\put(321,717){\line(-1, 0){195}}
\put(213,783){\line( 0,-1){ 30}}
\put(243,783){\line( 0,-1){ 30}}
\put(183,783){\line( 0,-1){ 30}}
\put(123,783){\line( 0,-1){ 30}}
\put(123,753){\line( 1, 0){195}}
\put(318,753){\line( 0, 1){ 30}}
\put(318,783){\line(-1, 0){195}}
\put(324,675){\makebox(0,0)[lb]{\raisebox{0pt}[0pt][0pt]{\twlrm
 \sc $b$}}}
\put(227,696){\makebox(0,0)[lb]{\raisebox{0pt}[0pt][0pt]{\twlrm
 \sc $u$}}}
\put(189,696){\makebox(0,0)[lb]{\raisebox{0pt}[0pt][0pt]{\twlrm
 \sc $u\-\lambda$}}}
\put(197,675){\makebox(0,0)[lb]{\raisebox{0pt}[0pt][0pt]{\twlrm
 \tiny $p(a,\!b,\!n)_{j,i}$}}}
\put(149,675){\makebox(0,0)[lb]{\raisebox{0pt}[0pt][0pt]{\twlrm
 \tiny $p(a,\!b,\!n)_{j,i\-1}$}}}
\put(236,675){\makebox(0,0)[lb]{\raisebox{0pt}[0pt][0pt]{\twlrm
 \tiny $p(a,\!b,\!n)_{j,i\+1}$}}}
\put(197,720){\makebox(0,0)[lb]{\raisebox{0pt}[0pt][0pt]{\twlrm
 \tiny $\beta(e,\!f,\!n)_i$}}}
\put(151,720){\makebox(0,0)[lb]{\raisebox{0pt}[0pt][0pt]{\twlrm
 \tiny $\beta(e,\!f,\!n)_{i\-1}$}}}
\put(236,720){\makebox(0,0)[lb]{\raisebox{0pt}[0pt][0pt]{\twlrm
 \tiny $\beta(e,\!f,\!n)_{i\+1}$}}}
\put(117,720){\makebox(0,0)[lb]{\raisebox{0pt}[0pt][0pt]{\twlrm
 \sc $e$}}}
\put(114,744){\makebox(0,0)[lb]{\raisebox{0pt}[0pt][0pt]{\twlrm
 \sc $e$}}}
\put(327,714){\makebox(0,0)[lb]{\raisebox{0pt}[0pt][0pt]{\twlrm
 \sc $f$}}}
\put(324,744){\makebox(0,0)[lb]{\raisebox{0pt}[0pt][0pt]{\twlrm
 \sc $f$}}}
\put(146,741){\makebox(0,0)[lb]{\raisebox{0pt}[0pt][0pt]{\twlrm
 \tiny $q(e,\!f,\!n)_{k,i\-1}$}}}
\put(194,741){\makebox(0,0)[lb]{\raisebox{0pt}[0pt][0pt]{\twlrm
 \tiny $q(e,\!f,\!n)_{k,i}$}}}
\put(235,741){\makebox(0,0)[lb]{\raisebox{0pt}[0pt][0pt]{\twlrm
 \tiny $q(e,\!f,\!n)_{k,i\+1}$}}}
\put(189,762){\makebox(0,0)[lb]{\raisebox{0pt}[0pt][0pt]{\twlrm
 \sc $v\-\lambda$}}}
\put(227,762){\makebox(0,0)[lb]{\raisebox{0pt}[0pt][0pt]{\twlrm
 \sc $v$}}}
\put(210,786){\makebox(0,0)[lb]{\raisebox{0pt}[0pt][0pt]{\twlrm
 \sc $c_i$}}}
\put(240,786){\makebox(0,0)[lb]{\raisebox{0pt}[0pt][0pt]{\twlrm
 \sc $c_{i\+1}$}}}
\put(177,786){\makebox(0,0)[lb]{\raisebox{0pt}[0pt][0pt]{\twlrm
 \sc $c_{i\-1}$}}}
\put(324,783){\makebox(0,0)[lb]{\raisebox{0pt}[0pt][0pt]{\twlrm
 \sc $c$}}}
\put(114,783){\makebox(0,0)[lb]{\raisebox{0pt}[0pt][0pt]{\twlrm
 \sc $d$}}}
\put(114,681){\makebox(0,0)[lb]{\raisebox{0pt}[0pt][0pt]{\twlrm
 \sc $a$}}}
\put( 60,762){\makebox(0,0)[lb]{\raisebox{0pt}[0pt][0pt]{\twlrm
 \sc $\disp{\sum_k^{P_{(e,\!f)}^{(n)}}} 
\!\phi^{(k,\beta)}_{(e,\!f,\!n)}$}}}
\put( 60,699){\makebox(0,0)[lb]{\raisebox{0pt}[0pt][0pt]{\twlrm
 \sc $\disp{\sum_j^{P_{(a,\!b)}^{(n)}}} 
\!\phi^{(j,\alpha)}_{(a,\!b,\!n)}$}}}
\put( 48,729){\makebox(0,0)[lb]{\raisebox{0pt}[0pt][0pt]{\twlrm
 \sc $$}}}
\end{picture}\hspace*{2.2cm}
\label{eq:func3}\eea\vspace*{1.2cm}\newline 
This relation and (\ref{eq:func2}) imply
(\ref{eq:mbynfusionprop1}). Moreover, (\ref{eq:mbynfusionprop2}) follows
from 
(\ref{eq:mbynfusionprop1}) because of the symmetry $\wt  Wabcdu =\wt
Wadcbu$. 

By repeated use of (\ref{eq:YBR}), and with the help of the Lemma~2, we
obtain the
following theorem:
\begin{theorem} For a triple of positive integers $m,n,l$, the fused face
weights
(\ref{eq:mnfusion}) satisfy the Yang-Baxter equation
\bea &&\sum_{(\eta_1,\eta_2,\eta_3)}\!\sum_g \mbox{\small $
   \Wfln a{u}bgf{\mh\alpha\mh}{\eta_2}{\mh\eta_1\mh}{\rho}
   \Wfml g{v\-u}bcd{\mh\eta_2\mh}{\beta}{\mh\gamma\mh}{\eta_3}
   \Wf fvgde{\mh\eta_1\mh}{\eta_3}{\mh\mu\mh}{\nu}
$}
\hspace{1.1cm}\no\\&& \label{eq:fuYBR}\\
&&\hspace{-0.2cm}=\sum_{(\eta_1,\eta_2,\eta_3)}\!\sum_g \mbox{\small $
    \Wfln gucde{\mh\eta_1\mh}{\gamma}{\mh\mu\mh}{\eta_2}
    \Wfml f{v\-u}age{\mh\rho\mh}{\eta_3}{\mh\eta_2\mh}{\nu}
    \Wf avbcg{\mh\alpha\mh}{\beta}{\mh\eta_1\mh}{\eta_3}
$.}
\no\eea

\end{theorem}
 
By Lemma~1 the weights $W_{m\times n}$
(\ref{eq:mnfusion}) have zeros independent of the spins $a,b,c,d$ and bond
variables
$\alpha,\beta,\mu,\nu$. To remove these zeros we replace the $(M,N)$ fused
weight by
\bea
\wf {W_{m\times n}}{u}bcd{\alpha}{\nu}{\beta}{\mu}\rightarrow  
\wf {W_{m\times n}}{u}bcd{\alpha}{\nu}{\beta}{\mu}
\disp{\prod_{k=0}^{n-2}\prod_{j=0}^{m-1}{{\sin \lambda }\over 
                            {\sin [u+(k-j) \lambda ]}}}.
\eea\bigskip  
By construction it is obvious that 
$\wf {W_{m\times n}}{u}bcd{\alpha}{\nu}{\beta}{\mu}$ vanishes unless
\bea A^{(n)}_{a,b}\not=0\and \alpha=1,2,\cdots,A^{(n)}_{a,b} \no\\
A^{(n)}_{d,c}\not=0\and \beta=1,2,\cdots,A^{(n)}_{d,c} \no\\
A^{(m)}_{d,a}\not=0\and \mu=1,2,\cdots,A^{(m)}_{d,a} \no\\
A^{(m)}_{c,b}\not=0\and
\nu=1,2,\cdots,A^{(m)}_{c,b} 
\eea 
where the fused adjacency matrices are given by (\ref{eq:adjfusion}). In
particular,
\be
\wf {W_{m\times n}}{u}bcd{\alpha}{\nu}{\beta}{\mu} =0 \h
\as\;\mbox{if}\quad n=h-1
\quad{\rm or}\quad m=h-1\;. 
\ee

%% file: Fusion4.tex
\section{Row Transfer Matrix Fusion Hierarchy}
\label{Fusionhierarchy}
\setcounter{equation}{0}

Suppose that $\mbox{\boldmath $a$}(\mbox{\boldmath $\alpha$})$ and
$\mbox{\boldmath $b$}(\mbox{\boldmath $\beta$})$ are  allowed spin (bond)
configurations  of two consecutive rows of a lattice with $N$ 
columns and periodic  boundary conditions. The elements of the fused row
transfer
matrices  ${\bf T}^{(m,n)}(u)$ of the fused \ade models are given
\vspace*{-0.9cm}by 
\begin{eqnarray}
\langle\mbox{\boldmath $a,\alpha$}|{\bf T}^{m,n}(u)|\mbox{\boldmath
$b,\beta$}
 \rangle =
\prod_{j=1}^N\!\sum_{\{\eta_j\}} \Wf {a_j}u{b_j}{b_{j\+1}}{a_{j\+1}}{\mh\eta_j
        \mh}{\beta_j}{\mh\eta_{j\+1}\mh}{\alpha_j}=
\setlength{\unitlength}{0.0115in}%
\begin{picture}(48,90)(60,760)
\put( 75,780){\line( 0,-1){ 36}}
\put( 75,744){\line( 1, 0){ 30}}
\put(105,744){\line( 0, 1){ 36}}
\put(105,780){\line(-1, 0){ 30}}
\multiput(105,810)(0.00000,-7.82609){12}{\line( 0,-1){  3.913}}
\multiput( 75,810)(0.00000,-8.28571){11}{\line( 0,-1){  4.143}}
\put( 87,759){\makebox(0,0)[lb]{\raisebox{0pt}[0pt][0pt]{\twlrm\sc $ u$}}}
\put( 64,756){\makebox(0,0)[lb]{\raisebox{0pt}[0pt][0pt]{\twlrm \sc
$\alpha_j$}}}
\put(108,756){\makebox(0,0)[lb]{\raisebox{0pt}[0pt][0pt]{\twlrm \sc
$\beta_j$}}}
\put(108,774){\makebox(0,0)[lb]{\raisebox{0pt}[0pt][0pt]{\twlrm \sc
$b_{j\+1}$}}}
\put(108,738){\makebox(0,0)[lb]{\raisebox{0pt}[0pt][0pt]{\twlrm \sc
$b_j$}}}
\put( 64,738){\makebox(0,0)[lb]{\raisebox{0pt}[0pt][0pt]{\twlrm \sc
$a_j$}}}
\put( 60,774){\makebox(0,0)[lb]{\raisebox{0pt}[0pt][0pt]{\twlrm \sc
$a_{\!j\+1}$}}}
\end{picture}  
\eea\vspace*{0.3cm}\newline where $a_{N+1}=a_1$, $b_{N+1}=b_1$ and
$\eta_{N+1}=\eta_1$. Specifically, the Yang-Baxter equations
(\ref{eq:fuYBR}) imply
the commutation relations
\begin{equation} [{\bf T}^{m,n}(u),{\bf T}^{m,n'}(v)] = 0.
\label{eq:rowcommute}
\end{equation} 
Thus if $m$ is held fixed we obtain a hierarchy of commuting
families of  transfer matrices. These transfer matrices
satisfy the following remarkable functional equations:

\begin{theorem}[Fusion Hierarchy]
Let us define 
\be 
{\bf T}^{m,n}_k={\bf T}^{m,n}(u+k\lambda),\;\; {\bf T}^{m,0}_0=f^m_{-1}{\bf
I},
  \;\;f^m_n=[s^m_n]^N\;
\ee 
and
\be s_k^n =\prod_{j=0}^{n-1}
             {{\sin [u+(k-j) \lambda ]}\over {\sin \lambda }} .
\ee  
Then
\begin{equation} 
{\bf T}^{m,n}_0{\bf T}^{m,1}_n=f^m_n {\bf
T}^{m,n-1}_0+f^m_{n-1}{\bf T}^{m,n+1}_0
\label{eq:functionrelation} 
\end{equation}  
where the hierarchy closes at fusion level $h-1$ with 
\begin{equation}
{\bf T}^{p,h-1}=0.\label{eq:closure}
\end{equation}
\end{theorem}

\begin{theorem}[TBA Hierarchy]
If we further define
\be {\bf t}^{m,n}_0={{\bf T}^{m,n+1}_0{\bf T}^{m,n-1}_1\over
f^m_{-1}f^m_n}.
\ee 
Then the thermodynamic Bethe ansatz equations 
\be {\bf t}^{m,n}_0{\bf t}^{m,n}_1=({\bf I}+{\bf t}^{m,n+1}_0)
                               ({\bf I}+{\bf t}^{m,n-1}_1)
\label{eq:TBA}
\ee 
hold where
\be {\bf t}^{m,0}_0={\bf t}^{m,h-2}_0=0 .
\ee
\end{theorem}

 The main purpose of this section is to prove these theorems.  Clearly, the
functional
equations for the $D_L$ and $E_L$ models are the same as those for the
$A_L$ models.
In the
$A_L$ case the fusion hierarchy of functional equations was obtained by
Bazhanov and
Reshetikhin \cite{BaRe:89}. 
Although intertwiners can be constructed \cite{PeZh:93} between the row
transfer
matrices of the $D$ or $E$ models and an associated $A$ model, these
intertwiners
do not relate all eigenvalues. Rather, only a subset of common eigenvalues
are
intertwined.  As a consequence, the functional relations of the $D_L$ and
$E_L$ models cannot be obtained from those of the $A_L$ models using
intertwiners
alone. Instead it is necessary to prove these functional equations directly
for the $D_L$ and $E_L$ models as is done here. 
 
In Section~3 we described fusion of the \ade models corresponding to
the symmetric representation of the tensor products of
$n$ elementary blocks. To prove the theorems we need the fusion procedure 
corresponding to antisymmetric
representations. We therefore now describe the antisymmetric fusion of the
tensor
product of
$2$ elementary blocks. The symmetric and antisymmetric fusion
procedures are orthogonal to each other in the sense that

\be
\sum_{c\in antisy} \sum_{e\in sy}
\onebytwoface abdd{\;e}{\;c}{\;\;u}{\mh\;\;\;\; u\+\lambda}{\h}\h=0.
\label{eq:orth}
\ee 
\vspace{.3in}

\noindent
From 
(\ref{eq:1by2fu2})--(\ref{eq:1by2fu4}) we can indeed see that
(\ref{eq:orth}) holds
where the antisymmetric sum is defined by
\be
\!\!\sum_{c\in antisy} 
\onebytwoface abdd{\;e}{\;c}{\;\;u}{\mh\;\;\;\; u\+\lambda}{\h}
=\!\!\left\{
\begin{array}{ll}
\;\onebytwoface {a}{b}{{L}}{{L}}{\;e}{L\-2}{\;\;u}{\mh\;\;\;\;
u\+\lambda}{\h}\h
                                   & \mbox{$d={L}$ for $D_L$} \\&\\&\\
\;\onebytwoface {a}{b}{L}{L}{\;e}{L\-3}{\;\;u}{\mh\;\;\;\;
u\+\lambda}{\h}\h
                                   & \mbox{$d=L$ for $E_L$} \\&\\&\\

\vspace{.5in}
\onebytwoface {a}{b}{L\!\-\!2}{L\!\-\!2}{\;e}{L\!\-\!3}{\;\;u}{\mh\;\;\;\;
u\+\lambda}{\h}\h- 
  \onebytwoface
{a}{b}{L\!\-\!2}{L\!\-\!2}{\;e}{L\!\-\!1}{\;\;u}{\mh\;\;\;\;
u\+\lambda}{\h}\h- 
  \onebytwoface {a}{b}{L\!\-\!2}{L\!\-\!2}{\;e}{{L}}{\;\;u}{\mh\;\;\;\;
u\+\lambda}{\h}
                                                  \h  & \mbox{$d=L\!\-\!2$
for
$D_L$} \\&\\&\\

\vspace{.5in}
\onebytwoface {a}{b}{L\!\-\!3}{L\!\-\!3}{\;e}{\;L}{\;\;u}{\mh\;\;\;\;
u\+\lambda}{\h}\h- 
  \onebytwoface
{a}{b}{L\!\-\!3}{L\!\-\!3}{\;e}{L\!\-\!4}{\;\;u}{\mh\;\;\;\;
u\+\lambda}{\h}\h- 
  \onebytwoface
{a}{b}{L\!\-\!3}{L\!\-\!3}{\;e}{L\!\-\!2}{\;\;u}{\mh\;\;\;\;
u\+\lambda}{\h}
                                                  \h  & \mbox{$d=L\!\-\!3$
for
$E_L$} \\&\\&\\

\vspace{.5in}
\onebytwoface abdd{\;e}{d\-1}{\;\;u}{\mh\;\;\;\; u\+\lambda}{\h}\h
-\onebytwoface
abdd{\;e}{d\+1}{\;\;u}{\mh\;\;\;\; u\+\lambda}{\h}\h
                             & \mbox{otherwise.} \\&
\end{array} \right. \label{eq:antisym}
\ee 

Furthermore (\ref{eq:orth}) implies that
\be
\sum_{c\in antisy}
\onebytwoface abdd{\;e}{\;c}{\;\;u}{\mh\;\;\;\; u\+\lambda}{\h}\h=0\h\h
\mbox{unless $a=b$}.
\ee 
\vspace{.3in}

\noindent
Hence, for the $D_L$ models, we can construct the antisymmetric fusion by
\be
\onebytwoface
aabb{}{\;\v\hbox{\circle{4}}}{\;\;u}{\mh\;\;\;\;u\+\lambda}{\h}
=\left\{ \begin{array}{ll}
 -\disp{\sum_{c\in antisy}}
  \onebytwoface {1}{1}{2}{2}{\;2}{\;c}{\;\;u}{\mh\;\;\;\; u\+\lambda}{\h}\h

                                            & \mbox{$a=1,b=2$} \\&\\ &\\
 \;\;\disp{\sum_{c\in antisy}}\;\;
\;\onebytwoface {{L}}{{L}}{L\-2}{L\-2}{L\-2}{\;c}{\;\;u}{\mh\;\;\;\;
u\+\lambda}{\h}\h
                                                    & \mbox{$a={L},b=L\-2$}
\\&\\&\\
\;\; \disp{\sum_{c\in antisy}} A_{b,c}\!
\onebytwoface aabb{a\-1}{\;c}{\;\;u}{\mh\;\;\;\; u\+\lambda}{\h}\h
                                                 & \mbox{otherwise. }\\ &
\end{array} \right. \label{eq:antifusionD}
\ee    

\bigskip
\noindent
Similarly, the antisymmetric fusion for the $E_L$ models is given by 
\be
\onebytwoface
aabb{}{\;\v\hbox{\circle{4}}}{\;\;u}{\mh\;\;\;\;u\+\lambda}{\h}
=\left\{ \begin{array}{ll}
\;-\disp{\sum_{c\in antisy}}\h
  \onebytwoface {1}{1}{2}{2}{\;2}{\;c}{\;\;u}{\mh\;\;\;\; u\+\lambda}{\h}\h

                                              & \mbox{$a=1,b=2$} \\&\\ &\\
\;-\disp{\sum_{c\in antisy}}\h
\onebytwoface {L\-3}{L\-3}{L\-2}{L\-2}{L\-4}{\;c}{\;\;u}{\mh\;\;\;\;
u\+\lambda}{\h}\h
                                                 & \mbox{$a=L\-3,b=L\-2$}\\
&\\&\\
\;\;\;\disp{\sum_{c\in antisy}}\h
\onebytwoface {L\-4}{L\-4}{L\-3}{L\-3}{L\-3}{\;c}{\;\;u}{\mh\;\;\;\;
u\+\lambda}{\h}\h 
                                                  \h & 
\mbox{$a=L\-4,b=L\-3$}\\
&\\&\\
\;\;-\disp{\sum_{c\in antisy}}\h
\onebytwoface LL{L\-3}{L\-3}{L\-3}{\;c}{\;\;u}{\mh\;\;\;\;
u\+\lambda}{\h}\h 
                                                   &  \mbox{$a=L,b=L\-3$}\\
&\\&\\
\h\disp{\sum_{c\in antisy}}\h
\onebytwoface aabb{a\-1}{\;c}{\;\;u}{\mh\;\;\;\; u\+\lambda}{\h}\h
                                                 & \mbox{otherwise.}\\ &
\end{array} \right. \label{eq:antifusionE}
\ee
   
\vspace{.2in}
By direct calculation we have
\be {S_a\over S_b}
\onebytwoface
acbb{}{\;\v\hbox{\circle{4}}}{\;\;u}{\mh\;\;\;\;u\+\lambda}{\h}
=\delta_{a,c} {S_a\over S_b}
\onebytwoface
aabb{}{\;\v\hbox{\circle{4}}}{\;\;u}{\mh\;\;\;\;u\+\lambda}{\h}
=\delta_{a,c} s_1^1 s^1_{-1} \label{eq:antifusion1}
\ee
\vspace{.6in}

{\sl Proof of Theorem 2:} For simplicity we prove the functional equations
only for
the case of $m=1$. The general case can be proved similarly. Representing
${\bf T}^{1,n}_0 {\bf T}^{1,1}_n$ graphically as

\be
\setlength{\unitlength}{0.0110in}%
\begin{picture}(450,230)(10,580)
\put(240,700){\oval(320, 40)[tr]}
\put(240,700){\oval(320, 40)[tl]}
\put(120,660){\circle*{5}}
\put(160,660){\circle*{5}}
\put(120,620){\circle*{5}}
\put(160,620){\circle*{5}}
\put(160,780){\circle*{5}}
\put(120,780){\circle*{5}}
\put(320,780){\circle*{5}}
\put(160,740){\circle*{5}}
\put(120,740){\circle*{5}}
\put(400,620){\circle*{5}}
\put(360,620){\circle*{5}}
\put(360,660){\circle*{5}}
\put(400,660){\circle*{5}}
\put(360,740){\circle*{5}}
\put(400,740){\circle*{5}}
\put(400,780){\circle*{5}}
\put(360,780){\circle*{5}}
\put(320,740){\circle*{5}}
\put(400,700){\circle*{5}}
\put(320,660){\circle*{5}}
\put(120,660){\line( 0,-1){ 40}}
\put(160,655){\line( 0,-1){ 40}}
\put(360,780){\line( 1, 0){ 80}}
\put(440,780){\line( 0,-1){160}}
\put(440,620){\line(-1, 0){ 80}}
\put(360,740){\line( 1, 0){ 80}}
\put(440,740){\line( 0,-1){ 80}}
\put(440,660){\line(-1, 0){ 80}}
\put(360,660){\line( 0,-1){ 40}}
\put(360,620){\line( 1, 0){ 40}}
\put(400,620){\line( 0, 1){160}}
\put(360,700){\line( 1, 0){ 80}}
\put(120,780){\line( 0,-1){ 60}}
\put(360,780){\line(-1, 0){280}}
\put( 80,780){\line( 0,-1){160}}
\put( 80,620){\line( 1, 0){280}}
\put( 80,700){\line( 1, 0){320}}
\put( 80,660){\line( 1, 0){280}}
\put(320,780){\line( 0,-1){ 60}}
\put(320,700){\line( 0,-1){ 80}}
\put( 80,740){\line( 1, 0){280}}
\put(120,685){\line( 0,-1){ 30}}
\put(120,700){\line( 0,-1){  5}}
\put(160,700){\line( 0,-1){  5}}
\put(160,680){\line( 0,-1){ 25}}
\put(360,660){\line( 0, 1){ 20}}
\put(360,700){\line( 0,-1){  5}}
\put(160,780){\line( 0,-1){ 60}}
\put(360,780){\line( 0,-1){ 60}}
\multiput(440,780)(0.00000,7.69231){7}{\line( 0, 1){  3.846}}
\multiput(440,830)(8.57143,0.00000){4}{\line( 1, 0){  4.286}}
\multiput(470,830)(0.00000,-7.93651){32}{\line( 0,-1){  3.968}}
\multiput(470,580)(-8.57143,0.00000){4}{\line(-1, 0){  4.286}}
\multiput(440,580)(0.00000,7.27273){6}{\line( 0, 1){  3.636}}
\multiput( 80,620)(0.00000,-7.27273){6}{\line( 0,-1){  3.636}}
\multiput( 80,580)(-8.57143,0.00000){4}{\line(-1, 0){  4.286}}
\multiput( 50,580)(0.00000,8.23529){9}{\line( 0, 1){  4.118}}
\multiput( 50,750)(0.00000,8.09524){11}{\line( 0, 1){  4.048}}
\multiput( 50,835)(8.57143,0.00000){4}{\line( 1, 0){  4.286}}
\multiput( 80,835)(0.00000,-8.46154){7}{\line( 0,-1){  4.231}}
\put(100,635){\makebox(0,0)[lb]{\raisebox{0pt}[0pt][0pt]{\twlrm 
\scriptsize $u$}}}
\put(130,635){\makebox(0,0)[lb]{\raisebox{0pt}[0pt][0pt]{\twlrm 
\scriptsize $u+\lambda$}}}
\put( 70,695){\makebox(0,0)[lb]{\raisebox{0pt}[0pt][0pt]{\twlrm 
$a$}}}
\put(445,695){\makebox(0,0)[lb]{\raisebox{0pt}[0pt][0pt]{\twlrm 
$b$}}}
\put(405,637){\makebox(0,0)[lb]{\raisebox{0pt}[0pt][0pt]{\twlrm 
\scriptsize $u+n\lambda$}}}
\put(405,680){\makebox(0,0)[lb]{\raisebox{0pt}[0pt][0pt]{\twlrm 
$b'$}}}
\put( 95,710){\makebox(0,0)[lb]{\raisebox{0pt}[0pt][0pt]{\twlrm 
\scriptsize $p(a,b',n)_{i,2}$}}}
\put(155,710){\makebox(0,0)[lb]{\raisebox{0pt}[0pt][0pt]{\twlrm 
\scriptsize $p(a,b',n)_{i,3}$}}}
\put(330,710){\makebox(0,0)[lb]{\raisebox{0pt}[0pt][0pt]{\twlrm 
\scriptsize $p(a,b',n)_{i,n}$}}}
\put(100,687){\makebox(0,0)[lb]{\raisebox{0pt}[0pt][0pt]{\twlrm 
\scriptsize $\mu(a,b',n)_2$}}}
\put(155,687){\makebox(0,0)[lb]{\raisebox{0pt}[0pt][0pt]{\twlrm 
\scriptsize $\mu(a,b',n)_3$}}}
\put(335,687){\makebox(0,0)[lb]{\raisebox{0pt}[0pt][0pt]{\twlrm 
\scriptsize $\mu(a,b',n)_{n}$}}}
\put( 10,739){\makebox(0,0)[lb]{\raisebox{0pt}[0pt][0pt]{\twlrm 
$\disp{\sum_{i=1}^{P_{(a,b')}^{(n)}}}$}}}
\put( 10,689){\makebox(0,0)[lb]{\raisebox{0pt}[0pt][0pt]{\twlrm
 $\disp{\sum_{\mu=1}^{A^{(n)}_{(a,b')}}}$}}}
\put( 15,657){\makebox(0,0)[lb]{\raisebox{0pt}[0pt][0pt]{\twlrm 
$\times\phi^{(i,\mu)}_{(a,b',n)}$}}}
\end{picture} 
\label{eq:t2proof1}
\ee 
and inserting (\ref{eq:g3p4}) we obtain the sum of
two terms:

\be
\setlength{\unitlength}{0.0110in}%
\begin{picture}(435,240)(30,550)
\put(180,620){\circle*{5}}
\put(140,620){\circle*{5}}
\put(140,740){\circle*{5}}
\put(180,740){\circle*{5}}
\put(340,740){\circle*{5}}
\put(340,780){\circle*{5}}
\put(180,780){\circle*{5}}
\put(140,780){\circle*{5}}
\put(140,580){\circle*{5}}
\put(180,580){\circle*{5}}
\put(340,580){\circle*{5}}
\put(380,740){\circle*{5}}
\put(420,740){\circle*{5}}
\put(420,700){\circle*{5}}
\put(420,660){\circle*{5}}
\put(420,620){\circle*{5}}
\put(380,620){\circle*{5}}
\put(340,620){\circle*{5}}
\put(100,700){\line( 0, 1){ 80}}
\put(100,780){\line( 1, 0){280}}
\put(380,780){\line( 0,-1){ 80}}
\put(380,700){\line(-1, 0){ 40}}
\put(340,700){\line( 0, 1){ 80}}
\put(340,780){\line( 0,-1){ 40}}
\put(340,740){\line( 1, 0){ 40}}
\put(380,740){\line(-1, 0){280}}
\put(100,660){\line( 1, 0){280}}
\put(380,660){\line( 0,-1){ 80}}
\put(380,580){\line(-1, 0){280}}
\put(100,580){\line( 0, 1){ 80}}
\put(100,620){\line( 1, 0){280}}
\put(380,620){\line( 0,-1){  5}}
\put(340,660){\line( 0,-1){ 80}}
\put(140,660){\line( 0,-1){ 80}}
\put(180,660){\line( 0,-1){ 80}}
\put(140,780){\line( 0,-1){ 80}}
\put(180,780){\line( 0,-1){ 80}}
\put(100,700){\line( 1, 0){240}}
\multiput(180,580)(0.00000,-8.57143){4}{\line( 0,-1){  4.286}}
\multiput(140,580)(0.00000,-8.57143){4}{\line( 0,-1){  4.286}}
\multiput(100,580)(0.00000,-8.57143){4}{\line( 0,-1){  4.286}}
\multiput(380,580)(0.00000,-8.57143){4}{\line( 0,-1){  4.286}}
\multiput(180,805)(0.00000,-8.57143){4}{\line( 0,-1){  4.286}}
\multiput(140,805)(0.00000,-8.57143){4}{\line( 0,-1){  4.286}}
\multiput(100,805)(0.00000,-8.57143){4}{\line( 0,-1){  4.286}}
\multiput(380,800)(0.00000,-8.57143){4}{\line( 0,-1){  4.286}}
\put(420,700){\line(1,-1){ 40}}
\put(380,660){\line( 1, 0){ 40}}
\put(420,660){\line(-1, 1){ 40}}
\put(460,700){\line(-1, 0){ 40}}
\put(420,700){\line(-1, 0){ 40}}
\put(380,700){\line( 0, 1){ 80}}
\put(380,780){\line( 1, 0){ 80}}
\put(460,780){\line( 0,-1){ 80}}
\put(460,700){\line(-1, 0){ 40}}
\put(420,700){\line( 0, 1){ 80}}
\put(380,740){\line( 1, 0){ 80}}
\put(420,580){\line(-1, 0){ 40}}
\put(380,580){\line( 0, 1){ 80}}
\put(380,660){\line( 1, 0){ 80}}
\put(460,660){\line( 0,-1){ 80}}
\put(460,580){\line(-1, 0){ 40}}
\put(420,580){\line( 0, 1){ 80}}
\put(380,620){\line( 1, 0){ 80}}
\multiput(420,580)(0.00000,-8.57143){4}{\line( 0,-1){  4.286}}
\multiput(460,580)(0.00000,-8.57143){4}{\line( 0,-1){  4.286}}
\multiput(420,810)(0.00000,-8.57143){4}{\line( 0,-1){  4.286}}
\multiput(460,810)(0.00000,-8.57143){4}{\line( 0,-1){  4.286}}
\put( 90,700){\makebox(0,0)[lb]{\raisebox{0pt}[0pt][0pt]{\twlrm $a$}}}
\put( 90,655){\makebox(0,0)[lb]{\raisebox{0pt}[0pt][0pt]{\twlrm $a$}}}
\put( 10,730){\makebox(0,0)[lb]{\raisebox{0pt}[0pt][0pt]{\twlrm 
$\disp{\sum_{i=1}^{P_{(a,b')}^{(n)}}}$}}}
\put(10,677){\makebox(0,0)[lb]{\raisebox{0pt}[0pt][0pt]{\twlrm 
$\disp{\sum_{\mu=1}^{A^{(n)}_{(a,b')}}}\phi^{(i,\mu)}_{(a,b',n)}$}}}
\put(15,642){\makebox(0,0)[lb]{\raisebox{0pt}[0pt][0pt]{\twlrm 
$(2\cos \lambda)^{-1}$}}}
\put(105,665){\makebox(0,0)[lb]{\raisebox{0pt}[0pt][0pt]{\twlrm 
\scriptsize $\mu(a,b',n)_2$}}}
\put(170,665){\makebox(0,0)[lb]{\raisebox{0pt}[0pt][0pt]{\twlrm 
\scriptsize $\mu(a,b',n)_3$}}}
\put(430,660){\makebox(0,0)[lb]{\raisebox{0pt}[0pt][0pt]{\twlrm $b'$}}}
\put(465,695){\makebox(0,0)[lb]{\raisebox{0pt}[0pt][0pt]{\twlrm $b$}}}
\put(465,655){\makebox(0,0)[lb]{\raisebox{0pt}[0pt][0pt]{\twlrm $b$}}}
\put(415,675){\makebox(0,0)[lb]{\raisebox{0pt}[0pt][0pt]{\twlrm 
$-\lambda$}}}
\put(345,665){\makebox(0,0)[lb]{\raisebox{0pt}[0pt][0pt]{\twlrm 
\scriptsize $\mu(a,b',n)_{n}$}}}
\put(335,687){\makebox(0,0)[lb]{\raisebox{0pt}[0pt][0pt]{\twlrm 
\scriptsize $p(a,b',n)_{i,n}$}}}
\put(105,687){\makebox(0,0)[lb]{\raisebox{0pt}[0pt][0pt]{\twlrm 
\scriptsize $p(a,b',n)_{i,2}$}}}
\put(180,687){\makebox(0,0)[lb]{\raisebox{0pt}[0pt][0pt]{\twlrm 
\scriptsize $p(a,b',n)_{i,3}$}}}
\end{picture}
\label{eq:t2proof2}\ee and
\be
\setlength{\unitlength}{0.0110in}%
\begin{picture}(435,240)(30,550)
\put(180,620){\circle*{5}}
\put(140,620){\circle*{5}}
\put(140,740){\circle*{5}}
\put(180,740){\circle*{5}}
\put(340,740){\circle*{5}}
\put(340,780){\circle*{5}}
\put(180,780){\circle*{5}}
\put(140,780){\circle*{5}}
\put(140,580){\circle*{5}}
\put(180,580){\circle*{5}}
\put(340,580){\circle*{5}}
\put(380,740){\circle*{5}}
\put(420,740){\circle*{5}}
\put(420,700){\circle*{5}}
\put(420,660){\circle*{5}}
\put(420,620){\circle*{5}}
\put(380,620){\circle*{5}}
\put(340,620){\circle*{5}}
\put(100,700){\line( 0, 1){ 80}}
\put(100,780){\line( 1, 0){280}}
\put(380,780){\line( 0,-1){ 80}}
\put(380,700){\line(-1, 0){ 40}}
\put(340,700){\line( 0, 1){ 80}}
\put(340,780){\line( 0,-1){ 40}}
\put(340,740){\line( 1, 0){ 40}}
\put(380,740){\line(-1, 0){280}}
\put(100,660){\line( 1, 0){280}}
\put(380,660){\line( 0,-1){ 80}}
\put(380,580){\line(-1, 0){280}}
\put(100,580){\line( 0, 1){ 80}}
\put(100,620){\line( 1, 0){280}}
\put(380,620){\line( 0,-1){  5}}
\put(340,660){\line( 0,-1){ 80}}
\put(140,660){\line( 0,-1){ 80}}
\put(180,660){\line( 0,-1){ 80}}
\put(140,780){\line( 0,-1){ 80}}
\put(180,780){\line( 0,-1){ 80}}
\put(100,700){\line( 1, 0){240}}
\multiput(180,580)(0.00000,-8.57143){4}{\line( 0,-1){  4.286}}
\multiput(140,580)(0.00000,-8.57143){4}{\line( 0,-1){  4.286}}
\multiput(100,580)(0.00000,-8.57143){4}{\line( 0,-1){  4.286}}
\multiput(380,580)(0.00000,-8.57143){4}{\line( 0,-1){  4.286}}
\multiput(180,805)(0.00000,-8.57143){4}{\line( 0,-1){  4.286}}
\multiput(140,805)(0.00000,-8.57143){4}{\line( 0,-1){  4.286}}
\multiput(100,805)(0.00000,-8.57143){4}{\line( 0,-1){  4.286}}
\multiput(380,800)(0.00000,-8.57143){4}{\line( 0,-1){  4.286}}
\put(420,700){\line(1,-1){ 40}}
\put(380,660){\line( 1, 0){ 40}}
\put(420,660){\line(-1, 1){ 40}}
\put(460,700){\line(-1, 0){ 40}}
\put(420,700){\line(-1, 0){ 40}}
\put(380,700){\line( 0, 1){ 80}}
\put(380,780){\line( 1, 0){ 80}}
\put(460,780){\line( 0,-1){ 80}}
\put(460,700){\line(-1, 0){ 40}}
\put(420,700){\line( 0, 1){ 80}}
\put(380,740){\line( 1, 0){ 80}}
\put(420,580){\line(-1, 0){ 40}}
\put(380,580){\line( 0, 1){ 80}}
\put(380,660){\line( 1, 0){ 80}}
\put(460,660){\line( 0,-1){ 80}}
\put(460,580){\line(-1, 0){ 40}}
\put(420,580){\line( 0, 1){ 80}}
\put(380,620){\line( 1, 0){ 80}}
\multiput(420,580)(0.00000,-8.57143){4}{\line( 0,-1){  4.286}}
\multiput(460,580)(0.00000,-8.57143){4}{\line( 0,-1){  4.286}}
\multiput(420,810)(0.00000,-8.57143){4}{\line( 0,-1){  4.286}}
\multiput(460,810)(0.00000,-8.57143){4}{\line( 0,-1){  4.286}}
\put( 90,700){\makebox(0,0)[lb]{\raisebox{0pt}[0pt][0pt]{\twlrm $a$}}}
\put( 90,655){\makebox(0,0)[lb]{\raisebox{0pt}[0pt][0pt]{\twlrm $a$}}}
\put( 10,733){\makebox(0,0)[lb]{\raisebox{0pt}[0pt][0pt]{\twlrm 
$\disp{\sum_{i=1}^{P_{(a,b')}^{(n)}}}$}}}
\put( 10,677){\makebox(0,0)[lb]{\raisebox{0pt}[0pt][0pt]{\twlrm 
$\disp{\sum_{\mu=1}^{A^{(n)}_{(a,b')}}}\phi^{(i,\mu)}_{(a,b',n)}$}}}
\put( 15,642){\makebox(0,0)[lb]{\raisebox{0pt}[0pt][0pt]{\twlrm 
$(2\cos \lambda)^{-1}$}}}
\put(105,665){\makebox(0,0)[lb]{\raisebox{0pt}[0pt][0pt]{\twlrm 
\scriptsize $\mu(a,b',n)_2$}}}
\put(170,665){\makebox(0,0)[lb]{\raisebox{0pt}[0pt][0pt]{\twlrm 
\scriptsize $\mu(a,b',n)_3$}}}
\put(430,660){\makebox(0,0)[lb]{\raisebox{0pt}[0pt][0pt]{\twlrm 
$b'$}}}
\put(465,695){\makebox(0,0)[lb]{\raisebox{0pt}[0pt][0pt]{\twlrm 
$b$}}}
\put(465,655){\makebox(0,0)[lb]{\raisebox{0pt}[0pt][0pt]{\twlrm 
$b$}}}
\put(415,675){\makebox(0,0)[lb]{\raisebox{0pt}[0pt][0pt]{\twlrm 
$\lambda$}}}
\put(345,665){\makebox(0,0)[lb]{\raisebox{0pt}[0pt][0pt]{\twlrm 
\scriptsize $\mu(a,b',n)_{n}$}}}
\put(335,687){\makebox(0,0)[lb]{\raisebox{0pt}[0pt][0pt]{\twlrm 
\scriptsize $p(a,b',n)_{i,n}$}}}
\put(105,687){\makebox(0,0)[lb]{\raisebox{0pt}[0pt][0pt]{\twlrm 
\scriptsize $p(a,b',n)_{i,2}$}}}
\put(180,687){\makebox(0,0)[lb]{\raisebox{0pt}[0pt][0pt]{\twlrm 
\scriptsize $p(a,b',n)_{i,3}$}}}
\end{picture} 
\label{eq:t2proof3}\ee

But now, by (\ref{eq:g1p2}), the second term (\ref{eq:t2proof3}) vanishes
unless
$p(a,b',n)_{i,n}=b$. In this case we can choose an
independent path with $\mu(a,b',n)_{n}=b$ so (\ref{eq:t2proof3}) becomes
\be
\setlength{\unitlength}{0.0110in}%
\begin{picture}(435,260)(30,550)
\put(180,620){\circle*{5}}
\put(140,620){\circle*{5}}
\put(140,740){\circle*{5}}
\put(180,740){\circle*{5}}
\put(340,740){\circle*{5}}
\put(340,780){\circle*{5}}
\put(180,780){\circle*{5}}
\put(140,780){\circle*{5}}
\put(140,580){\circle*{5}}
\put(180,580){\circle*{5}}
\put(340,580){\circle*{5}}
\put(380,740){\circle*{5}}
\put(420,740){\circle*{5}}
\put(420,700){\circle*{5}}
\put(420,660){\circle*{5}}
\put(420,620){\circle*{5}}
\put(380,620){\circle*{5}}
\put(340,620){\circle*{5}}
\put(100,700){\line( 0, 1){ 80}}
\put(100,780){\line( 1, 0){280}}
\put(380,780){\line( 0,-1){ 80}}
\put(380,700){\line(-1, 0){ 40}}
\put(340,700){\line( 0, 1){ 80}}
\put(340,780){\line( 0,-1){ 40}}
\put(340,740){\line( 1, 0){ 40}}
\put(380,740){\line(-1, 0){280}}
\put(100,660){\line( 1, 0){280}}
\put(380,660){\line( 0,-1){ 80}}
\put(380,580){\line(-1, 0){280}}
\put(100,580){\line( 0, 1){ 80}}
\put(100,620){\line( 1, 0){280}}
\put(380,620){\line( 0,-1){  5}}
\put(340,660){\line( 0,-1){ 80}}
\put(140,660){\line( 0,-1){ 80}}
\put(180,660){\line( 0,-1){ 80}}
\put(140,780){\line( 0,-1){ 80}}
\put(180,780){\line( 0,-1){ 80}}
\put(100,700){\line( 1, 0){240}}
\multiput(180,580)(0.00000,-8.57143){4}{\line( 0,-1){  4.286}}
\multiput(140,580)(0.00000,-8.57143){4}{\line( 0,-1){  4.286}}
\multiput(100,580)(0.00000,-8.57143){4}{\line( 0,-1){  4.286}}
\multiput(380,580)(0.00000,-8.57143){4}{\line( 0,-1){  4.286}}
\multiput(180,805)(0.00000,-8.57143){4}{\line( 0,-1){  4.286}}
\multiput(140,805)(0.00000,-8.57143){4}{\line( 0,-1){  4.286}}
\multiput(100,805)(0.00000,-8.57143){4}{\line( 0,-1){  4.286}}
\multiput(380,800)(0.00000,-8.57143){4}{\line( 0,-1){  4.286}}
\put(420,700){\line(-1,-1){ 40}}
\put(380,660){\line( 1, 0){ 40}}
\put(420,660){\line(1, 1){ 40}}
\put(460,700){\line(-1, 0){ 40}}
\put(420,700){\line(-1, 0){ 40}}
\put(380,700){\line( 0, 1){ 80}}
\put(380,780){\line( 1, 0){ 80}}
\put(460,780){\line( 0,-1){ 80}}
\put(460,700){\line(-1, 0){ 40}}
\put(420,700){\line( 0, 1){ 80}}
\put(380,740){\line( 1, 0){ 80}}
\put(420,580){\line(-1, 0){ 40}}
\put(380,580){\line( 0, 1){ 80}}
\put(380,660){\line( 1, 0){ 80}}
\put(460,660){\line( 0,-1){ 80}}
\put(460,580){\line(-1, 0){ 40}}
\put(420,580){\line( 0, 1){ 80}}
\put(380,620){\line( 1, 0){ 80}}
\multiput(420,580)(0.00000,-8.57143){4}{\line( 0,-1){  4.286}}
\multiput(460,580)(0.00000,-8.57143){4}{\line( 0,-1){  4.286}}
\multiput(420,810)(0.00000,-8.57143){4}{\line( 0,-1){  4.286}}
\multiput(460,810)(0.00000,-8.57143){4}{\line( 0,-1){  4.286}}
\put( 90,700){\makebox(0,0)[lb]{\raisebox{0pt}[0pt][0pt]{\twlrm $a$}}}
\put( 90,655){\makebox(0,0)[lb]{\raisebox{0pt}[0pt][0pt]{\twlrm $a$}}}
\put( 10,728){\makebox(0,0)[lb]{\raisebox{0pt}[0pt][0pt]{\twlrm 
$\disp{\sum_{i=1}^{P_{(a,b')}^{(n)}}}$}}}
\put( 10,674){\makebox(0,0)[lb]{\raisebox{0pt}[0pt][0pt]{\twlrm 
$\disp{\sum_{\mu=1}^{A^{(n)}_{(a,b')}}}\phi^{(i,\mu)}_{(a,b',n)}$}}}
\put( 15,639){\makebox(0,0)[lb]{\raisebox{0pt}[0pt][0pt]{\twlrm 
$(2\cos \lambda)^{-1}$}}}
\put(105,665){\makebox(0,0)[lb]{\raisebox{0pt}[0pt][0pt]{\twlrm 
\scriptsize $\mu(a,b',n)_2$}}}
\put(170,665){\makebox(0,0)[lb]{\raisebox{0pt}[0pt][0pt]{\twlrm 
\scriptsize $\mu(a,b',n)_3$}}}
\put(430,660){\makebox(0,0)[lb]{\raisebox{0pt}[0pt][0pt]{\twlrm 
$b'$}}}
\put(465,695){\makebox(0,0)[lb]{\raisebox{0pt}[0pt][0pt]{\twlrm 
$b$}}}
\put(465,655){\makebox(0,0)[lb]{\raisebox{0pt}[0pt][0pt]{\twlrm 
$b$}}}
\put(415,675){\makebox(0,0)[lb]{\raisebox{0pt}[0pt][0pt]{\twlrm 
$\lambda$}}}
\put(335,665){\makebox(0,0)[lb]{\raisebox{0pt}[0pt][0pt]{\twlrm 
\scriptsize $\mu(a,b',n)_{n}$}}}
\put(338,687){\makebox(0,0)[lb]{\raisebox{0pt}[0pt][0pt]{\twlrm 
\scriptsize $p(a,b',n)_{i,n}$}}}
\put(105,687){\makebox(0,0)[lb]{\raisebox{0pt}[0pt][0pt]{\twlrm 
\scriptsize $p(a,b',n)_{i,2}$}}}
\put(180,687){\makebox(0,0)[lb]{\raisebox{0pt}[0pt][0pt]{\twlrm 
\scriptsize $p(a,b',n)_{i,3}$}}}
\end{picture} 
\label{eq:t2proof4}\ee 
Using (\ref{eq:g1p2})--(\ref{eq:g1p3}), (\ref{eq:g3p3})--(\ref{eq:g3p8})
and
(\ref{eq:antifusionD})--(\ref{eq:antifusionE}), this can be
reduced to
\be
\setlength{\unitlength}{0.0110in}%
\begin{picture}(475,260)(36,550)
\put(460,620){\oval( 80, 20)[tr]}
\put(460,620){\oval( 80, 20)[tl]}
\put(460,740){\oval( 80, 20)[tr]}
\put(460,740){\oval( 80, 20)[tl]}
\put(460,580){\oval( 80, 20)[tr]}
\put(460,580){\oval( 80, 20)[tl]}
\put(460,700){\oval( 80, 20)[tr]}
\put(460,700){\oval( 80, 20)[tl]}
\put(180,620){\circle*{5}}
\put(140,620){\circle*{5}}
\put(140,740){\circle*{5}}
\put(180,740){\circle*{5}}
\put(340,740){\circle*{5}}
\put(340,780){\circle*{5}}
\put(180,780){\circle*{5}}
\put(140,780){\circle*{5}}
\put(140,580){\circle*{5}}
\put(180,580){\circle*{5}}
\put(340,580){\circle*{5}}
\put(340,620){\circle*{5}}
\put(460,660){\circle{8}}
\put(460,740){\circle{8}}
\put(460,620){\circle{8}}
\put(460,580){\circle{8}}
\put(460,780){\circle{8}}
\put(100,700){\line( 0, 1){ 80}}
\put(100,780){\line( 1, 0){280}}
\put(380,780){\line( 0,-1){ 80}}
\put(380,700){\line(-1, 0){ 40}}
\put(340,700){\line( 0, 1){ 80}}
\put(340,780){\line( 0,-1){ 40}}
\put(340,740){\line( 1, 0){ 40}}
\put(380,740){\line(-1, 0){280}}
\put(100,660){\line( 1, 0){280}}
\put(380,660){\line( 0,-1){ 80}}
\put(380,580){\line(-1, 0){280}}
\put(100,580){\line( 0, 1){ 80}}
\put(100,620){\line( 1, 0){280}}
\put(380,620){\line( 0,-1){  5}}
\put(340,660){\line( 0,-1){ 80}}
\put(140,660){\line( 0,-1){ 80}}
\put(180,660){\line( 0,-1){ 80}}
\put(140,780){\line( 0,-1){ 80}}
\put(180,780){\line( 0,-1){ 80}}
\put(100,700){\line( 1, 0){240}}
\multiput(180,580)(0.00000,-8.57143){4}{\line( 0,-1){  4.286}}
\multiput(140,580)(0.00000,-8.57143){4}{\line( 0,-1){  4.286}}
\multiput(100,580)(0.00000,-8.57143){4}{\line( 0,-1){  4.286}}
\multiput(380,580)(0.00000,-8.57143){4}{\line( 0,-1){  4.286}}
\multiput(180,805)(0.00000,-8.57143){4}{\line( 0,-1){  4.286}}
\multiput(140,805)(0.00000,-8.57143){4}{\line( 0,-1){  4.286}}
\multiput(100,805)(0.00000,-8.57143){4}{\line( 0,-1){  4.286}}
\multiput(380,800)(0.00000,-8.57143){4}{\line( 0,-1){  4.286}}
\multiput(420,580)(0.00000,-8.57143){4}{\line( 0,-1){  4.286}}
\multiput(460,580)(0.00000,-8.57143){4}{\line( 0,-1){  4.286}}
\multiput(420,810)(0.00000,-8.57143){4}{\line( 0,-1){  4.286}}
\multiput(460,810)(0.00000,-8.57143){4}{\line( 0,-1){  4.286}}
\multiput(500,580)(0.00000,-8.57143){4}{\line( 0,-1){  4.286}}
\multiput(500,810)(0.00000,-8.57143){4}{\line( 0,-1){  4.286}}
\put(420,740){\line( 1, 0){ 80}}
\put(420,620){\line( 1, 0){ 80}}
\put(420,780){\line( 0,-1){ 80}}
\put(420,700){\line( 1, 0){ 80}}
\put(500,700){\line( 0, 1){ 80}}
\put(500,780){\line(-1, 0){ 80}}
\put(420,660){\line( 0,-1){ 80}}
\put(420,580){\line( 1, 0){ 80}}
\put(500,580){\line( 0, 1){ 80}}
\put(500,660){\line(-1, 0){ 80}}
\put(460,655){\line( 0,-1){ 25}}
\put(460,615){\line( 0,-1){ 25}}
\put(460,735){\line( 0,-1){ 25}}
\put(460,775){\line( 0,-1){ 25}}
\put( 90,700){\makebox(0,0)[lb]{\raisebox{0pt}[0pt][0pt]{\twlrm $a$}}}
\put( 90,655){\makebox(0,0)[lb]{\raisebox{0pt}[0pt][0pt]{\twlrm $a$}}}
\put( 30,690){\makebox(0,0)[lb]{\raisebox{0pt}[0pt][0pt]{\twlrm 
$\disp{\sum_{i=1}^{P_{(a,b)}^{(n-1)}}}$}}}
\put( 30,614){\makebox(0,0)[lb]{\raisebox{0pt}[0pt][0pt]{\twlrm 
$\disp{\sum_{r=1}^{A^{(n-1)}_{(a,b)}}}$}}}
\put( 25,654){\makebox(0,0)[lb]{\raisebox{0pt}[0pt][0pt]{\twlrm 
$\times\phi^{(i,r)}_{(a,b,n\!-\!1)}$}}}
\put(285,665){\makebox(0,0)[lb]{\raisebox{0pt}[0pt][0pt]{\twlrm 
\scriptsize $r(a,b,n-1)_{n-2}$}}}
\put(275,687){\makebox(0,0)[lb]{\raisebox{0pt}[0pt][0pt]{\twlrm 
\scriptsize $p(a,b,n-1)_{i,n-2}$}}}
\put(180,687){\makebox(0,0)[lb]{\raisebox{0pt}[0pt][0pt]{\twlrm 
\scriptsize $p(a,b,n\-1)_{i,3}$}}}
\put(170,665){\makebox(0,0)[lb]{\raisebox{0pt}[0pt][0pt]{\twlrm 
\scriptsize $r(a,b,n\-1)_3$}}}
\put(105,687){\makebox(0,0)[lb]{\raisebox{0pt}[0pt][0pt]{\twlrm 
\scriptsize $p(a,b,n\-1)_{i,2}$}}}
\put(105,665){\makebox(0,0)[lb]{\raisebox{0pt}[0pt][0pt]{\twlrm 
\scriptsize $r(a,b,n\-1)_2$}}}
\put(380,685){\makebox(0,0)[lb]{\raisebox{0pt}[0pt][0pt]{\twlrm $b$}}}
\put(115,593){\makebox(0,0)[lb]{\raisebox{0pt}[0pt][0pt]{\twlrm $u$}}}
\put(342,595){\makebox(0,0)[lb]{\raisebox{0pt}[0pt][0pt]{\twlrm 
\tiny $u$+($n$-$2)\lambda$}}}
\put(422,595){\makebox(0,0)[lb]{\raisebox{0pt}[0pt][0pt]{\twlrm 
\tiny $u$+($n$-$1)\lambda$}}}
\put(468,595){\makebox(0,0)[lb]{\raisebox{0pt}[0pt][0pt]{\twlrm 
\scriptsize $u+n\lambda$}}}
\put(415,665){\makebox(0,0)[lb]{\raisebox{0pt}[0pt][0pt]{\twlrm $b$}}}
\put(505,655){\makebox(0,0)[lb]{\raisebox{0pt}[0pt][0pt]{\twlrm $b$}}}
\put(505,690){\makebox(0,0)[lb]{\raisebox{0pt}[0pt][0pt]{\twlrm $b$}}}
\put(385,660){\makebox(0,0)[lb]{\raisebox{0pt}[0pt][0pt]{\twlrm $b$}}}
\put(415,685){\makebox(0,0)[lb]{\raisebox{0pt}[0pt][0pt]{\twlrm $b$}}}
\put(505,735){\makebox(0,0)[lb]{\raisebox{0pt}[0pt][0pt]{\twlrm $c$}}}
\put(410,735){\makebox(0,0)[lb]{\raisebox{0pt}[0pt][0pt]{\twlrm $c$}}}
\put(410,610){\makebox(0,0)[lb]{\raisebox{0pt}[0pt][0pt]{\twlrm $d$}}}
\put(505,610){\makebox(0,0)[lb]{\raisebox{0pt}[0pt][0pt]{\twlrm $d$}}}
\end{picture}
\ee

By virtue of (\ref{eq:antifusion1}) this gives the first term $f^1_n {\bf
T}^{1,n-1}_0$ in the fusion hierarchy. From  the push through property
(\ref{eq:func3}) and (\ref{eq:p2})  of the $1\times 2$ fusion we can see
that the
path of $3$ steps 
$(\mu(a,b',n)_{n},b',b)$ in (\ref{eq:t2proof2}) satisfies the properties 
(\ref{eq:1by2fu1})--(\ref{eq:1by2fu4}). This together with the push through
property
(\ref{eq:func3}) ensures that the path of $n+1$ steps from $a$  to $b$ to
satisfies
(\ref{eq:projprop2})--(\ref{eq:projprop5}). Applying the push through
property
(\ref{eq:func3}) to the $n+1$ blocks we obtain the level $n+1$ fusion
transfer
matrix given by the second term $f^1_{n-1}{\bf T}^{1,n+1}_0$.

{\sl Proof of Theorem 3:} Following Kl\"umper and Pearce \cite{KlPe:92} the
functional equations
\be 
{\bf T}^{m,n}_0{\bf T}^{m,n}_1=f^m_{-1}f^m_n {\bf I}+
     {\bf T}^{m,n+1}_0{\bf T}^{m,n-1}_1 
\ee 
are derived by substituting the fusion hierarchy
(\ref{eq:functionrelation}) into the
identity
\be {\bf T}^{m,n}_0({\bf T}^{m,n-1}_1{\bf T}^{m,1}_n)= ({\bf T}^{m,n}_0{\bf
T}^{m,1}_n){\bf T}^{m,n-1}_1\; .
\ee 
This then yields
\bea 
{\bf t}^{m,n}_0{\bf t}^{m,n}_1&=&{({\bf T}^{m,n-1}_1{\bf T}^{m,n-1}_2)
({\bf
T}^{m,n+1}_0{\bf T}^{m,n+1}_1)\over f^m_0f^m_nf^m_{-1}f^m_{n+1}}
\no\\ &=&\left({\bf I}+{{\bf T}^{m,n}_1{\bf T}^{m,n-2}_2\over
f^m_0f^m_n}\right)
\left({\bf I}+{{\bf T}^{m,n+2}_0{\bf T}^{m,n}_1\over
f^m_{-1}f^m_{n+1}}\right)
\no\\ &=&({\bf I}+{\bf t}^{m,n-1}_1)({\bf I}+{\bf t}^{m,n+1}_0).   
\eea

The functional equations (\ref{eq:TBA}) are 
identical in form  to the equations of the thermodynamic Bethe ansatz 
\cite{Za:91,KlMe:90,KlMe:91,Martins:91}.  The fusion hierarchy for the
$A_L$ has  been solved \cite{KlPe:92} for the finite-size corrections 
and hence the central charges, scaling dimensions and critical exponents. 
A similar analysis can be carried out for the $D_L$ and $E_L$ models.

The functional equations of the elliptic $D_L$ models can be obtained
by straightforwardly replacing the $\sin u$ functions with the elliptic
functions $h(u)$ in Theorems {\sl 2} and {\sl 3}.   
The functional equations of the elliptic $A$ model are given in
\cite{BaRe:89}. Here  we have shown that the functional equations
of the elliptic $D_L$ model are identical in form to those of the 
elliptic $A_{2L+3}$ model.

%% file: Fusion5.tex
\section{Intertwiners and Symmetric Fused Weights}
\setcounter{equation}{0}

Here we extend the \ade intertwiners constructed in
\cite{PeZh:93} to the fused \ade models. We build symmetric fused
face weights and generalize the intertwining relation to apply directly to
the
symmetric face weights. We also construct the intertwiners between the row
transfer matrices of the fused \ade models.

\subsection{Intertwiners}
Let $A$ and $G$ be adjacency matrices of an $A$
and a $D$ or $E$ model respectively. These are 
square matrices with nonnegative integer elements. Then the adjacency
matrix 
$C$ is said to intertwine $A$ and $G$ if
\begin{equation}
   A C =C G. \label{eq:adjinter}
\end{equation}
In general $C$ is a rectangular matrix with nonnegative integer elements. 
Similarly, there is an intertwining relation 
between the symmetric face weights $W^A$ of $A$ model and the symmetric
face 
weights $W^G$ of the $D$ or $E$ models if
\cite{PeZh:93}, 
\begin{equation}
\setlength{\unitlength}{0.0115in}%
\begin{picture}(234,87)(45,739)
\thicklines
\put(111,774){\circle*{6}}
\multiput( 66,789)(-0.42857,-0.42857){8}{\makebox(0.4444,0.6667){\sevrm .}}
\put( 63,786){\vector(-1,-1){0}}
\multiput( 66,759)(0.42857,-0.42857){8}{\makebox(0.4444,0.6667){\sevrm .}}
\put( 69,756){\vector( 1,-1){0}}
\put( 81,804){\line( 1, 0){ 30}}
\put(111,804){\line( 1,-1){ 30}}
\put(141,774){\line(-1, 0){ 30}}
\put(141,774){\line(-1,-1){ 30}}
\put(111,744){\line(-1, 0){ 30}}
\put( 81,804){\line( 1,-1){ 30}}
\put(111,774){\line(-1,-1){ 30}}
\put( 81,804){\line(-1,-1){ 30}}
\put( 51,774){\line( 1,-1){ 30}}
\put(126,774){\vector( 1, 0){  3}}
\put( 96,804){\vector( 1, 0){  3}}
\put( 99,744){\vector( 1, 0){  3}}
\multiput( 96,759)(-0.42857,-0.42857){8}{\makebox(0.4444,0.6667){\sevrm .}}
\put( 93,756){\vector(-1,-1){0}}
\multiput( 96,789)(0.42857,-0.42857){8}{\makebox(0.4444,0.6667){\sevrm .}}
\put( 99,786){\vector( 1,-1){0}}
\multiput(126,789)(0.42857,-0.42857){8}{\makebox(0.4444,0.6667){\sevrm .}}
\put(129,786){\vector( 1,-1){0}}
\multiput(126,759)(-0.42857,-0.42857){8}{\makebox(0.4444,0.6667){\sevrm .}}
\put(123,756){\vector(-1,-1){0}}
\put(111,732){\makebox(0,0)[lb]{\raisebox{0pt}[0pt][0pt]{\twlrm \sc $b'$}}}
\put(144,768){\makebox(0,0)[lb]{\raisebox{0pt}[0pt][0pt]{\twlrm \sc $c'$}}}
\put(114,804){\makebox(0,0)[lb]{\raisebox{0pt}[0pt][0pt]{\twlrm \sc $d'$}}}
\put( 81,807){\makebox(0,0)[lb]{\raisebox{0pt}[0pt][0pt]{\twlrm \sc $d$}}}
\put( 72,771){\makebox(0,0)[lb]{\raisebox{0pt}[0pt][0pt]{\twlrm \sc
$W^A$}}}
\put( 45,771){\makebox(0,0)[lb]{\raisebox{0pt}[0pt][0pt]{\twlrm \sc $a$}}}
\put( 78,732){\makebox(0,0)[lb]{\raisebox{0pt}[0pt][0pt]{\twlrm \sc $b$}}}
\put(208,773){\circle*{6}}
\multiput(192,789)(-0.42857,-0.42857){8}{\makebox(0.4444,0.6667){\sevrm .}}
\put(189,786){\vector(-1,-1){0}}
\multiput(222,789)(-0.42857,-0.42857){8}{\makebox(0.4444,0.6667){\sevrm .}}
\put(219,786){\vector(-1,-1){0}}
\multiput(222,759)(0.42857,-0.42857){8}{\makebox(0.4444,0.6667){\sevrm .}}
\put(225,756){\vector( 1,-1){0}}
\multiput(192,759)(0.42857,-0.42857){8}{\makebox(0.4444,0.6667){\sevrm .}}
\put(195,756){\vector( 1,-1){0}}
\put(237,804){\line( 1,-1){ 30}}
\put(267,774){\line(-1,-1){ 30}}
\put(237,804){\line(-1,-1){ 30}}
\put(207,774){\line( 1,-1){ 30}}
\multiput(252,759)(-0.42857,-0.42857){8}{\makebox(0.4444,0.6667){\sevrm .}}
\put(249,756){\vector(-1,-1){0}}
\multiput(252,789)(0.42857,-0.42857){8}{\makebox(0.4444,0.6667){\sevrm .}}
\put(255,786){\vector( 1,-1){0}}
\put(237,804){\line(-1, 0){ 30}}
\put(207,804){\line(-1,-1){ 30}}
\put(177,774){\line( 1, 0){ 30}}
\put(177,774){\line( 1,-1){ 30}}
\put(207,744){\line( 1, 0){ 30}}
\put(222,804){\vector( 1, 0){  3}}
\put(222,744){\vector( 1, 0){  3}}
\put(192,774){\vector( 1, 0){  3}}
\put(228,771){\makebox(0,0)[lb]{\raisebox{0pt}[0pt][0pt]{\twlrm \sc
$W^G$}}}
\put(240,729){\makebox(0,0)[lb]{\raisebox{0pt}[0pt][0pt]{\twlrm \sc $b'$}}}
\put(204,729){\makebox(0,0)[lb]{\raisebox{0pt}[0pt][0pt]{\twlrm \sc $b$}}}
\put(171,768){\makebox(0,0)[lb]{\raisebox{0pt}[0pt][0pt]{\twlrm \sc $a$}}}
\put(204,807){\makebox(0,0)[lb]{\raisebox{0pt}[0pt][0pt]{\twlrm \sc $d$}}}
\put(240,804){\makebox(0,0)[lb]{\raisebox{0pt}[0pt][0pt]{\twlrm \sc $d'$}}}
\put(270,771){\makebox(0,0)[lb]{\raisebox{0pt}[0pt][0pt]{\twlrm \sc $c'$}}}
\put(159,768){\makebox(0,0)[lb]{\raisebox{0pt}[0pt][0pt]{\twlrm \sc $=$}}}
\end{picture}
\label{eq:faceinter}
\end{equation}
where 
\begin{equation}
\cella abcd{\nu}{c_1}{c_2}
\end{equation}

\vspace{.15in}\noindent
is a family of 
cells labelled by four bond variables. Here the cells vanish 
unless the spins $d,a$ are adjacent sites of $A$, the spins $c,b$ are 
adjacent sites of $G$ and the spins $a,b$ and $d,c$ are adjacent sites of 
the intertwining graph $C$. The bond variables $c_1(c_2)=1,2,\cdots,
C_{a,b}(C_{d,c})$, $\nu =1,2,\cdots,G_{c,b}$. These cells satisfy two 
unitarity conditions which can be written in the \vspace{-.4in} form
\begin{equation}
\begin{picture}(300,70)(50,25)
\thicklines
\put(80,30){$\disp{\sum_{(b,c_1,\nu_2)}}$}
\put(150,50){\vector(1,-1){10}} \put(150,50){\vector(-1,-1){10}}
\put(130,30){\line(1,1){10}}    \put(170,30){\line(-1,1){10}}
\put(130,30){\vector(1,-1){10}} \put(170,30){\vector(-1,-1){10}}
\put(150,10){\line(1,1){10}}    \put(150,10){\line(-1,1){10}}
\put(167,23){\tiny $b$} \put(150,53){\tiny $c$} \put(150,5){\tiny $a$} 
\put(125,30){\tiny $d$}
\multiput(159,15)(15,0){2}{\tiny $c_1$} 
\multiput(162,39)(8.5,0){2}{\tiny $\nu_2$} 
\put(131,41){\tiny $c_2$} \put(134,15){\tiny $\nu_1$} 
\put(201,41){\tiny $c_2'$} \put(200,15){\tiny $\nu_1'$} 
\put(170,30){\circle*{2}}
\put(190,50){\vector(1,-1){10}} \put(190,50){\vector(-1,-1){10}}
\put(170,30){\line(1,1){10}}    \put(210,30){\line(-1,1){10}}
\put(170,30){\vector(1,-1){10}} \put(210,30){\vector(-1,-1){10}}
\put(190,10){\line(1,1){10}}    \put(190,10){\line(-1,1){10}}  
\put(190,53){\tiny $c$} \put(190,5){\tiny $a$} \put(215,30){\tiny $d'$}
\put(170,30){\circle*{2}}
\put(233,30){$= \delta_{d,d'} \delta_{\nu_1,\nu_1'} \delta_{c_2,c_2'}\; $}
\end{picture} \label{eq:unitary1}
\end{equation}

\vspace{-.3in}
\begin{equation}
\begin{picture}(300,70)(48,25)
\thicklines
\put(80,30){$\disp{\sum_{(b,c_1,\nu_2)}}$}
\put(150,50){\vector(1,-1){10}} \put(150,50){\line(-1,-1){10}}
\put(130,30){\vector(1,1){10}}  \put(170,30){\line(-1,1){10}}
\put(130,30){\vector(1,-1){10}} \put(170,30){\line(-1,-1){10}}
\put(150,10){\vector(1,1){10}}  \put(150,10){\line(-1,1){10}}
\put(167,23){\tiny $b$} \put(150,53){\tiny $c$} \put(150,5){\tiny $a$} 
\put(125,30){\tiny $d$}
\multiput(159,15)(15,0){2}{\tiny $c_1$} \multiput(162,39)(8.5,0){2}{\tiny
$\nu_2$} 
\put(131,41){\tiny $c_2$} \put(134,15){\tiny $\nu_1$} 
\put(201,41){\tiny $c_2'$} \put(200,15){\tiny $\nu_1'$} 
\put(170,30){\circle*{2}}
\put(190,50){\line(1,-1){10}} \put(190,50){\vector(-1,-1){10}}
\put(170,30){\line(1,1){10}}  \put(210,30){\vector(-1,1){10}}
\put(170,30){\line(1,-1){10}} \put(210,30){\vector(-1,-1){10}}
\put(190,10){\line(1,1){10}}  \put(190,10){\vector(-1,1){10}}
\put(190,53){\tiny $c$} \put(190,5){\tiny $a$} \put(213,30){\tiny $d'$}
\put(223,30){$\disp{S_b S_d\over S_c S_a}\;\;=\delta_{d,d'} 
     \delta_{\nu_1,\nu_1'} \delta_{c_2,c_2'}\; .$}
\end{picture}
\label{eq:unitary2}
\end{equation}

\vspace{.3in}
Using the adjacency intertwining relation (\ref{eq:adjinter}) and the
fusion rules (\ref{eq:adjfusion}) it follows that the same intertwining 
relations hold between the fused adjacency matrices, that is,
\begin{equation}
 A^{(n)} C=C G^{(n)}. \label{eq:adjintern}
\end{equation} 
We therefore expect to find fused cells that intertwine between the fused
face
weights.

Let us perform the following gauge transformations for the cells
\bea
&&\cella abcd{}{}{}\;\; \mapsto \cella abcd{}{}{}
\;\;\sqrt{S^G_c \over S^A_a}{f^G_c\over f^A_a}\label{eq:cellgauge1}\\
&& \no  \\
&&\cellp badc\;\; \mapsto \cellp badc
\;\;\sqrt{S^A_d \over S^G_b}{f^A_d\over f^G_a} \label{eq:cellgauge2}
\eea
Here we do not need the bond variables because they take the value $1$ for
unfused face weights. The transformed cells can be fused in the same way as
the A 
models. The level $n$ fusion of the transformed cells
\vspace*{-1.0cm}(\ref{eq:cellgauge1}) is given by
\be
\setlength{\unitlength}{0.0110in}%
\begin{picture}(70,79)(100,735)
\put(100,760){\line( 0,-1){ 30}}
\put(100,730){\line( 1, 0){ 60}}
\put(160,730){\line( 0, 1){ 30}}
\put(160,760){\line(-1, 0){ 60}}
\put(100,760){\vector( 1, 0){ 35}}
\put(100,730){\vector( 1, 0){ 35}}
\put(100,730){\vector( 0, 1){ 20}}
\put(160,730){\vector( 0, 1){ 20}}
\put(130,765){\makebox(0,0)[lb]{\raisebox{0pt}[0pt][0pt]{\twlrm $\mu$}}}
\put( 95,765){\makebox(0,0)[lb]{\raisebox{0pt}[0pt][0pt]{\twlrm $c$}}}
\put( 95,715){\makebox(0,0)[lb]{\raisebox{0pt}[0pt][0pt]{\twlrm $d$}}}
\put(160,715){\makebox(0,0)[lb]{\raisebox{0pt}[0pt][0pt]{\twlrm $a$}}}
\put(165,765){\makebox(0,0)[lb]{\raisebox{0pt}[0pt][0pt]{\twlrm $b$}}}
\put(130,742){\sc $C_n$}
\end{picture}=
\setlength{\unitlength}{0.0110in}%
\begin{picture}(275,54)(90,725)
\put(130,720){\circle*{5}}
\put(160,720){\circle*{5}}
\put(330,720){\circle*{5}}
\put(100,750){\line( 0,-1){ 30}}
\put(100,720){\line( 1, 0){260}}
\put(360,720){\line( 0, 1){ 30}}
\put(360,750){\line(-1, 0){260}}
\put(130,750){\line( 0,-1){ 30}}
\put(130,720){\line( 1, 0){ 30}}
\put(160,720){\line( 0, 1){ 30}}
\put(330,750){\line( 0,-1){ 30}}
\put(100,750){\vector( 1, 0){ 17}}
\put(100,720){\vector( 1, 0){ 17}}
\put(130,720){\vector( 1, 0){ 17}}
\put(130,750){\vector( 1, 0){ 17}}
\put(330,750){\vector( 1, 0){ 17}}
\put(330,720){\vector( 1, 0){ 17}}
\put(100,730){\vector( 0,1){ 11}}
\put(130,730){\vector( 0,1){ 11}}
\put(160,730){\vector( 0,1){ 11}}
\put(330,730){\vector( 0,1){ 11}}
\put(360,730){\vector( 0,1){ 11}}
\put(160,755){\makebox(0,0)[lb]{\raisebox{0pt}[0pt][0pt]{\twlrm $c_2$}}}
\put( 90,710){\makebox(0,0)[lb]{\raisebox{0pt}[0pt][0pt]{\twlrm $d$}}}
\put( 90,750){\makebox(0,0)[lb]{\raisebox{0pt}[0pt][0pt]{\twlrm $c$}}}
\put(320,755){\makebox(0,0)[lb]{\raisebox{0pt}[0pt][0pt]{\twlrm
$c_{n-1}$}}}
\put(365,710){\makebox(0,0)[lb]{\raisebox{0pt}[0pt][0pt]{\twlrm $a$}}}
\put(365,750){\makebox(0,0)[lb]{\raisebox{0pt}[0pt][0pt]{\twlrm $b$}}}
\put(125,755){\makebox(0,0)[lb]{\raisebox{0pt}[0pt][0pt]{\twlrm $c_1$}}}
\end{picture}
\label{eq:cellfusion1}\ee 
\newline 
where the solid circles indicate a summation over all possible paths
$p(d,a,n)$
of the $A$ model. The fused cell  
satisfies the same properties with respect to the path $p(c,b,n)$ of the
$G$ model
as does the operator $P$ presented in 
(\ref{eq:projprop2})--(\ref{eq:projprop5}).
We can therefore restrict our attention to the independent  paths from $c$
to $b$ of
the $G$ model with  
$c_i=\mu(c,b,n)_{i+1}$. Applying the intertwining relation
(\ref{eq:faceinter})
to the $m\times n$ blocks we therefore obtain the intertwining relation
between the fused weights \vspace*{-0.3cm}$W^A_{m,n}$ and $W^G_{m,n}$ 
given \vspace{-0.3cm}by  (\ref{eq:mnfusion})
\newpage
\begin{equation}
\setlength{\unitlength}{0.0125in}%
\begin{picture}(226,82)(150,720)
\thicklines
\put(216,762){\circle*{7}}
\put(291,736){\circle*{7}}
\put(216,723){\line( 0, 1){ 40}}
\put(216,762){\line( 1, 1){ 17.500}}
\put(233,780){\line(-1, 0){ 71}}
\put(162,780){\line(-1,-1){ 17.500}}
\put(145,762){\line( 1, 0){ 71}}
\put(216,762){\line( 1, 1){ 17.500}}
\put(233,780){\line( 0,-1){ 44}}
\put(233,736){\line(-1,-1){ 17}}
\put(216,719){\line(-1, 0){ 71}}
\put(145,719){\line( 0, 1){ 43}}
\put(291,780){\line( 0,-1){ 43}}
\put(291,736){\line(-1,-1){ 17.500}}
\put(273,719){\line( 1, 0){ 71}}
\put(344,719){\line( 1, 1){ 17.500}}
\put(362,736){\line(-1, 0){ 71}}
\put(291,736){\line(-1,-1){ 17.500}}
\put(273,719){\line( 0, 1){ 43}}
\put(273,762){\line( 1, 1){ 18}}
\put(291,780){\line( 1, 0){ 71}}
\put(362,780){\line( 0,-1){ 44}}
\put(273,754){\vector( 0,-1){ 18}}
\put(145,754){\vector( 0,-1){ 18}}
\put(216,754){\vector( 0,-1){ 18}}
\put(233,767){\vector( 0,-1){ 18}}
\put(362,767){\vector( 0,-1){ 18}}
\put(291,767){\vector( 0,-1){ 18}}
\put(162,719){\vector( 1, 0){ 18}}
\put(167,762){\vector( 1, 0){ 17}}
\put(180,780){\vector( 1, 0){ 18}}
\put(309,780){\vector( 1, 0){ 17}}
\put(304,736){\vector( 1, 0){ 18}}
\put(291,719){\vector( 1, 0){ 18}}
\put(229,732){\vector( 1, 1){0}}
\put(287,733){\vector( 1, 1){0}}
\put(357,732){\vector( 1, 1){0}}
\put(158,776){\vector( 1, 1){0}}
\put(229,776){\vector( 1, 1){0}}
\put(287,777){\vector( 1, 1){0}}
\put(216,762){\line( 0,-1){ 43}}
\multiput(187,769)(125,-44){2}{\tiny $C_n$}
\multiput(218,748)(59,0){2}{\tiny $C_m$}
\put(175,741){\makebox(0,0)[lb]{\raisebox{0pt}[0pt][0pt]{\elvrm \sc
$W^A_{m,n}$}}}
\put(317,754){\makebox(0,0)[lb]{\raisebox{0pt}[0pt][0pt]{\elvrm \sc
$W^G_{m,n}$}}}
\put(140,710){\makebox(0,0)[lb]{\raisebox{0pt}[0pt][0pt]{\elvrm $a$}}}
\put(211,710){\makebox(0,0)[lb]{\raisebox{0pt}[0pt][0pt]{\elvrm $b$}}}
\put(233,727){\makebox(0,0)[lb]{\raisebox{0pt}[0pt][0pt]{\elvrm $b'$}}}
\put(140,762){\makebox(0,0)[lb]{\raisebox{0pt}[0pt][0pt]{\elvrm $d$}}}
\put(348,710){\makebox(0,0)[lb]{\raisebox{0pt}[0pt][0pt]{\elvrm $b$}}}
\put(264,758){\makebox(0,0)[lb]{\raisebox{0pt}[0pt][0pt]{\elvrm $d$}}}
\put(269,710){\makebox(0,0)[lb]{\raisebox{0pt}[0pt][0pt]{\elvrm $a$}}}
\put(366,727){\makebox(0,0)[lb]{\raisebox{0pt}[0pt][0pt]{\elvrm $b'$}}}
\put(366,776){\makebox(0,0)[lb]{\raisebox{0pt}[0pt][0pt]{\elvrm $c'$}}}
\put(287,784){\makebox(0,0)[lb]{\raisebox{0pt}[0pt][0pt]{\elvrm $d'$}}}
\put(162,784){\makebox(0,0)[lb]{\raisebox{0pt}[0pt][0pt]{\elvrm $d'$}}}
\put(233,780){\makebox(0,0)[lb]{\raisebox{0pt}[0pt][0pt]{\elvrm $c'$}}}
\put(251,745){\makebox(0,0)[lb]{\raisebox{0pt}[0pt][0pt]{\elvrm $=$}}}
\put(325,785){\makebox(0,0)[lb]{\raisebox{0pt}[0pt][0pt]{\twlrm $\mu$}}}
\put(368,750){\makebox(0,0)[lb]{\raisebox{0pt}[0pt][0pt]{\twlrm $\nu$}}}
\put(200,785){\makebox(0,0)[lb]{\raisebox{0pt}[0pt][0pt]{\twlrm $\mu$}}}
\put(235,760){\makebox(0,0)[lb]{\raisebox{0pt}[0pt][0pt]{\twlrm $\nu$}}}
\end{picture}\label{eq:unsyminter1}
\end{equation}
\newline 
Here summation is implied over each of the inner bond and spin variables.
Alternatively, we can fuse the transformed cells (\ref{eq:cellgauge2})
\vspace*{-1.0cm} giving
\be
\setlength{\unitlength}{0.0110in}%
\begin{picture}(70,79)(100,735)
\put(100,760){\vector( 0,-1){ 20}}
\put(100,730){\line( 1, 0){ 60}}
\put(160,760){\vector( 0, -1){ 20}}
\put(160,760){\line(-1, 0){ 60}}
\put(100,760){\vector( 1, 0){ 35}}
\put(100,730){\vector( 1, 0){ 35}}
\put(100,730){\line( 0, 1){ 30}}
\put(160,730){\line( 0, 1){ 30}}
\put(130,713){\makebox(0,0)[lb]{\raisebox{0pt}[0pt][0pt]{\twlrm $\mu$}}}
\put( 95,765){\makebox(0,0)[lb]{\raisebox{0pt}[0pt][0pt]{\twlrm $c$}}}
\put( 95,715){\makebox(0,0)[lb]{\raisebox{0pt}[0pt][0pt]{\twlrm $d$}}}
\put(160,715){\makebox(0,0)[lb]{\raisebox{0pt}[0pt][0pt]{\twlrm $a$}}}
\put(165,765){\makebox(0,0)[lb]{\raisebox{0pt}[0pt][0pt]{\twlrm $b$}}}
\put(130,742){\sc $C^T_n$}
\end{picture}=\sum_i^{P_{(d,a)}^{(n)}} \phi_{(d,a,n)}^{(i,\mu)}
\setlength{\unitlength}{0.0110in}%
\begin{picture}(275,54)(90,725)
\put(113,713){\tiny $p(\!d,\!a,\!n\!)_{i,\!2}$}
\put(156,713){\tiny $p(\!d,\!a,\!n\!)_{i,\!3}$}
\put(302,713){\tiny $p(\!d,\!a,\!n\!)_{i,\!n}$}
\put(100,750){\line( 0,-1){ 30}}
\put(100,720){\line( 1, 0){260}}
\put(360,720){\line( 0, 1){ 30}}
\put(360,750){\line(-1, 0){260}}
\put(130,750){\line( 0,-1){ 30}}
\put(130,720){\line( 1, 0){ 30}}
\put(160,720){\line( 0, 1){ 30}}
\put(330,750){\line( 0,-1){ 30}}
\put(100,750){\vector( 1, 0){ 17}}
\put(100,720){\vector( 1, 0){ 17}}
\put(130,720){\vector( 1, 0){ 17}}
\put(130,750){\vector( 1, 0){ 17}}
\put(330,750){\vector( 1, 0){ 17}}
\put(330,720){\vector( 1, 0){ 17}}
\multiput(100,749)(30,0){3}{\vector( 0,-1){ 15}}
\multiput(360,749)(-30,0){2}{\vector( 0,-1){ 15}}
\put(160,755){\makebox(0,0)[lb]{\raisebox{0pt}[0pt][0pt]{\twlrm $c_2$}}}
\put( 90,710){\makebox(0,0)[lb]{\raisebox{0pt}[0pt][0pt]{\twlrm $d$}}}
\put( 90,750){\makebox(0,0)[lb]{\raisebox{0pt}[0pt][0pt]{\twlrm $c$}}}
\put(320,755){\makebox(0,0)[lb]{\raisebox{0pt}[0pt][0pt]{\twlrm
$c_{n-1}$}}}
\put(365,710){\makebox(0,0)[lb]{\raisebox{0pt}[0pt][0pt]{\twlrm $a$}}}
\put(365,750){\makebox(0,0)[lb]{\raisebox{0pt}[0pt][0pt]{\twlrm $b$}}}
\put(125,755){\makebox(0,0)[lb]{\raisebox{0pt}[0pt][0pt]{\twlrm $c_1$}}}
\end{picture}
\label{eq:cellfusion2}
\ee
\newline 
The path $p(c,b,n)$ in the fused cell is satisfies the same properties as 
the operator $P$ of the $A$ model presented in (\ref{eq:projprop2}) and 
(\ref{eq:projprop5}). That is, it is independent of  $c_1,c_2,
\cdots,c_{n-1}$ if the fused cell is nonzero. We thus have another
intertwining 
relation for $W^A_{m,n}$ and \vspace*{-0.3cm}$W^G_{m,n}$
\begin{equation}
\setlength{\unitlength}{0.0125in}%
\begin{picture}(226,82)(150,730)
\thicklines
\put(216,762){\circle*{7}}
\put(291,736){\circle*{7}}
\put(216,723){\line( 0, 1){ 39}}
\put(216,762){\line( 1, 1){ 17.500}}
\put(233,780){\line(-1, 0){ 71}}
\put(162,780){\line(-1,-1){ 17.500}}
\put(145,762){\line( 1, 0){ 71}}
\put(216,762){\line( 1, 1){ 17.500}}
\put(233,780){\line( 0,-1){ 44}}
\put(233,736){\line(-1,-1){ 17}}
\put(216,719){\line(-1, 0){ 71}}
\put(145,719){\line( 0, 1){ 43}}
\put(291,776){\line( 0,-1){ 40}}
\put(291,736){\line(-1,-1){ 17.500}}
\put(273,719){\line( 1, 0){ 71}}
\put(344,719){\line( 1, 1){ 17.500}}
\put(362,736){\line(-1, 0){ 71}}
\put(291,736){\line(-1,-1){ 17.500}}
\put(273,719){\line( 0, 1){ 43}}
\put(273,762){\line( 1, 1){ 18}}
\put(291,780){\line( 1, 0){ 71}}
\put(362,780){\line( 0,-1){ 44}}
\put(273,754){\vector( 0,-1){ 18}}
\put(145,754){\vector( 0,-1){ 18}}
\put(216,754){\vector( 0,-1){ 18}}
\put(233,767){\vector( 0,-1){ 18}}
\put(362,767){\vector( 0,-1){ 18}}
\put(291,767){\vector( 0,-1){ 18}}
\put(162,719){\vector( 1, 0){ 18}}
\put(167,762){\vector( 1, 0){ 17}}
\put(180,780){\vector( 1, 0){ 18}}
\put(309,780){\vector( 1, 0){ 17}}
\put(304,736){\vector( 1, 0){ 18}}
\put(291,719){\vector( 1, 0){ 18}}
\put(220,723){\vector( -1, -1){0}}
\put(278,724){\vector( -1, -1){0}}
\put(348,723){\vector( -1, -1){0}}
\put(149,767){\vector( -1, -1){0}}
\put(220,767){\vector( -1, -1){0}}
\put(278,768){\vector( -1, -1){0}}
\put(216,762){\line( 0,-1){ 43}}
\multiput(187,769)(125,-44){2}{\tiny $C^T_n$}
\multiput(218,748)(59,0){2}{\tiny $C^T_m$}
\put(175,741){\makebox(0,0)[lb]{\raisebox{0pt}[0pt][0pt]{\elvrm \sc
$W^G_{m,n}$}}}
\put(317,754){\makebox(0,0)[lb]{\raisebox{0pt}[0pt][0pt]{\elvrm \sc
$W^A_{m,n}$}}}
\put(140,710){\makebox(0,0)[lb]{\raisebox{0pt}[0pt][0pt]{\elvrm $a$}}}
\put(211,710){\makebox(0,0)[lb]{\raisebox{0pt}[0pt][0pt]{\elvrm $b$}}}
\put(233,727){\makebox(0,0)[lb]{\raisebox{0pt}[0pt][0pt]{\elvrm $b'$}}}
\put(140,762){\makebox(0,0)[lb]{\raisebox{0pt}[0pt][0pt]{\elvrm $d$}}}
\put(348,710){\makebox(0,0)[lb]{\raisebox{0pt}[0pt][0pt]{\elvrm $b$}}}
\put(264,758){\makebox(0,0)[lb]{\raisebox{0pt}[0pt][0pt]{\elvrm $d$}}}
\put(269,710){\makebox(0,0)[lb]{\raisebox{0pt}[0pt][0pt]{\elvrm $a$}}}
\put(366,727){\makebox(0,0)[lb]{\raisebox{0pt}[0pt][0pt]{\elvrm $b'$}}}
\put(366,776){\makebox(0,0)[lb]{\raisebox{0pt}[0pt][0pt]{\elvrm $c'$}}}
\put(287,784){\makebox(0,0)[lb]{\raisebox{0pt}[0pt][0pt]{\elvrm $d'$}}}
\put(162,784){\makebox(0,0)[lb]{\raisebox{0pt}[0pt][0pt]{\elvrm $d'$}}}
\put(233,780){\makebox(0,0)[lb]{\raisebox{0pt}[0pt][0pt]{\elvrm $c'$}}}
\put(251,745){\makebox(0,0)[lb]{\raisebox{0pt}[0pt][0pt]{\elvrm $=$}}}
\put(325,785){\makebox(0,0)[lb]{\raisebox{0pt}[0pt][0pt]{\twlrm $\mu$}}}
\put(368,750){\makebox(0,0)[lb]{\raisebox{0pt}[0pt][0pt]{\twlrm $\nu$}}}
\put(200,785){\makebox(0,0)[lb]{\raisebox{0pt}[0pt][0pt]{\twlrm $\mu$}}}
\put(235,760){\makebox(0,0)[lb]{\raisebox{0pt}[0pt][0pt]{\twlrm $\nu$}}}
\end{picture}\label{eq:unsyminter2}
\end{equation}
\newline 
where again summations are implied over inner bond and spin variables.
From the unitarity conditions (\ref{eq:unitary1})--(\ref{eq:unitary2}) 
it is easy to check the unitarity conditions for the fused 
\vspace*{-1.0cm}cells
\bea
&&\sum_{\mu,b,c_1}
\setlength{\unitlength}{0.0090in}%
\begin{picture}(221,108)(60,731)
\thicklines
\put(135,780){\line( 1,-1){ 30}}
\put(165,750){\line(-1,-1){ 60}}
\put(105,690){\line(-1, 1){ 30}}
\put( 75,720){\line( 1, 1){ 60}}
\multiput(150,765)(0.42857,-0.42857){8}{\makebox(0.4444,0.6667){\sevrm .}}
\put(153,762){\vector( 1,-1){0}}
\multiput( 90,705)(0.42857,-0.42857){8}{\makebox(0.4444,0.6667){\sevrm .}}
\put( 93,702){\vector( 1,-1){0}}
\multiput(102,747)(-0.42857,-0.42857){8}{\makebox(0.4444,0.6667){\sevrm .}}
\put( 99,744){\vector(-1,-1){0}}
\multiput(132,717)(-0.42857,-0.42857){8}{\makebox(0.4444,0.6667){\sevrm .}}
\put(129,714){\vector(-1,-1){0}}
\put(195,780){\line(-1,-1){ 30}}
\put(165,750){\line( 1,-1){ 60}}
\put(225,690){\line( 1, 1){ 30}}
\put(255,720){\line(-1, 1){ 60}}
\multiput(180,765)(-0.42857,-0.42857){8}{\makebox(0.4444,0.6667){\sevrm .}}
\put(177,762){\vector(-1,-1){0}}
\multiput(240,705)(-0.42857,-0.42857){8}{\makebox(0.4444,0.6667){\sevrm .}}
\put(237,702){\vector(-1,-1){0}}
\multiput(228,747)(0.42857,-0.42857){8}{\makebox(0.4444,0.6667){\sevrm .}}
\put(231,744){\vector( 1,-1){0}}
\multiput(198,717)(0.42857,-0.42857){8}{\makebox(0.4444,0.6667){\sevrm .}}
\put(201,714){\vector( 1,-1){0}}
\put(132,783){\makebox(0,0)[lb]{\raisebox{0pt}[0pt][0pt]{\twlrm \sc $c$}}}
\put(195,783){\makebox(0,0)[lb]{\raisebox{0pt}[0pt][0pt]{\twlrm \sc $c$}}}
\put(169,767){\makebox(0,0)[lb]{\raisebox{0pt}[0pt][0pt]{\twlrm \sc
$c_1$}}}
\put(152,767){\makebox(0,0)[lb]{\raisebox{0pt}[0pt][0pt]{\twlrm \sc
$c_1$}}}
\put(261,717){\makebox(0,0)[lb]{\raisebox{0pt}[0pt][0pt]{\twlrm \sc $d'$}}}
\put( 63,714){\makebox(0,0)[lb]{\raisebox{0pt}[0pt][0pt]{\twlrm \sc $d$}}}
\put(135,708){\makebox(0,0)[lb]{\raisebox{0pt}[0pt][0pt]{\twlrm \sc
$\mu$}}}
\put(183,708){\makebox(0,0)[lb]{\raisebox{0pt}[0pt][0pt]{\twlrm \sc
$\mu$}}}
\put( 93,753){\makebox(0,0)[lb]{\raisebox{0pt}[0pt][0pt]{\twlrm \sc
$\nu'$}}}
\put(120,732){\makebox(0,0)[lb]{\raisebox{0pt}[0pt][0pt]{\twlrm \sc
$C^1_n$}}}
\put(201,729){\makebox(0,0)[lb]{\raisebox{0pt}[0pt][0pt]{\twlrm \sc
$C^2_n$}}}
\put(228,753){\makebox(0,0)[lb]{\raisebox{0pt}[0pt][0pt]{\twlrm \sc
$\nu$}}}
\put(222,681){\makebox(0,0)[lb]{\raisebox{0pt}[0pt][0pt]{\twlrm \sc $a$}}}
\put(102,681){\makebox(0,0)[lb]{\raisebox{0pt}[0pt][0pt]{\twlrm \sc $a$}}}
\put(80,698){\makebox(0,0)[lb]{\raisebox{0pt}[0pt][0pt]{\twlrm \sc $c_2$}}}
\put(238,698){\makebox(0,0)[lb]{\raisebox{0pt}[0pt][0pt]{\twlrm \sc
$c_2$}}}
\put(161,737){\makebox(0,0)[lb]{\raisebox{0pt}[0pt][0pt]{\twlrm \sc $b$}}}
\end{picture}
=\delta_{\nu,\nu'}\delta_{d,d'} \label{eq:fuunitary1}\\
&&\no\\
&&\sum_{\mu,b,c_1}
\setlength{\unitlength}{0.0090in}%
\begin{picture}(225,111)(73,751)
\thicklines
\put( 87,759){\line( 1, 1){ 30}}
\put(117,789){\line( 1,-1){ 60}}
\put(177,729){\line(-1,-1){ 30}}
\put(147,699){\line(-1, 1){ 60}}
\multiput(102,774)(0.42857,0.42857){8}{\makebox(0.4444,0.6667){\sevrm .}}
\put(105,777){\vector( 1, 1){0}}
\multiput(162,714)(0.42857,0.42857){8}{\makebox(0.4444,0.6667){\sevrm .}}
\put(165,717){\vector( 1, 1){0}}
\multiput(120,726)(0.42857,-0.42857){8}{\makebox(0.4444,0.6667){\sevrm .}}
\put(123,723){\vector( 1,-1){0}}
\multiput(150,756)(0.42857,-0.42857){8}{\makebox(0.4444,0.6667){\sevrm .}}
\put(153,753){\vector( 1,-1){0}}
\put(267,759){\line(-1, 1){ 30}}
\put(237,789){\line(-1,-1){ 60}}
\put(177,729){\line( 1,-1){ 30}}
\put(207,699){\line( 1, 1){ 60}}
\multiput(252,774)(-0.42857,0.42857){8}{\makebox(0.4444,0.6667){\sevrm .}}
\put(249,777){\vector(-1, 1){0}}
\multiput(192,714)(-0.42857,0.42857){8}{\makebox(0.4444,0.6667){\sevrm .}}
\put(189,717){\vector(-1, 1){0}}
\multiput(234,726)(-0.42857,-0.42857){8}{\makebox(0.4444,0.6667){\sevrm .}}
\put(231,723){\vector(-1,-1){0}}
\multiput(204,756)(-0.42857,-0.42857){8}{\makebox(0.4444,0.6667){\sevrm .}}
\put(201,753){\vector(-1,-1){0}}
\put( 78,753){\makebox(0,0)[lb]{\raisebox{0pt}[0pt][0pt]{\twlrm \sc $d$}}}
\put(86,775){\makebox(0,0)[lb]{\raisebox{0pt}[0pt][0pt]{\twlrm \sc $c_2$}}}
\put(256,775){\makebox(0,0)[lb]{\raisebox{0pt}[0pt][0pt]{\twlrm \sc
$c_2$}}}
\put(273,753){\makebox(0,0)[lb]{\raisebox{0pt}[0pt][0pt]{\twlrm \sc $d'$}}}
\put(123,738){\makebox(0,0)[lb]{\raisebox{0pt}[0pt][0pt]{\twlrm \sc
$C_n^1$}}}
\put(204,687){\makebox(0,0)[lb]{\raisebox{0pt}[0pt][0pt]{\twlrm \sc $a$}}}
\put(176,734){\makebox(0,0)[lb]{\raisebox{0pt}[0pt][0pt]{\twlrm \sc $b$}}}
\put(161,705){\makebox(0,0)[lb]{\raisebox{0pt}[0pt][0pt]{\twlrm \sc
$c_1$}}}
\put(186,705){\makebox(0,0)[lb]{\raisebox{0pt}[0pt][0pt]{\twlrm \sc
$c_1$}}}
\put(147,762){\makebox(0,0)[lb]{\raisebox{0pt}[0pt][0pt]{\twlrm \sc
$\mu$}}}
\put(195,762){\makebox(0,0)[lb]{\raisebox{0pt}[0pt][0pt]{\twlrm \sc
$\mu$}}}
\put(105,723){\makebox(0,0)[lb]{\raisebox{0pt}[0pt][0pt]{\twlrm \sc
$\nu$}}}
\put(147,687){\makebox(0,0)[lb]{\raisebox{0pt}[0pt][0pt]{\twlrm \sc $a$}}}
\put(213,738){\makebox(0,0)[lb]{\raisebox{0pt}[0pt][0pt]{\twlrm \sc
$C^2_n$}}}
\put(231,792){\makebox(0,0)[lb]{\raisebox{0pt}[0pt][0pt]{\twlrm \sc $c$}}}
\put(120,792){\makebox(0,0)[lb]{\raisebox{0pt}[0pt][0pt]{\twlrm \sc $c$}}}
\put(243,726){\makebox(0,0)[lb]{\raisebox{0pt}[0pt][0pt]{\twlrm \sc
$\nu'$}}}
\end{picture}{S_b S_d\over S_c S_a}=\delta_{d,d'} 
     \delta_{\nu,\nu'} \label{eq:fuunitary2}
\eea\vspace*{0.7cm}\newline 
where $(C^1_n,C^2_n)$ is $(C_n,C^T_n)$ or $(C^T_n,C_n)$.
The bond variables $\nu=\nu'=1$ for $(C^1_n,C^2_n)=(C_n,C^T_n)$ 
because there is only one 
independent path between two spins of the fused $A$ models. In such cases
we  
discard the bond variable between adjacent spins.

The fused cells are given by both (\ref{eq:cellfusion1}) and 
(\ref{eq:cellfusion2}). They give the intertwining relations 
(\ref{eq:unsyminter1}) and (\ref{eq:unsyminter2}) respectively, either of
which
can be taken as the intertwining relation between the fused face weights of
the $A$
and $D$ or $E$ models. However, the fused cells (\ref{eq:cellfusion1}) and 
(\ref{eq:cellfusion2}) are independent. Since we need both the fused cells
and their
conjugates, the fused weights of the $D$ or $E$ models cannot
be obtained from those of the $A$ model and the fused cells
(\ref{eq:cellfusion1})
alone.

\subsection{Symmetric weights}

The fused face weights given by (\ref{eq:mnfusion}) are not 
symmetric, that is,
\be
\wf {W^s_{m\times n}}{u}bcd{\alpha}{\nu}{\beta}{\mu}
\not= {W^s_{n\times m}\mbox{$\left(\matrix{
d&\mu&a\cr\beta&&\alpha\cr c&\nu&b\cr}\biggm|\mbox{$u$}\right).$}}
\ee
To symmetrize the fused face weights we need to apply a gauge 
transformation.

Although the operator $P(n-1,u,a,b)$ does not have an inverse matrix,  
the $A^{(n)}_{(a,b)}\times A^{(n)}_{(a,b)}$ matrix 
$\wp(n-1,u,a,b)$ is nonsingular. This can be shown using intertwiners.
Specifically, from (\ref{eq:faceinter}), 
(\ref{eq:cellfusion1})--(\ref{eq:cellfusion2}) and the properties 
(\ref{eq:projprop2})--(\ref{eq:projprop5}) we have an intertwining relation
between the operators $\wp^A$ and the $\wp^G$,
\bea
&&\setlength{\unitlength}{0.0100in}%
\begin{picture}(189,105)(6,690)
\thicklines
\put( 60,705){\line( 1, 5){ 15.577}}
\put( 75,783){\line( 1,-5){ 15.577}}
\put( 90,705){\line(-1, 0){ 30}}
\multiput(180,705)(0,75){2}{\vector(1, 0){17}}
\multiput(180,780)(24,0){2}{\vector(0,-1){39}}
\put(180,780){\line( 0,-1){ 75}}
\put(180,705){\line( 1, 0){ 24}}
\put(204,705){\line( 0, 1){ 75}}
\put(204,780){\line(-1, 0){ 24}}
\put( 72,690){\makebox(0,0)[lb]{\raisebox{0pt}[0pt][0pt]{\twlrm \sc $b$}}}
\put( 72,789){\makebox(0,0)[lb]{\raisebox{0pt}[0pt][0pt]{\twlrm \sc $a$}}}
\put( 67,729){\makebox(0,0)[lb]{\raisebox{0pt}[0pt][0pt]{\twlrm \tiny
$\wp^A$}}}
\put(204,690){\makebox(0,0)[lb]{\raisebox{0pt}[0pt][0pt]{\twlrm \sc $b'$}}}
\put(177,786){\makebox(0,0)[lb]{\raisebox{0pt}[0pt][0pt]{\twlrm \sc $a$}}}
\put(204,783){\makebox(0,0)[lb]{\raisebox{0pt}[0pt][0pt]{\twlrm \sc $a'$}}}
\put(177,690){\makebox(0,0)[lb]{\raisebox{0pt}[0pt][0pt]{\twlrm \sc $b$}}}
\put(186,729){\makebox(0,0)[lb]{\raisebox{0pt}[0pt][0pt]{\twlrm \tiny
$C_n$}}}
\put(211,729){\makebox(0,0)[lb]{\raisebox{0pt}[0pt][0pt]{\twlrm 
\tiny $f_G^{\!-\!1}\!(\!a',\!\nu,\!b'\!)$}}}
\put(139,732){\makebox(0,0)[lb]{\raisebox{0pt}[0pt][0pt]{\twlrm 
\tiny $f_A\!(\!b,\!1,\!a\!)$}}}
\put( 87,732){\makebox(0,0)[lb]{\raisebox{0pt}[0pt][0pt]{\twlrm 
\tiny $f_A^{\!-\!1}\!(\!b,\!1,\!a\!)$}}}
\put( 21,732){\makebox(0,0)[lb]{\raisebox{0pt}[0pt][0pt]{\twlrm 
\tiny $f_A(\!b,\!1,\!a\!)$}}}
\put(205,736){\makebox(0,0)[lb]{\raisebox{0pt}[0pt][0pt]{\twlrm \sc
$\nu$}}}
\end{picture}  \\
&&=\sum_\mu\setlength{\unitlength}{0.0100in}%
\begin{picture}(225,102)(15,755)
\thicklines
\put(234,726){\line(-1, 5){ 15.577}}
\put(219,804){\line(-1,-5){ 15.577}}
\put(204,726){\line( 1, 0){ 30}}
\put(159,756){\makebox(0,0)[lb]{\raisebox{0pt}[0pt][0pt]{\twlrm 
\tiny $f_G\!(\!a'\!,\!\mu,\!b'\!)$}}}
\put(234,759){\makebox(0,0)[lb]{\raisebox{0pt}[0pt][0pt]{\twlrm 
\tiny $f_G^{\!-\!1}\!(\!a'\!,\!\nu,\!b'\!)$}}}
\put(211,750){\makebox(0,0)[lb]{\raisebox{0pt}[0pt][0pt]{\twlrm 
\tiny $\wp^G$}}}
\put(216,714){\makebox(0,0)[lb]{\raisebox{0pt}[0pt][0pt]{\twlrm \sc $b'$}}}
\put(216,807){\makebox(0,0)[lb]{\raisebox{0pt}[0pt][0pt]{\twlrm \sc $a'$}}}
\put(203,774){\makebox(0,0)[lb]{\raisebox{0pt}[0pt][0pt]{\twlrm \sc
$\mu$}}}
\put(228,777){\makebox(0,0)[lb]{\raisebox{0pt}[0pt][0pt]{\twlrm \sc
$\nu$}}}
\put( 96,804){\line( 0,-1){ 75}}
\multiput(72,804)(24,0){2}{\vector( 0,-1){39}}
\multiput(72,804)(0,-75){2}{\vector(1,0){17}}
\put( 96,729){\line(-1, 0){ 24}}
\put( 72,729){\line( 0, 1){ 75}}
\put( 72,804){\line( 1, 0){ 24}}
\put( 99,759){\makebox(0,0)[lb]{\raisebox{0pt}[0pt][0pt]{\twlrm 
\tiny $f_G\!(\!a'\!,\!\mu,\!b'\!)$}}}
\put( 78,756){\makebox(0,0)[lb]{\raisebox{0pt}[0pt][0pt]{\twlrm 
\tiny $C^T_n$}}}
\put(  25,756){\makebox(0,0)[lb]{\raisebox{0pt}[0pt][0pt]{\twlrm 
\tiny $f_A^{\!-\!1}\!(\!b,\!1,\!a\!)$}}}
\put( 69,807){\makebox(0,0)[lb]{\raisebox{0pt}[0pt][0pt]{\twlrm \sc $a$}}}
\put( 96,807){\makebox(0,0)[lb]{\raisebox{0pt}[0pt][0pt]{\twlrm \sc $a'$}}}
\put( 72,714){\makebox(0,0)[lb]{\raisebox{0pt}[0pt][0pt]{\twlrm \sc $b$}}}
\put( 96,714){\makebox(0,0)[lb]{\raisebox{0pt}[0pt][0pt]{\twlrm \sc $b'$}}}
\put( 100,777){\makebox(0,0)[lb]{\raisebox{0pt}[0pt][0pt]{\twlrm \sc
$\mu$}}}
\end{picture}\no
\eea\vspace*{0.2cm}\newline
Here we have expressed the operator 
$\wp(n-1,-(1-n)\lambda,a,b)_{\mu(a,b,n)}^{\nu(a,b,n)}$ graphically as
a triangle with 
\bea
&&f_A(a,\mu,b)=\prod_{i=1}^{n} \sqrt{S^A_{\mu(a,b,n)_i}}
f^A_{\mu(a,b,n)_i},
\no\\
&&f_G(a,\mu,b)=\prod_{i=1}^{n} \sqrt{S^G_{\mu(a,b,n)_i}}
f^G_{\mu(a,b,n)_i}.
\no\eea
From these equations, and with the help of 
(\ref{eq:fuunitary1})--(\ref{eq:fuunitary2}), we can easily obtain
\be
\setlength{\unitlength}{0.0120in}%
\begin{picture}(261,105)(21,711)
\thicklines
\put( 57,726){\line(-1, 5){ 15.577}}
\put( 42,804){\line(-1,-5){ 15.577}}
\put( 27,726){\line( 1, 0){ 30}}
\put( 34,750){\makebox(0,0)[lb]{\raisebox{0pt}[0pt][0pt]{\twlrm \tiny
$\wp^G$}}}
\put( 39,714){\makebox(0,0)[lb]{\raisebox{0pt}[0pt][0pt]{\twlrm \sc $b'$}}}
\put( 39,807){\makebox(0,0)[lb]{\raisebox{0pt}[0pt][0pt]{\twlrm \sc $a'$}}}
\put( 25,777){\makebox(0,0)[lb]{\raisebox{0pt}[0pt][0pt]{\twlrm \sc
$\mu$}}}
\put( 51,777){\makebox(0,0)[lb]{\raisebox{0pt}[0pt][0pt]{\twlrm \sc
$\nu$}}}
\put(210,726){\line( 1, 5){ 15.577}}
\put(225,804){\line( 1,-5){ 15.577}}
\put(240,726){\line(-1, 0){ 30}}
\put(252,801){\line( 0,-1){ 75}}
\put(252,726){\line( 1, 0){ 24}}
\put(276,726){\line( 0, 1){ 75}}
\put(276,801){\line(-1, 0){ 24}}
\put(141,801){\line( 0,-1){ 75}}
\put(141,726){\line( 1, 0){ 24}}
\put(165,726){\line( 0, 1){ 75}}
\put(165,801){\line(-1, 0){ 24}}
\put(222,810){\makebox(0,0)[lb]{\raisebox{0pt}[0pt][0pt]{\twlrm \sc $a$}}}
\put(217,750){\makebox(0,0)[lb]{\raisebox{0pt}[0pt][0pt]{\twlrm \tiny
$\wp^A$}}}
\put(222,711){\makebox(0,0)[lb]{\raisebox{0pt}[0pt][0pt]{\twlrm \sc $b$}}}
\put(276,711){\makebox(0,0)[lb]{\raisebox{0pt}[0pt][0pt]{\twlrm \sc $b'$}}}
\put(249,807){\makebox(0,0)[lb]{\raisebox{0pt}[0pt][0pt]{\twlrm \sc $a$}}}
\put(276,804){\makebox(0,0)[lb]{\raisebox{0pt}[0pt][0pt]{\twlrm \sc
$\ol{a}'$}}}
\put(249,711){\makebox(0,0)[lb]{\raisebox{0pt}[0pt][0pt]{\twlrm \sc $b$}}}
\put(258,750){\makebox(0,0)[lb]{\raisebox{0pt}[0pt][0pt]{\twlrm \tiny
$C_n$}}}
\put(278,765){\makebox(0,0)[lb]{\raisebox{0pt}[0pt][0pt]{\twlrm \sc
$\nu$}}}
\put(174,750){\makebox(0,0)[lb]{\raisebox{0pt}[0pt][0pt]{\twlrm 
\tiny $f^2\!(\!b,\!1,\!a\!)$}}}
\put(168,804){\makebox(0,0)[lb]{\raisebox{0pt}[0pt][0pt]{\twlrm \sc $a$}}}
\put(141,804){\makebox(0,0)[lb]{\raisebox{0pt}[0pt][0pt]{\twlrm \sc $a'$}}}
\put(165,711){\makebox(0,0)[lb]{\raisebox{0pt}[0pt][0pt]{\twlrm \sc $b$}}}
\put(141,711){\makebox(0,0)[lb]{\raisebox{0pt}[0pt][0pt]{\twlrm \sc $b'$}}}
\put(147,750){\makebox(0,0)[lb]{\raisebox{0pt}[0pt][0pt]{\twlrm \tiny
$C_n$}}}
\put(133,765){\makebox(0,0)[lb]{\raisebox{0pt}[0pt][0pt]{\twlrm \sc
$\mu$}}}
\put( 90,750){\makebox(0,0)[lb]{\raisebox{0pt}[0pt][0pt]{\twlrm  
{\tiny $f^{\!-\!2}\!(\!a'\!,\!\mu,\!b'\!)$}}}}
\put( 60,750){\makebox(0,0)[lb]{\raisebox{0pt}[0pt][0pt]{\twlrm \sc
$=\disp{\sum_b}$}}}
\put( 5,750){\sc $\delta_{a'\!,\!\ol{a}\!'}$}
\put(264,801){\vector( 1, 0){  6}}
\put(264,726){\vector( 1, 0){  6}}
\put(252,768){\vector( 0,-1){  6}}
\put(276,768){\vector( 0,-1){  6}}
\put(141,768){\vector( 0,-1){  6}}
\put(165,768){\vector( 0,-1){  6}}
\put(156,801){\vector( -1, 0){  6}}
\put(156,726){\vector( -1, 0){  6}}
\end{picture}
\ee
and the inverse of the operator $\wp$
\be
\setlength{\unitlength}{0.0120in}%
\begin{picture}(312,105)(10,708)
\thicklines
\put(243,765){\vector( 0,-1){  6}}
\put(255,798){\vector( 1, 0){  6}}
\put(255,723){\vector( 1, 0){  6}}
\put(243,798){\line( 0,-1){ 75}}
\put(243,723){\line( 1, 0){ 24}}
\put(267,723){\line( 0, 1){ 75}}
\put(267,798){\line(-1, 0){ 24}}
\put(267,765){\vector( 0,-1){  6}}
\put(165,723){\line( 1, 5){ 15.577}}
\put(180,801){\line( 1,-5){ 15.577}}
\put(195,723){\line(-1, 0){ 30}}
\put(141,798){\vector(-1, 0){  6}}
\put(141,723){\vector(-1, 0){  6}}
\put(153,765){\vector( 0,-1){  6}}
\put(129,765){\vector( 0,-1){  6}}
\put(153,798){\line( 0,-1){ 75}}
\put(153,723){\line(-1, 0){ 24}}
\put(129,723){\line( 0, 1){ 75}}
\put(129,798){\line( 1, 0){ 24}}
\put( 60,801){\line(-1,-5){ 15.577}}
\put( 45,723){\line(-1, 5){ 15.577}}
\put( 30,801){\line( 1, 0){ 30}}
\put(270,747){\makebox(0,0)[lb]{\raisebox{0pt}[0pt][0pt]{\twlrm 
\tiny $f^2(\!\ol{a}\!'\!,\!\mu,\!b'\!)$}}}
\put(270,759){\makebox(0,0)[lb]{\raisebox{0pt}[0pt][0pt]{\twlrm \sc
$\mu$}}}
\put(249,747){\makebox(0,0)[lb]{\raisebox{0pt}[0pt][0pt]{\twlrm \tiny
$C^T_n$}}}
\put(200,747){\makebox(0,0)[lb]{\raisebox{0pt}[0pt][0pt]{\twlrm 
\tiny $f^{-2}\!(\!b,\!1,\!a\!)$}}}
\put(177,807){\makebox(0,0)[lb]{\raisebox{0pt}[0pt][0pt]{\twlrm \sc $a$}}}
\put(177,708){\makebox(0,0)[lb]{\raisebox{0pt}[0pt][0pt]{\twlrm \sc $b$}}}
\put(129,708){\makebox(0,0)[lb]{\raisebox{0pt}[0pt][0pt]{\twlrm \sc $b'$}}}
\put(156,804){\makebox(0,0)[lb]{\raisebox{0pt}[0pt][0pt]{\twlrm \sc $a$}}}
\put(129,801){\makebox(0,0)[lb]{\raisebox{0pt}[0pt][0pt]{\twlrm \sc $a'$}}}
\put(156,708){\makebox(0,0)[lb]{\raisebox{0pt}[0pt][0pt]{\twlrm \sc $b$}}}
\put(122,759){\makebox(0,0)[lb]{\raisebox{0pt}[0pt][0pt]{\twlrm \sc
$\nu$}}}
\put(135,747){\makebox(0,0)[lb]{\raisebox{0pt}[0pt][0pt]{\twlrm \tiny
$C^T_n$}}}
\put(170,745){\makebox(0,0)[lb]{\raisebox{0pt}[0pt][0pt]{\twlrm \tiny 
$\wp^{\!{A}^{\!-1\!}\!}$}}}
\put( 42,711){\makebox(0,0)[lb]{\raisebox{0pt}[0pt][0pt]{\twlrm \sc $b'$}}}
\put( 42,804){\makebox(0,0)[lb]{\raisebox{0pt}[0pt][0pt]{\twlrm \sc $a'$}}}
\put(270,801){\makebox(0,0)[lb]{\raisebox{0pt}[0pt][0pt]{\twlrm \sc
$\ol{a}\!'$}}}
\put(243,801){\makebox(0,0)[lb]{\raisebox{0pt}[0pt][0pt]{\twlrm \sc $a$}}}
\put(243,708){\makebox(0,0)[lb]{\raisebox{0pt}[0pt][0pt]{\twlrm \sc $b$}}}
\put(267,708){\makebox(0,0)[lb]{\raisebox{0pt}[0pt][0pt]{\twlrm \sc $b'$}}}
\put( 35,770){\makebox(0,0)[lb]{\raisebox{0pt}[0pt][0pt]{\twlrm \tiny 
$\wp^{\!{G}^{\!-1\!}\!}$}}}
\put( 54,750){\makebox(0,0)[lb]{\raisebox{0pt}[0pt][0pt]{\twlrm \sc
$\mu$}}}
\put( 28,750){\makebox(0,0)[lb]{\raisebox{0pt}[0pt][0pt]{\twlrm \sc
$\nu$}}}
\put(67,753){\makebox(0,0)[lb]{\raisebox{0pt}[0pt][0pt]{\twlrm \sc 
$\delta_{a'\!,\!\ol{a}\!'}=\disp{\sum_b}$}}}
\end{picture}
\ee
As a result we have shown that $\wp(n-1,-(n-1)\lambda,a,b)$ is nonsingular
\be
\setlength{\unitlength}{0.0120in}%
\begin{picture}(261,102)(99,708)
\thicklines
\put(165,723){\line( 1, 5){ 15.577}}
\put(180,801){\line( 1,-5){ 15.577}}
\put(195,723){\line(-1, 0){ 30}}
\put(135,801){\line(-1,-5){ 15.577}}
\put(120,723){\line(-1, 5){ 15.577}}
\put(105,801){\line( 1, 0){ 30}}
\put(177,708){\makebox(0,0)[lb]{\raisebox{0pt}[0pt][0pt]{\twlrm \sc $a$}}}
\put(174,747){\makebox(0,0)[lb]{\raisebox{0pt}[0pt][0pt]{\twlrm \tiny
$\wp^G$}}}
\put(195,750){\makebox(0,0)[lb]{\raisebox{0pt}[0pt][0pt]{\twlrm \sc
$\nu$}}}
\put(160,750){\makebox(0,0)[lb]{\raisebox{0pt}[0pt][0pt]{\twlrm \sc
$\beta$}}}
\put(112,777){\makebox(0,0)[lb]{\raisebox{0pt}[0pt][0pt]{\twlrm \tiny
$\wp^{\!G^{\!-\!1}}$}}}
\put( 104,750){\makebox(0,0)[lb]{\raisebox{0pt}[0pt][0pt]{\twlrm \sc
$\mu$}}}
\put(81,756){\makebox(0,0)[lb]{\raisebox{0pt}[0pt][0pt]{\twlrm \sc
$\disp{\sum_{\beta}}$}}}
\put(117,711){\makebox(0,0)[lb]{\raisebox{0pt}[0pt][0pt]{\twlrm \sc $a$}}}
\put(120,806){\makebox(0,0)[lb]{\raisebox{0pt}[0pt][0pt]{\twlrm \sc $b$}}}
\put(129,750){\makebox(0,0)[lb]{\raisebox{0pt}[0pt][0pt]{\twlrm \sc
$\beta$}}}
\put(177,806){\makebox(0,0)[lb]{\raisebox{0pt}[0pt][0pt]{\twlrm \sc $b$}}}
\put(282,726){\line(-1, 5){ 15.577}}
\put(267,804){\line(-1,-5){ 15.577}}
\put(252,726){\line( 1, 0){ 30}}
\put(312,804){\line( 1,-5){ 15.577}}
\put(327,726){\line( 1, 5){ 15.577}}
\put(342,804){\line(-1, 0){ 30}}
\put(270,711){\makebox(0,0)[lb]{\raisebox{0pt}[0pt][0pt]{\twlrm \sc $a$}}}
\put(330,714){\makebox(0,0)[lb]{\raisebox{0pt}[0pt][0pt]{\twlrm \sc $a$}}}
\put(327,806){\makebox(0,0)[lb]{\raisebox{0pt}[0pt][0pt]{\twlrm \sc $b$}}}
\put(270,806){\makebox(0,0)[lb]{\raisebox{0pt}[0pt][0pt]{\twlrm \sc $b$}}}
\put(249,755){\makebox(0,0)[lb]{\raisebox{0pt}[0pt][0pt]{\twlrm \sc
$\mu$}}}
\put(261,750){\makebox(0,0)[lb]{\raisebox{0pt}[0pt][0pt]{\twlrm \tiny
$\wp^G$}}}
\put(279,755){\makebox(0,0)[lb]{\raisebox{0pt}[0pt][0pt]{\twlrm \sc
$\beta$}}}
\put(312,755){\makebox(0,0)[lb]{\raisebox{0pt}[0pt][0pt]{\twlrm \sc
$\beta$}}}
\put(318,777){\makebox(0,0)[lb]{\raisebox{0pt}[0pt][0pt]{\twlrm \tiny
$\wp^{\!G^{\!-\!1}}$}}}
\put(336,753){\makebox(0,0)[lb]{\raisebox{0pt}[0pt][0pt]{\twlrm \sc
$\nu$}}}
\put(210,757){\makebox(0,0)[lb]{\raisebox{0pt}[0pt][0pt]{\twlrm \sc $=$}}}
\put(218,756){\makebox(0,0)[lb]{\raisebox{0pt}[0pt][0pt]{\twlrm \sc
$\disp{\sum_{\beta}}$}}}
\put(360,756){\makebox(0,0)[lb]{\raisebox{0pt}[0pt][0pt]{\twlrm 
$=\delta_{\mu,\nu}$ .}}}
\end{picture}
\ee
We use the square root of this operator to build the symmetric
face weights from the unsymmetric ones. 

\begin{theorem}

Define the $A^{(n)}_{(a,b)}\times A^{(n)}_{(a,b)}$ matrix
\begin{equation}
G(a,b,n)=\sqrt{F(a,b,n)\wp(n-1,-(n-1)\lambda,a,b)F(a,b,n)} ,
\label{eq:gauge}\end{equation}
where $ F(a,b,n)$ is the diagonal matrix 
\be
F(a,b,n)=\mbox{Diag}\;[\;f(a,1,b),\cdots , f(a,A^{(n)}_{(a,b)},b)\;].  
\ee
Then the symmetric weights
\be \mbox{\small $
\wf {W^s_{m\times n}}{u}bcd{\alpha}{\nu}{\beta}{\mu}=
\displaystyle{\sum_{\alpha',\nu',\beta',\mu'}
{G(d,a,n)_{\mu,\mu'}G(a,b,n)_{\alpha,\alpha'}\over 
G(c,b,n)_{\nu',\nu}G(d,c,n)_{\beta',\beta}}}\;
\wf {W^s_{m\times n}}{u}bcd{\alpha'}{\nu'}{\beta'}{\mu'}$}
\label{eq:symmetricweight}
\ee
satisfy
\bea
&&\wf {W^s_{m\times n}}{u}bcd{\alpha}{\nu}{\beta}{\mu}
={W^s_{n\times m}\mbox{$\left(\matrix{
d&\mu&a\cr\beta&&\alpha\cr c&\nu&b\cr}\biggm|\mbox{$u$}\right)$}}
                                        \label{eq:facesymmetry1}\\
&&\no\\
&&=\wf {W^s_{n\times m}}{u\-(n\-m)\lambda}dcb{\mu}{\beta}{\nu}{\alpha}.
\label{eq:facesymmetry2}\eea
\end{theorem}

{\sl Proof:} The symmetry (\ref{eq:facesymmetry2}) is implied by 
the symmetry of the elementary face  
\be 
\face abcdu =\face adcbu .
\ee
\vspace{0.2cm}\newline
The symmetry (\ref{eq:facesymmetry1}) follows from the equality

\begin{eqnarray}
& &\setlength{\unitlength}{0.0110in}%
\begin{picture}(260,174)(45,580)
\thicklines
\put(250,610){\line(-6,-1){ 90}}
\put(160,595){\line( 6,-1){ 90}}
\put(250,580){\line( 0, 1){ 30}}
\put(150,740){\line(-1,-6){ 20}}
\put(130,620){\line(-1, 6){ 20}}
\put(110,740){\line( 1, 0){ 40}}
\put(160,740){\line( 0,-1){120}}
\put(160,620){\line( 1, 0){ 90}}
\put(250,620){\line( 0, 1){120}}
\put(250,740){\line(-1, 0){ 90}}
\put(255,595){\makebox(0,0)[lb]{\raisebox{0pt}[0pt][0pt]{\twlrm $b$}}}
\put(185,580){\makebox(0,0)[lb]{\raisebox{0pt}[0pt][0pt]{\twlrm $\alpha$}}}
\put(180,608){\makebox(0,0)[lb]{\raisebox{0pt}[0pt][0pt]{\twlrm
$\alpha'$}}}
\put(200,625){\makebox(0,0)[lb]{\raisebox{0pt}[0pt][0pt]{\twlrm
$\alpha'$}}}
\put(240,675){\makebox(0,0)[lb]{\raisebox{0pt}[0pt][0pt]{\twlrm $\nu$}}}
\put(200,675){\makebox(0,0)[lb]{\raisebox{0pt}[0pt][0pt]{\twlrm $u$}}}
\put(200,743){\makebox(0,0)[lb]{\raisebox{0pt}[0pt][0pt]{\twlrm $\beta$}}}
\put(165,675){\makebox(0,0)[lb]{\raisebox{0pt}[0pt][0pt]{\twlrm $\mu'$}}}
\put(255,740){\makebox(0,0)[lb]{\raisebox{0pt}[0pt][0pt]{\twlrm $c$}}}
\put(150,615){\makebox(0,0)[lb]{\raisebox{0pt}[0pt][0pt]{\twlrm $a$}}}
\put(150,595){\makebox(0,0)[lb]{\raisebox{0pt}[0pt][0pt]{\twlrm $a$}}}
\put(130,610){\makebox(0,0)[lb]{\raisebox{0pt}[0pt][0pt]{\twlrm $a$}}}
\put(140,645){\makebox(0,0)[lb]{\raisebox{0pt}[0pt][0pt]{\twlrm $\mu'$}}}
\put(115,645){\makebox(0,0)[lb]{\raisebox{0pt}[0pt][0pt]{\twlrm $\mu$}}}
\put(165,745){\makebox(0,0)[lb]{\raisebox{0pt}[0pt][0pt]{\twlrm $d$}}}
\put(130,745){\makebox(0,0)[lb]{\raisebox{0pt}[0pt][0pt]{\twlrm $d$}}}
\put(260,620){\makebox(0,0)[lb]{\raisebox{0pt}[0pt][0pt]{\twlrm $b$}}}
\put(  15,675){\makebox(0,0)[lb]{\raisebox{0pt}[0pt][0pt]{\twlrm 
$\disp{\sum_{\alpha'=1}^{A^{(n)}_{(a,b)}}}$}}}
\put(265,675){\makebox(0,0)[lb]{\raisebox{0pt}[0pt][0pt]{\twlrm 
${f(d,\mu,a)f(a,\alpha,b)\over f(d,\beta,c)f(c,\nu,b)}$}}}
\put( 60,675){\makebox(0,0)[lb]{\raisebox{0pt}[0pt][0pt]{\twlrm 
$\disp{\sum_{\mu'=1}^{A^{(m)}_{(d,a)}}}$}}}
\end{picture}  \nonumber \\
& &\nonumber \\
& &\setlength{\unitlength}{0.0110in}%
\begin{picture}(235,154)(235,600)
\thicklines
\put(340,740){\line( 1, 0){120}}
\put(460,740){\line( 0,-1){ 90}}
\put(460,650){\line(-1, 0){120}}
\put(340,650){\line( 0, 1){ 90}}
\put(305,740){\line( 1,-6){ 15}}
\put(320,650){\line( 1, 6){ 15}}
\put(335,740){\line(-1, 0){ 30}}
\put(460,600){\line(-6, 1){120}}
\put(340,620){\line( 6, 1){120}}
\put(460,640){\line( 0,-1){ 40}}
\put(470,740){\makebox(0,0)[lb]{\raisebox{0pt}[0pt][0pt]{\twlrm $a$}}}
\put(470,740){\makebox(0,0)[lb]{\raisebox{0pt}[0pt][0pt]{\twlrm $a$}}}
\put(335,635){\makebox(0,0)[lb]{\raisebox{0pt}[0pt][0pt]{\twlrm $c$}}}
\put(315,635){\makebox(0,0)[lb]{\raisebox{0pt}[0pt][0pt]{\twlrm $c$}}}
\put(330,615){\makebox(0,0)[lb]{\raisebox{0pt}[0pt][0pt]{\twlrm $c$}}}
\put(340,745){\makebox(0,0)[lb]{\raisebox{0pt}[0pt][0pt]{\twlrm $d$}}}
\put(315,745){\makebox(0,0)[lb]{\raisebox{0pt}[0pt][0pt]{\twlrm $d$}}}
\put(465,615){\makebox(0,0)[lb]{\raisebox{0pt}[0pt][0pt]{\twlrm $b$}}}
\put(470,645){\makebox(0,0)[lb]{\raisebox{0pt}[0pt][0pt]{\twlrm $b$}}}
\put(380,605){\makebox(0,0)[lb]{\raisebox{0pt}[0pt][0pt]{\twlrm $\nu$}}}
\put(345,690){\makebox(0,0)[lb]{\raisebox{0pt}[0pt][0pt]{\twlrm $\beta'$}}}
\put(390,652){\makebox(0,0)[lb]{\raisebox{0pt}[0pt][0pt]{\twlrm $\nu'$}}}
\put(375,627){\makebox(0,0)[lb]{\raisebox{0pt}[0pt][0pt]{\twlrm $\nu'$}}}
\put(465,690){\makebox(0,0)[lb]{\raisebox{0pt}[0pt][0pt]{\twlrm 
${f(d,\beta,c)f(c,\nu,b)\over f(d,\mu,a)f(a,\alpha,b)}$}}}
\put(390,695){\makebox(0,0)[lb]{\raisebox{0pt}[0pt][0pt]{\twlrm $u$}}}
\put(450,690){\makebox(0,0)[lb]{\raisebox{0pt}[0pt][0pt]{\twlrm $\alpha$}}}
\put(395,730){\makebox(0,0)[lb]{\raisebox{0pt}[0pt][0pt]{\twlrm $\mu$}}}
\put(325,670){\makebox(0,0)[lb]{\raisebox{0pt}[0pt][0pt]{\twlrm $\beta'$}}}
\put(306,670){\makebox(0,0)[lb]{\raisebox{0pt}[0pt][0pt]{\twlrm $\beta$}}}
\put(210,680){\makebox(0,0)[lb]{\raisebox{0pt}[0pt][0pt]{\twlrm 
$=\disp{\sum_{\nu'=1}^{A^{(m)}_{(c,b)}}}$}}}
\put(265,680){\makebox(0,0)[lb]{\raisebox{0pt}[0pt][0pt]{\twlrm 
$\disp{\sum_{\beta'=1}^{A^{(n)}_{(d,c)}}}$}}}
\end{picture}
\end{eqnarray}

\vspace{.1in}\noindent
which follows from the Yang-Baxter equation (\ref{eq:YBR}) and
\be
 \face abcdu ={S_c \over S_a }
            \face cbadu .
\ee

\vspace{.2in}
It should be noted that this gauge transformation is different from that
used by
Date et al  \cite{DJKMO:88} for the fused $A$ models.
Obviously, the symmetric fused weights $W^{s,A}_{m\times n}$ 
and $W^{s,G}_{m\times n}$ satisfy the intertwining 
relations (\ref{eq:unsyminter1}) and (\ref{eq:unsyminter2}) with the cells
replaced by 
\be
\sum_\nu {G^A(d,a,n)\over G^G(c,b,n)_{\nu,\mu}}\h
\setlength{\unitlength}{0.0110in}%
\begin{picture}(70,79)(100,735)
\put(100,760){\line( 0,-1){ 30}}
\put(100,730){\line( 1, 0){ 60}}
\put(160,730){\line( 0, 1){ 30}}
\put(160,760){\line(-1, 0){ 60}}
\put(100,760){\vector( 1, 0){ 35}}
\put(100,730){\vector( 1, 0){ 35}}
\put(100,730){\vector( 0, 1){ 20}}
\put(160,730){\vector( 0, 1){ 20}}
\put(130,765){\makebox(0,0)[lb]{\raisebox{0pt}[0pt][0pt]{\twlrm $\nu$}}}
\put( 95,765){\makebox(0,0)[lb]{\raisebox{0pt}[0pt][0pt]{\twlrm $c$}}}
\put( 95,715){\makebox(0,0)[lb]{\raisebox{0pt}[0pt][0pt]{\twlrm $d$}}}
\put(160,715){\makebox(0,0)[lb]{\raisebox{0pt}[0pt][0pt]{\twlrm $a$}}}
\put(165,765){\makebox(0,0)[lb]{\raisebox{0pt}[0pt][0pt]{\twlrm $b$}}}
\put(130,742){\sc $C_n$}
\end{picture}
\ee
and
\be
\sum_\nu { G^G(d,a,n)_{\mu,\nu}\over G^A(a,b,n)}\h
\setlength{\unitlength}{0.0110in}%
\begin{picture}(70,59)(100,735)
\put(100,760){\vector( 0,-1){ 20}}
\put(100,730){\line( 1, 0){ 60}}
\put(160,760){\vector( 0, -1){ 20}}
\put(160,760){\line(-1, 0){ 60}}
\put(100,760){\vector( 1, 0){ 35}}
\put(100,730){\vector( 1, 0){ 35}}
\put(100,730){\line( 0, 1){ 30}}
\put(160,730){\line( 0, 1){ 30}}
\put(130,713){\makebox(0,0)[lb]{\raisebox{0pt}[0pt][0pt]{\twlrm $\nu$}}}
\put( 95,765){\makebox(0,0)[lb]{\raisebox{0pt}[0pt][0pt]{\twlrm $c$}}}
\put( 95,715){\makebox(0,0)[lb]{\raisebox{0pt}[0pt][0pt]{\twlrm $d$}}}
\put(160,715){\makebox(0,0)[lb]{\raisebox{0pt}[0pt][0pt]{\twlrm $a$}}}
\put(165,765){\makebox(0,0)[lb]{\raisebox{0pt}[0pt][0pt]{\twlrm $b$}}}
\put(130,742){\sc $C^T_n$}
\end{picture} 
\ee
\bigskip

\subsection{Row Transfer Matrix Intertwiners}

In this section we study intertwiners relating the row transfer
matrices of the fused \ade models. 

It is easy to show \cite{PeZh:93} that the
adjacency intertwining relation (\ref{eq:adjintern})
\begin{equation}
   A^{(n)}\stackrel{C}{\sim}G^{(n)} \label{rel:adjinter}
\end{equation}
is an equivalence relation among symmetric matrices. 
The existence of an intertwiner reflects a symmetry relating the two
graphs associated with $A^{(n)}$ and $G^{(n)}$. In
particular, we observe that the intertwining relation implies that
\begin{equation}
   [C C^T,A^{(n)}]=[C^T C,G^{(n)}]=0
\end{equation}
so that the symmetry operators $C C^T$ and $C^T C$ commute with $A^{(n)}$
and 
$G^{(n)}$ respectively and their eigenvalues can be used as quantum numbers
labelling the eigenvectors of $A^{(n)}$ and $G^{(n)}$.  

  The above properties of intertwiners at the adjacency matrix level
carry over to those at the row transfer matrix level. Let us introduce
cell row transfer matrices with fused cells $C_n,C^T_n$
\begin{equation}
\langle\mbox{\boldmath $a$}|{\cal C}_{(n)}|\mbox{\boldmath$b$}\rangle,
\langle\mbox{\boldmath $a$}|{\cal C}_{(n)}^T|\mbox{\boldmath$b$}\rangle\;=
\begin{picture}(300,40)(92,26)
\put(20,25){ }
\put(95,43){\tiny $b_1$} \put(118,43){\tiny $b_2$} \put(278,43){\tiny
$b_{N+1}$} 
\put(95,16){\tiny $a_1$} \put(118,16){\tiny $a_2$} \put(278,16){\tiny
$a_{N+1}$} 
\multiput(100,40)(20,0){9}{\vector(1,0){10}}   
\multiput(120,40)(20,0){9}{\line(-1,0){10}}
\multiput(100,40)(20,0){5}{\line(0,-1){10}}  
\multiput(100,20)(20,0){5}{\vector(0,1){10}}
\multiput(100,20)(20,0){9}{\vector(1,0){10}}   
\multiput(120,20)(20,0){9}{\line(-1,0){10}}
\multiput(280,40)(-20,0){3}{\line(0,-1){10}} 
\multiput(280,20)(-20,0){3}{\vector(0,1){10}}
\put(210,26){\sc $C_n,C^T_n$}
\end{picture} \label{eq:rowinter}
\end{equation}

\bigskip\smallskip\noindent 
where \mbox{\boldmath{$a$}} and \mbox{\boldmath{$b$}} are allowed row 
configurations of the graphs $A^{(n)}$ and $G^{(n)}$ with periodic boundary
conditions $a_{N+1}=a_1$ and $b_{N+1}=b_1$. In general, the row intertwiner
is a rectangular matrix. Using the cell intertwiner
relations (\ref{eq:unsyminter1}) and (\ref{eq:unsyminter2}) it can be shown
that 
\be {\cal A}^{(n)}(u){\cal C}_{(n)}={\cal C}_{(n)}{\cal G}^{(n)}(u)\;\;{\rm
or}\;\; {\cal A}^{(n)}(u)\stackrel{{\cal C}_{(n)}}{\sim}{\cal G}^{(n)}(u),
\label{eq:traninter}
\ee 
where ${\cal
A}^{(n)}={\bf T}^{(n,m)}_A (u)$ and 
${\cal G}^{(n)}={\bf T}^{(n,m)}_G(u)$ are the row transfer matrices of two
fused
models. This intertwining relation can be pictured as follows 
\begin{eqnarray}
& &\begin{picture}(300,40)(60,10)
\put(95,43){\tiny $a'_1$} \put(118,43){\tiny $a'_2$} \put(278,43){\tiny
$a'_{N+1}=a'_1$} 
\put(95,-4){\tiny $a_1$} \put(118,-4){\tiny $a_2$} \put(278,-4){\tiny
$a_{N+1}=a_1$} 
\multiput(93,18)(190,0){2}{\tiny $b$}
\multiput(100,40)(20,0){9}{\vector(1,0){10}}
\multiput(120,40)(20,0){9}{\line(-1,0){10}}
\multiput(100,40)(20,0){5}{\line(0,-1){10}} 
\multiput(100,20)(20,0){5}{\vector(0,1){10}}
\multiput(100,20)(20,0){9}{\vector(1,0){10}}   
\multiput(120,20)(20,0){9}{\line(-1,0){10}}
\multiput(280,40)(-20,0){3}{\line(0,-1){10}} 
\multiput(280,20)(-20,0){3}{\vector(0,1){10}}
\multiput(100,20)(20,0){5}{\circle*{3}}  
\multiput(280,20)(-20,0){3}{\circle*{3}}

\multiput(100,20)(20,0){5}{\vector(0,-1){10}} 
\multiput(100,0)(20,0){5}{\line(0,1){10}}
\multiput(100,0)(20,0){9}{\vector(1,0){10}}   
\multiput(120,0)(20,0){9}{\line(-1,0){10}}
\multiput(280,20)(-20,0){3}{\vector(0,-1){10}} 
\multiput(280,0)(-20,0){3}{\line(0,1){10}}
\put(207,7){\tiny {${\cal A}^{(n)}(u)$}} 
\put(207,27){\tiny {$C_{n}$}} 
\end{picture} 
  \nonumber \\  \label{eq:tranintergraph} \\
& & \begin{picture}(300,40)(60,10)
\put(80,20){=}
\put(95,43){\tiny $a'_1$} \put(118,43){\tiny $a'_2$} \put(278,43){\tiny
$a'_{N+1}=a'_1$} 
\put(95,-4){\tiny $a_1$} \put(118,-4){\tiny $a_2$} \put(278,-4){\tiny
$a_{N+1}=a_1$} 
\multiput(93,18)(190,0){2}{\tiny $b$}
\multiput(100,40)(20,0){9}{\vector(1,0){10}}
\multiput(120,40)(20,0){9}{\line(-1,0){10}}
\multiput(100,40)(20,0){5}{\vector(0,-1){10}} 
\multiput(100,20)(20,0){5}{\line(0,1){10}}
\multiput(100,20)(20,0){9}{\vector(1,0){10}}   
\multiput(120,20)(20,0){9}{\line(-1,0){10}}
\multiput(280,40)(-20,0){3}{\vector(0,-1){10}} 
\multiput(280,20)(-20,0){3}{\line(0,1){10}}
\multiput(100,20)(20,0){5}{\circle*{3}}  
\multiput(280,20)(-20,0){3}{\circle*{3}}

\multiput(100,20)(20,0){5}{\line(0,-1){10}} 
\multiput(100,0)(20,0){5}{\vector(0,1){10}}
\multiput(100,0)(20,0){9}{\vector(1,0){10}}   
\multiput(120,0)(20,0){9}{\line(-1,0){10}}
\multiput(280,20)(-20,0){3}{\line(0,-1){10}} 
\multiput(280,0)(-20,0){3}{\vector(0,1){10}}
\put(207,27){\tiny {${\cal G}^{(n)}(u)$}} 
\put(207,7){\tiny {$C_{n}$}}
 \nonumber
\end{picture}
\end{eqnarray}

\bigskip\noindent
where a solid circle indicates a summation over the corresponding spin.

This intertwining 
relation  is (i)~reflexive, (ii)~symmetric and
(iii)~transitive, that is, the intertwining relation is again an
equivalence
relation
\begin{eqnarray}
&&\ \ \mbox{(i)}\quad {\cal A}^{(n)}(u)\stackrel{I}{\sim}
                                      {\cal A}^{(n)}(u)\nonumber\\
&&\ \mbox{(ii)}\quad {\cal A}^{(n)}(u)\stackrel{{\cal C}_{(n)}}
                                     {\sim}{\cal G}^{(n)}(u)\quad
\mbox{implies}\quad{\cal G}^{(n)}(u)\stackrel{{\cal C}_{(n)}^T}
                                    {\sim}{\cal A}^{(n)}(u)\\
&&\mbox{(iii)}\quad {\cal A}^{(n)}(u)\stackrel{{\cal C}_{(n)}}
                                   {\sim}{\cal B}^{(n)}(u)\quad
\mbox{and}\quad {\cal B}^{(n)}(u)\stackrel{{\cal C}'_{\!(n)}}
                                {\sim}{\cal G}^{(n)}(u)\nonumber\\
&&\phantom{(iii)A^{(n)}}\mbox{implies}\quad {\cal A}^{(n)}(u)
\stackrel{{\cal C}_{(n)}{\cal C}'_{\!(n)}}{\sim}{\cal G}^{(n)}(u).
\nonumber\end{eqnarray}
and moreover
\begin{equation}
   [{\cal C}_{(n)} {\cal C}_{(n)}^T\;,\;{\cal A}^{(n)}(u)]=
[{\cal C}_{(n)}^T {\cal C}_{(n)}\;,\;{\cal G}^{(n)}(u)]=0.
\end{equation}

\noindent
Hence the symmetry operators ${\cal C}_{(n)} {\cal C}_{(n)}^T$ and 
${\cal C}_{(n)}^T {\cal C}_{(n)}$ and the row transfer matrices 
${\cal A}^{(n)}(u)$ and ${\cal G}^{(n)}(u)$, respectively, have the 
same eigenvectors and can be simultaneously diagonalised. The eigenvectors
that 
are not annihilated by the symmetry operators give the eigenvalues that
are intertwined and are common to ${\cal A}^{(n)}(u)$ and ${\cal
G}^{(n)}(u)$.
Since 
\be 
{\cal C}_{(n)}{\cal C}_{(n)}^T\stackrel{{\cal C}_{(n)}}{\sim}
{\cal C}_{(n)}^T{\cal C}_{(n)}
\ee
it is precisely the nonzero eigenvalues of these symmetry operators that
are in
common. 

Let us now consider the $A_L$--$D_{(L+3)/2}$ fused models, with $L$ odd,
and
define the height reversal operators ${\cal R}_A$ and ${\cal R}_D$ 
for these models by the elements 
\be
\langle\mbox{\boldmath $a$}|{\cal R}_A|\mbox{\boldmath
$b$}\rangle=\prod_{j=1}^N \delta_{a_j,r(b_j)},
\qquad\qquad
\langle\mbox{\boldmath $a$}|{\cal R}_D|\mbox{\boldmath
$b$}\rangle=\prod_{j=1}^N \delta_{a_j,r(b_j)}
\ee
where for the $A$ models $r(b)=h-b$ and for the $D$ models 
\be
r(b)=\cases{b,&for $b=1,2,\cdots,(L-1)/2$\cr
(L+3)/2,&for $b=(L+1)/2$\cr 
(L+1)/2,&for $b=(L+3)/2$.}
\ee
These matrix
operators implement the $\Z_2$
symmetry of the models. It is easy to show that the fused cell row transfer
matrices satisfy  
\be
{\cal C}_{(n)}{\cal C}^T_{(n)}=I+{\cal R}_A,
\qquad\qquad {\cal C}_{(n)}^T{\cal C}_{(n)}=I+{\cal
R}_D.\label{symops}
\ee
The row transfer matrices 
of the fused $A$ and $D$ models commute with the corresponding height
reversal
operators. An  immediate consequence of this is that the eigenvalues of
${\cal
A}(u)^{(n)}= {\bf T}^{(n,n)}_A (u)$ and ${\cal D}(u)^{(n)}={\bf
T}^{(n,n)}_D
(u)$ are in  common if and only if the corresponding eigenvectors 
are even under the $\Z_2$ symmetry. In particular, since the largest
eigenvalue has an even eigenvector, the largest eigenvalue is in common and
hence the intertwined models have the same central charge 
\be
c={3n\over n+2}\left(1-{2(n+2)\over h(h-n)}\right).
\ee
Similarly, following Kl\"umper and Pearce \cite{KlPe:92}, it can be shown
that the
conformal weights 
\be
(\Delta_{r,s},\overline{\Delta}_{r,s})
\ee 
of the excited states are given by \cite{DJKMO:87},
\be
\Delta_{r,s}={[ht-(h-n)s]^2-n^2\over 4nh(h-n)}+
                {(s_0-1)(n-s_0+1)\over 2n(n+2)}
\ee
where $s$ and $r$ label the rows and columns of the Kac table and $s_0$ is
the
unique integer determined by $1\leq s_0\leq n+1$ and $s_0-1=\pm
(t-s)\ {\rm mod}\ 2n$. However, in contrast to the case of the $A$ models,
nondiagonal terms with
$\Delta_{r,s}\ne\overline{\Delta}_{r,s}$ occur for the $D$ models.

%% file: Fusion6.tex
\section{An example: $D_4$}
\setcounter{equation}{0}

In this section we find the $2\times 2$ fused face weights
of $D_4$ model and construct explicitly the intertwining relation between
the 
$A_5$ and $D_4$ models. The $D_4$ model is an interesting example because
it
corresponds to the three-state Potts model. 

The adjacency matrices for the fused $A_5$ and $D_4$ models are given by
the fusion rules (\ref{eq:adjfusion}). The adjacency graphs are thus as
shown in
Figure~4.

\begin{figure}[htb]
\begin{picture}(450,370)(-30,-30)
\put(-10,280){$A_5$:} 
\multiput(20,280)(20,0){4}{\line(1,0){20}}
\multiput(20,280)(20,0){5}{\circle*{3}} \put(20,285){\scriptsize 1}
\put(40,285){\scriptsize 2} 
\put(60,285){\scriptsize 3} \put(80,285){\scriptsize 4}
\put(100,285){\scriptsize 5}

\put(140,280){\vector(1,0){60}} \put(140,285){\scriptsize level 3 fusion
$A^{(3)}$}
\multiput(240,280)(20,0){4}{\line(1,0){20}}
\multiput(240,280)(20,0){5}{\circle*{3}} \put(240,285){\scriptsize 1}
\put(260,285){\scriptsize 4} 
\put(280,285){\scriptsize 3} \put(300,285){\scriptsize 2}
\put(320,285){\scriptsize 5}

\put(140,295){\vector(3,1){30}} \put(140,315){\scriptsize level 2 fusion
$A^{(2)}$}
\multiput(240,320)(20,0){2}{\line(1,0){20}}  \put(260,325){\circle{10}}
\put(320,320){\line(1,0){20}}              
\multiput(320,325)(20,0){2}{\circle{10}}
\multiput(240,320)(20,0){3}{\circle*{3}}
\multiput(320,320)(20,0){2}{\circle*{3}} 
\put(240,325){\scriptsize 1} \put(258,313){\scriptsize 3}
\put(280,325){\scriptsize 5} 
\put(318,313){\scriptsize 2} \put(338,313){\scriptsize 4} 

\put(140,265){\vector(3,-1){30}} \put(140,245){\scriptsize level 4 fusion
$A^{(4)}$}
\multiput(240,240)(60,0){2}{\line(1,0){20}} 
\multiput(240,240)(20,0){2}{\circle*{3}}
\multiput(300,240)(20,0){2}{\circle*{3}}
\put(350,240){\circle*{3}} \put(350,245){\circle{10}}
\put(240,245){\scriptsize 1} \put(260,245){\scriptsize 5} 
\put(300,245){\scriptsize 2} \put(320,245){\scriptsize 4}
\put(348,232){\scriptsize 3}

\put(-10,93){$D_4$:}
\put(20,95){\line(1,0){40}}  \multiput(20,95)(68,-28){2}{\circle*{3}}
\put(88,123){\circle*{3}}
\put(60,95){\circle*{3}} \put(60,95){\line(1,1){28}} 
\put(60,95){\line(1,-1){28}}
\put(15,95){\scriptsize $1'$} \put(88,125){\scriptsize $\ol{3}'$} 
\put(90,65){\scriptsize $3'$} 
\put(66,93){\scriptsize $2'$} 

\put(140,92){\vector(1,0){60}} \put(140,95){\scriptsize level 3 fusion
$D^{(3)}$}
\put(240,95){\line(1,0){40}}  \multiput(240,95)(68,-28){2}{\circle*{3}}
\put(308,123){\circle*{3}}
\put(280,95){\circle*{3}} \put(280,95){\line(1,1){28}} 
\put(280,95){\line(1,-1){28}}
\put(235,95){\scriptsize $1'$} \put(308,125){\scriptsize $\ol{3}'$} 
\put(310,65){\scriptsize $3'$} 
\put(286,93){\scriptsize $2'$} 

\put(140,150){\vector(3,1){30}} \put(140,166){\scriptsize level 2 fusion
$D^{(2)}$}
\put(240,180){\line(3,2){25}}  \put(240,180){\line(3,-2){25}}
\put(265,197){\line(0,-1){32}} 
\multiput(240,180)(25,-16){2}{\circle*{3}}    \put(265,197){\circle*{3}}
\put(235,180){\scriptsize $1'$} \put(267,160){\scriptsize $3'$}
\put(267,198){\scriptsize $\ol{3}'$} 
\put(300,180){\circle*{3}} \put(298,170){\scriptsize $2'$} 
\put(300,184){\circle{8}} 
\put(300,187){\circle{14}} 

\put(140,40){\vector(3,-1){30}} \put(140,23){\scriptsize level 4 fusion
$D^{(4)}$}
\multiput(240,25)(25,0){4}{\circle{10}} 
\multiput(240,20)(25,0){4}{\circle*{3}}
\put(238,11){\scriptsize $1'$} \put(263,11){\scriptsize $2'$}
\put(288,11){\scriptsize $3'$}
\put(313,11){\scriptsize $\ol{3}'$}
\end{picture}

\vspace{-.4in}
\caption{The adjacency graphs of the fused $A_5$ and $D_4$ models.}
\end{figure}

The adjacency graphs decompose into two groups for 
level 2 fusion. The symmetric $2\times 2$ fused face weights of the $A_5$
model are \cite{DJKMO:88}  
\begin{eqnarray}
& & \wt {W^A_{2 \times 2}}1313u
   ={{\cos (u+\lambda )\cos u }\over \sin \lambda },               \no\\
& & \no\\
& & \wt {W^A_{2 \times 2}}1333u
  = {{\sin (2 u )}\over {\sqrt{2} \sin \lambda }},                 \no\\
& & \no\\
& & \wt {W^A_{2 \times 2}}3131u
  = {{\sin (u+2 \lambda )\sin (u+ \lambda) }\over \sin^2 \lambda },\no\\
& & \no\\
& &  \wt {W^A_{2 \times 2}}3133u
  = {{\cos (u+2 \lambda )\cos (u+ \lambda) }\over \sin^2 \lambda },  \\
& & \no\\
& & \wt {W^A_{2 \times 2}}3333u
  ={{\cos \lambda }\over {\sin \lambda }},                        \no \\
& & \no\\
& & \wt {W^A_{2 \times 2}}1353u
  = {{\sin (u+ \lambda )\sin u }\over \sin \lambda },             \no \\
& & \no\\
& & \wt {W^A_{2 \times 2}}3135u
  = {{\sin (u- \lambda )\sin (u- 2 \lambda ) }\over \sin^2 \lambda }\no    
\end{eqnarray}
for group 1 and 

\bea
& & \wt {W^A_{2 \times 2}}2222u
  = {{\cos (u-\lambda )\cos u }\over {2 \sin^2 \lambda }},       \no \\
& & \no\\
& & \wt {W^A_{2 \times 2}}2224u
  = {{\sqrt{3} \sin (u-\lambda )\cos u }\over {2 \sin^2 \lambda }},\no\\
& & \no\\
& & \wt {W^A_{2 \times 2}}4422u
  = {{\sin (u-\lambda )\sin u }\over {2 \sin^2 \lambda }},            \\
& & \no\\
& & \wt {W^A_{2 \times 2}}4242u
  = {{ \cos (u-\lambda )\cos u }\over {2 \sin^2 \lambda }};        \no\\
& & \no\\
& & \wt {W^A_{2 \times 2}}2242u
  = -{{2\sqrt{3} \sin (u+2 \lambda )\sin u }\over { \sin^2 \lambda }}\no 
\end{eqnarray}
for group 2. These weights satisfy the following symmetries
\begin{eqnarray}
 \wt {W^A_{2 \times 2}}abcdu =
    \wt {W^A_{2 \times 2}}cbadu =
    \wt {W^A_{2 \times 2}}{6-a}{6-b}{6-c}{6-d}u  .
\eea

To obtain the fused cells we find a gauge transformation $g^A(a,b)$
for the $A$ paths and $g^D(a,b)$ for the $D$ paths such\vspace{-.3in} that
\be
\begin{picture}(40,40)(-5,19)
\setlength{\unitlength}{0.0110in}%
\thicklines
\multiput(5,5)(0,30){2}{\vector(1,0){13}} 
\multiput(25,5)(0,30){2}{\line(-1,0){10}} 
\multiput(5,5)(20,0){2}{\line(0,1){30}}   
\multiput(5,35)(20,0){2}{\vector(0,-1){18}} 
\put(1,2){\scriptsize $a$} \put(25,2){\scriptsize $b$} 
\put(25,35){\scriptsize $c$} \put(1,35){\scriptsize $d$} 
\put(26,18){\tiny $\mu$} \put(13,18){\tiny $$}
\end{picture} =\sum_{\mu'}g^A(d,a)\begin{picture}(40,40)(-5,19)
\setlength{\unitlength}{0.0110in}%
\thicklines
\multiput(5,5)(0,30){2}{\vector(1,0){13}} 
\multiput(25,5)(0,30){2}{\line(-1,0){10}} 
\multiput(5,5)(20,0){2}{\line(0,1){30}}   
\multiput(5,35)(20,0){2}{\vector(0,-1){18}} 
\put(1,2){\scriptsize $a$} \put(25,2){\scriptsize $b$} 
\put(25,35){\scriptsize $c$} \put(1,35){\scriptsize $d$} 
\put(26,18){\tiny $\mu'$} \put(11,18){\tiny $C_2$}
\end{picture} g^D(c,b)^{-1}_{\mu',\mu}=
\sum_{\mu'}g^D(d,a)_{\mu,\mu'}\begin{picture}(40,40)(-5,19)
\setlength{\unitlength}{0.0110in}%
\thicklines
\multiput(5,5)(0,30){2}{\line(1,0){10}} 
\multiput(25,5)(0,30){2}{\vector(-1,0){13}} 
\multiput(5,5)(20,0){2}{\line(0,1){30}}   
\multiput(5,35)(20,0){2}{\vector(0,-1){18}} 
\put(1,2){\scriptsize $a$} \put(25,2){\scriptsize $b$} 
\put(25,35){\scriptsize $c$} \put(1,35){\scriptsize $d$} 
\put(10.5,18){\tiny $C^t_2$} \put(-4,18){\tiny $\mu\!'$}  
\end{picture} g^A(c,b)^{-1}. 
\ee

\vspace{.2in}\noindent
The unitary
conditions then take the form given by
(\ref{eq:unitary1}) and (\ref{eq:unitary2}). Dividing the fused cells into
two groups, we find
\begin{eqnarray}
& & \begin{picture}(180,70)(0,10)
\multiput(30,60)(40,0){2}{\vector(1,0){10}}
 \multiput(50,60)(40,0){2}{\line(-1,0){10}}
\multiput(30,40)(40,0){2}{\vector(1,0){10}}
 \multiput(50,40)(40,0){2}{\line(-1,0){10}}
\multiput(30,60)(40,0){2}{\vector(0,-1){10}}
 \multiput(30,40)(40,0){2}{\line(0,1){10}}
\multiput(50,60)(40,0){2}{\vector(0,-1){10}}
 \multiput(50,40)(40,0){2}{\line(0,1){10}}
\multiput(30,20)(40,0){2}{\vector(1,0){10}}
 \multiput(50,20)(40,0){2}{\line(-1,0){10}}
\multiput(50,20)(40,0){2}{\vector(0,-1){10}}
 \multiput(50,0)(40,0){2}{\line(-1,0){10}}
\multiput(30,20)(40,0){2}{\vector(0,-1){10}}
 \multiput(30,0)(40,0){2}{\line(0,1){10}}
\multiput(30,0)(40,0){2}{\vector(1,0){10}}
 \multiput(50,0)(40,0){2}{\line(0,1){10}}
\put(15,30){$\left( \begin{array}{c} \\ \\ \end{array} \right.$} 
\put(90,30){$ \left. \begin{array}{c}\\ \\ \end{array} \right)\;\;$=}
\put(130,30){${1 \over \sqrt{2}} \left(\begin{array}{cc}
      -1 & 1 \\
      1 & 1     \end{array} \right) $ }
\multiput(25,58)(40,0){2}{\tiny 3}\multiput(50,60)(0,-40){2}{\tiny $3'$} 
\multiput(25,38)(40,0){2}{\tiny 1}\multiput(90,60)(0,-40){2}{\tiny
$\ol{3}'$}
\multiput(50,38)(40,0){2}{\tiny $1'$}  
\multiput(25,18)(40,0){2}{\tiny 3}
\multiput(25,-2)(40,0){2}{\tiny 5}
\multiput(50,-2)(40,0){2}{\tiny $1'$}  
\end{picture}    \nonumber \\
& & \begin{picture}(180,70)(0,20)
\multiput(30,60)(40,0){2}{\vector(1,0){10}}
 \multiput(50,60)(40,0){2}{\line(-1,0){10}}
\multiput(30,40)(40,0){2}{\vector(1,0){10}}
 \multiput(50,40)(40,0){2}{\line(-1,0){10}}
\multiput(30,60)(40,0){2}{\vector(0,-1){10}}
 \multiput(30,40)(40,0){2}{\line(0,1){10}}
\multiput(50,60)(40,0){2}{\vector(0,-1){10}}
 \multiput(50,40)(40,0){2}{\line(0,1){10}}
\multiput(30,20)(40,0){2}{\vector(1,0){10}}
 \multiput(50,20)(40,0){2}{\line(-1,0){10}}
\multiput(50,20)(40,0){2}{\vector(0,-1){10}}
 \multiput(50,0)(40,0){2}{\line(0,1){10}}
\multiput(30,20)(40,0){2}{\vector(0,-1){10}}
 \multiput(30,0)(40,0){2}{\line(0,1){10}}
\multiput(30,0)(40,0){2}{\vector(1,0){10}}
 \multiput(50,0)(40,0){2}{\line(-1,0){10}}
\put(15,30){$\left( \begin{array}{c} \\ \\ \end{array} \right.$} 
\put(90,30){$ \left. \begin{array}{c} \\ \\ \end{array}\right)\;\;$=}
\put(130,30){$\left( \begin{array}{cc}
      -1 & 1 \\
      1 & 1     \end{array} \right) $ }
\multiput(50,60)(40,0){2}{\tiny $1'$}
\multiput(25,58)(40,0){2}{\tiny 1}\multiput(50,38)(0,-40){2}{\tiny $3'$} 
\multiput(25,38)(40,0){2}{\tiny 3}\multiput(90,38)(0,-40){2}{\tiny
$\ol{3}'$}
\multiput(50,18)(40,0){2}{\tiny $1'$}
\multiput(25,18)(40,0){2}{\tiny 5}
\multiput(25,-2)(40,0){2}{\tiny 3}  
\end{picture}     \\
& & \begin{picture}(30,50)(0,-30)
\multiput(30,-20)(40,0){2}{\vector(1,0){10}}
 \multiput(50,-20)(40,0){2}{\line(-1,0){10}}
\multiput(30,-40)(40,0){2}{\vector(1,0){10}}
 \multiput(50,-40)(40,0){2}{\line(-1,0){10}}
\multiput(30,-20)(40,0){2}{\vector(0,-1){10}}
 \multiput(30,-40)(40,0){2}{\line(0,1){10}}
\multiput(50,-20)(40,0){2}{\vector(0,-1){10}}
 \multiput(50,-40)(40,0){2}{\line(0,1){10}}
\put(15,-30){$\left( \begin{array}{c} \\ \end{array} \right.$} 
\put(90,-30){$ \left. \begin{array}{c} \\ \end{array}\right)\;\;$=}
\put(130,-30){$\left( \begin{array}{cc} 1 & 1 \end{array} \right) $ }
\multiput(25,-20)(40,0){2}{\tiny 3}   
  \multiput(25,-42)(40,0){2}{\tiny 3}
\multiput(52,-20)(40,-20){2}{\tiny $\ol{3}'$} 
    \multiput(52,-42)(40,20){2}{\tiny $3'$}
\end{picture}  \nonumber
\end{eqnarray} 
for the first group and 
\begin{eqnarray}
& & \begin{picture}(180,50)(0,20)
\multiput(30,60)(40,0){2}{\vector(1,0){10}}
 \multiput(50,60)(40,0){2}{\line(-1,0){10}}
\multiput(30,40)(40,0){2}{\vector(1,0){10}}
 \multiput(50,40)(40,0){2}{\line(-1,0){10}}
\multiput(30,60)(40,0){2}{\vector(0,-1){10}}
 \multiput(30,40)(40,0){2}{\line(0,1){10}}
\multiput(50,60)(40,0){2}{\vector(0,-1){10}}
 \multiput(50,40)(40,0){2}{\line(0,1){10}}
\multiput(30,20)(40,0){2}{\vector(1,0){10}}
 \multiput(50,20)(40,0){2}{\line(-1,0){10}}
\multiput(50,20)(40,0){2}{\vector(0,-1){10}}
 \multiput(50,0)(40,0){2}{\line(-1,0){10}}
\multiput(30,20)(40,0){2}{\vector(0,-1){10}}
 \multiput(30,0)(40,0){2}{\line(0,1){10}}
\multiput(30,0)(40,0){2}{\vector(1,0){10}}
 \multiput(50,0)(40,0){2}{\line(0,1){10}}
\put(15,30){$\left( \begin{array}{c} \\ \\ \end{array} \right.$} 
\put(90,30){$ \left. \begin{array}{c} \\ \\ \end{array}\right)\;\;$=}
\put(130,30){${1 \over \sqrt{2}} \left( \begin{array}{cc} 
            1 & 1 \\
           1 & -1 \end{array} \right) $ }
\multiput(25,58)(40,0){2}{\tiny 2}  \multiput(50,60)(40,0){2}{\tiny $2'$} 
\put(25,38){\tiny 2}    \put(65,38){\tiny 2}
\multiput(50,38)(40,0){2}{\tiny $2'$}
\put(52,50){\tiny $3'$}  \put(92,50){\tiny $\ol{3}'$} 
\put(52,10){\tiny $3'$}  \put(92,10){\tiny $\ol{3}'$} 
\multiput(25,18)(40,0){2}{\tiny 4}  \multiput(50,20)(40,0){2}{\tiny $2'$} 
\put(25,-2){\tiny 2}    \put(65,-2){\tiny 2}
\multiput(50,-2)(40,0){2}{\tiny $2'$}  
\end{picture} \nonumber \\
&&\begin{picture}(180,70)(0,30)
\multiput(30,60)(40,0){2}{\vector(1,0){10}}
 \multiput(50,60)(40,0){2}{\line(-1,0){10}}
\multiput(30,40)(40,0){2}{\vector(1,0){10}}
 \multiput(50,40)(40,0){2}{\line(-1,0){10}}
\multiput(30,60)(40,0){2}{\vector(0,-1){10}}
 \multiput(30,40)(40,0){2}{\line(0,1){10}}
\multiput(50,60)(40,0){2}{\vector(0,-1){10}}
 \multiput(50,40)(40,0){2}{\line(0,1){10}}
\multiput(30,20)(40,0){2}{\vector(1,0){10}}
 \multiput(50,20)(40,0){2}{\line(-1,0){10}}
\multiput(50,20)(40,0){2}{\vector(0,-1){10}}
 \multiput(50,0)(40,0){2}{\line(-1,0){10}}
\multiput(30,20)(40,0){2}{\vector(0,-1){10}}
 \multiput(30,0)(40,0){2}{\line(0,1){10}}
\multiput(30,0)(40,0){2}{\vector(1,0){10}}
 \multiput(50,0)(40,0){2}{\line(0,1){10}}
\put(15,30){$\left( \begin{array}{c} \\ \\ \end{array} \right.$} 
\put(90,30){$ \left. \begin{array}{c} \\ \\ \end{array} \right)\;\;$=}
\put(130,30){${1 \over \sqrt{2}} \left( \begin{array}{cc}
               1 & -1 \\
               1 & 1     \end{array} \right) $ }
\multiput(25,58)(40,0){2}{\tiny 2}\multiput(50,60)(40,0){2}{\tiny $2'$} 
\put(25,38){\tiny 4}    \put(65,38){\tiny 4}
\multiput(50,38)(40,0){2}{\tiny $2'$}
\put(52,50){\tiny $3'$}  \put(92,50){\tiny $\ol{3}'$} 
\put(52,10){\tiny $3'$}  \put(92,10){\tiny $\ol{3}'$} 
\multiput(25,18)(40,0){2}{\tiny 4}\multiput(50,20)(40,0){2}{\tiny $2'$} 
\put(25,-2){\tiny 4}    \put(65,-2){\tiny 4}
\multiput(50,-2)(40,0){2}{\tiny $2'$}  
\end{picture}
\end{eqnarray}

\vspace{.45in}\noindent
for the second group.

These cells satisfy the unitary conditions 
(\ref{eq:unitary1}) and (\ref{eq:unitary2}). Hence,
from the intertwining relation
the $2\times 2$ fused $D$ face weights must be given \cite{Roch:90} in
terms of the $A$ face weights by
\be
\setlength{\unitlength}{0.0110in}%
\begin{picture}(222,90)(66,747)
\thicklines
\put(225,795){\line( 1, 0){ 30}}
\put(255,795){\line( 1,-1){ 30}}
\put(285,765){\line(-1, 0){ 30}}
\put(285,765){\line(-1,-1){ 30}}
\put(255,735){\line(-1, 0){ 30}}
\put(225,795){\line( 1,-1){ 30}}
\put(255,765){\line(-1,-1){ 30}}
\put(225,795){\line(-1, 0){ 30}}
\put(195,795){\line(-1,-1){ 30}}
\put(165,765){\line( 1, 0){ 30}}
\put(165,765){\line( 1,-1){ 30}}
\put(195,735){\line( 1, 0){ 30}}
\put(225,795){\line(-1,-1){ 30}}
\put(195,765){\line( 1,-1){ 30}}
\put(225,738){\circle*{6}}
\put(255,765){\circle*{6}}
\put(196,764){\circle*{6}}
\put(105,795){\line( 1,-1){ 30}}
\put(135,765){\line(-1,-1){ 30}}
\put(105,795){\line(-1,-1){ 30}}
\put( 75,765){\line( 1,-1){ 30}}
\put(180,765){\vector(-1, 0){  3}}
\put(270,765){\vector( 1, 0){  3}}
\put(240,795){\vector( 1, 0){  3}}
\put(243,735){\vector( 1, 0){  3}}
\put(207,735){\vector(-1, 0){  3}}
\put(210,795){\vector(-1, 0){  3}}
\multiput(180,780)(-0.42857,-0.42857){8}{\makebox(0.4444,0.6667){\sevrm .}}
\put(177,777){\vector(-1,-1){0}}
\multiput(210,780)(-0.42857,-0.42857){8}{\makebox(0.4444,0.6667){\sevrm .}}
\put(207,777){\vector(-1,-1){0}}
\multiput(240,750)(-0.42857,-0.42857){8}{\makebox(0.4444,0.6667){\sevrm .}}
\put(237,747){\vector(-1,-1){0}}
\multiput(210,750)(0.42857,-0.42857){8}{\makebox(0.4444,0.6667){\sevrm .}}
\put(213,747){\vector( 1,-1){0}}
\multiput(180,750)(0.42857,-0.42857){8}{\makebox(0.4444,0.6667){\sevrm .}}
\put(183,747){\vector( 1,-1){0}}
\multiput(240,780)(0.42857,-0.42857){8}{\makebox(0.4444,0.6667){\sevrm .}}
\put(243,777){\vector( 1,-1){0}}
\multiput(270,780)(0.42857,-0.42857){8}{\makebox(0.4444,0.6667){\sevrm .}}
\put(273,777){\vector( 1,-1){0}}
\multiput(270,750)(-0.42857,-0.42857){8}{\makebox(0.4444,0.6667){\sevrm .}}
\put(267,747){\vector(-1,-1){0}}
\put( 66,759){\makebox(0,0)[lb]{\raisebox{0pt}[0pt][0pt]{\twlrm \sc $a'$}}}
\put(105,717){\makebox(0,0)[lb]{\raisebox{0pt}[0pt][0pt]{\twlrm \sc $b'$}}}
\put(105,795){\makebox(0,0)[lb]{\raisebox{0pt}[0pt][0pt]{\twlrm \sc $d'$}}}
\put(147,762){\makebox(0,0)[lb]{\raisebox{0pt}[0pt][0pt]{\twlrm \sc $=$}}}
\put(189,723){\makebox(0,0)[lb]{\raisebox{0pt}[0pt][0pt]{\twlrm \sc $b'$}}}
\put(255,723){\makebox(0,0)[lb]{\raisebox{0pt}[0pt][0pt]{\twlrm \sc $b'$}}}
\put(288,759){\makebox(0,0)[lb]{\raisebox{0pt}[0pt][0pt]{\twlrm \sc $c'$}}}
\put(186,792){\makebox(0,0)[lb]{\raisebox{0pt}[0pt][0pt]{\twlrm \sc $d'$}}}
\put(258,795){\makebox(0,0)[lb]{\raisebox{0pt}[0pt][0pt]{\twlrm \sc $d'$}}}
\put( 90,762){\makebox(0,0)[lb]{\raisebox{0pt}[0pt][0pt]{\twlrm \sc
$W^D_{\!2,2}$}}}
\put(210,762){\makebox(0,0)[lb]{\raisebox{0pt}[0pt][0pt]{\twlrm \sc
$W^A_{\!2,2}$}}}
\put(138,762){\makebox(0,0)[lb]{\raisebox{0pt}[0pt][0pt]{\twlrm \sc $c'$}}}
\put(156,762){\makebox(0,0)[lb]{\raisebox{0pt}[0pt][0pt]{\twlrm \sc $a'$}}}
\put(225,798){\makebox(0,0)[lb]{\raisebox{0pt}[0pt][0pt]{\twlrm \sc $d$}}}
\put(123,780){\makebox(0,0)[lb]{\raisebox{0pt}[0pt][0pt]{\twlrm \sc
$\beta$}}}
\put( 84,741){\makebox(0,0)[lb]{\raisebox{0pt}[0pt][0pt]{\twlrm \sc
$\alpha$}}}
\put( 81,780){\makebox(0,0)[lb]{\raisebox{0pt}[0pt][0pt]{\twlrm \sc
$\mu$}}}
\put(120,741){\makebox(0,0)[lb]{\raisebox{0pt}[0pt][0pt]{\twlrm \sc
$\nu$}}}
\put(273,744){\makebox(0,0)[lb]{\raisebox{0pt}[0pt][0pt]{\twlrm \sc
$\nu$}}}
\put(171,744){\makebox(0,0)[lb]{\raisebox{0pt}[0pt][0pt]{\twlrm \sc
$\alpha$}}}
\put(168,780){\makebox(0,0)[lb]{\raisebox{0pt}[0pt][0pt]{\twlrm \sc
$\mu$}}}
\put(270,780){\makebox(0,0)[lb]{\raisebox{0pt}[0pt][0pt]{\twlrm \sc
$\beta$}}}
\end{picture}
\ee 
\vspace*{1.0cm}

\noindent
independent of the spin $d$. Inserting the fused 
weights $W^A_{\!2,\!2}$ and fused cells
given above we find the fused face weights of the $D_4$ model.
Explicitly, for the first group the nonzero weights read 
\begin{eqnarray}
& & \wt{W^D_{2 \times 2}}ababu=
 {\cos \lambda \over \sin \lambda }      \\
& & \no\\
& & \wt{W^D_{2 \times 2}}abcbu=
 {\sin (2 \lambda -2 u) \over \sin \lambda }      \\
& & \no\\
& & \wt{W^D_{2 \times 2}}abacu=
 {\sin 2 u \over \sin \lambda }.      
\end{eqnarray}
where $a,b,c,d=1',3',\ol{3}'$ are distinct. For the second group the face
weights are
\begin{eqnarray}
& & W^D_{2 \times 2} \left( \begin{array}{ccc}  2' & \mu &  2'  \\
                                                \mu  &   &  \mu \\ 
                                                 2' & \mu &  2'      
\end{array} \left.\begin{array}{c}
                              \\ \\ \end{array}\!\!\! \right| u \right)
 =\sin 2 \lambda \left(1 - {\sin (2 u - \lambda ) \over \sin \lambda
}\right)    \\
& & \no\\
& & W^D_{2 \times 2} \left( \begin{array}{ccc}  2' & \mu  & 2'  \\
                                                \nu  &    & \nu \\ 
                                                  2' & \mu & 2'      
\end{array} \left.\begin{array}{c}
                              \\ \\ \end{array} \!\!\! \right| u \right)
 =\sin (2 \lambda )\left(1 + {\sin (2 u - \lambda ) \over \sin \lambda
}\right)   \\
& & \no\\
& & W^D_{2 \times 2} \left( \begin{array}{ccc}  2'  & \nu & 2'  \\
                                              \nu   &     & \mu \\ 
                                                2'  & \mu & 2' 
     \end{array} \left. \begin{array}{c}
                         \\ \\ \end{array}\!\!\!  \right| u \right)
 ={\cos u \cos (u - \lambda ) \over \sin \lambda }    \\
& & \no\\
& & W^D_{2 \times 2} \left( \begin{array}{ccc}  2'  & \nu & 2'  \\
                                               \mu   &   & \nu \\ 
                                                2'  & \mu & 2' 
     \end{array} \left. \begin{array}{c}
                     \\ \\ \end{array}\!\!\! \right| u \right)
 ={\sin u \sin ( \lambda -u) \over \sin \lambda } 
\end{eqnarray}
where the bond variables $\mu\not=\nu=3',\ol{3}'$. 
It can be directly verified that these fused weights
satisfy the Yang-Baxter equation. In fact, the first group gives precisely
the face
weights of the critical 3-state CSOS model \cite{PearS}.  The second group
gives the
weights of the 8-vertex model at the Ising decoupling point.

The $2\times 2$ fused face weights of the
$D_4$ model have been obtained here via the intertwining relation. However,
precisely
the same results are obtained by following the fusion procedure presented
in 
Sections~3 and 5. Although we have concentrated in this article on fusion
of the
classical \ade models, the affine \ade and dilute \ade models
\cite{WaNiSe:92,Roch:92}
can also be fused using these methods. Similarly, the methods are easily
extended to
fuse the elliptic off-critical $D$ models.